\documentclass{aastex}
\usepackage{spr-astr-addons}

\RequirePackage{color}

\begin{document}

\title{The effect of two-temperature post-shock accretion
flow on the linear polarization pulse in magnetic cataclysmic variables}
\shorttitle{mCV two-temperature flows}
\shortauthors{Sarty, Saxton and Wu}

\author{Gordon E. Sarty}
\affil{Department of Physics and Engineering Physics, University of Saskatchewan, 
Saskatoon, Saskatchewan S7N 5E2, Canada}
\and 
\author{Curtis J. Saxton} \and \author{Kinwah Wu}
\affil{Mullard Space Science Laboratory, University College London, 
  Holmbury St.~Mary, Dorking, Surrey RH5 6NT, UK}

\begin{abstract}
The temperatures of electrons and ions in the post-shock accretion region of a magnetic cataclysmic variable (mCV) will be equal at sufficiently high mass flow rates or for sufficiently weak magnetic fields.
At lower mass flow rates or in stronger magnetic fields, efficient cyclotron cooling will cool the electrons faster than the electrons
can cool the ions
and a two-temperature flow will result. Here we investigate the differences in polarized
radiation expected from mCV post-shock accretion columns modeled with one- and two-temperature hydrodynamics.
In an mCV model with one accretion region, a magnetic field \raisebox{-0.2em}{$\stackrel{>}{\sim}$} 30 MG and a specific mass flow rate of $\sim$0.5 g cm$^{-2}$ s$^{-1}$, along with a relatively generic geometric orientation of the system, we find that in the ultraviolet either a single linear polarization pulse per binary orbit or two pulses per binary orbit can be expected, depending on the accretion
column hydrodynamic structure (one- or two-temperature) modeled. Under conditions where the physical flow is two-temperature,
one pulse per orbit is predicted from a single accretion region where a one-temperature model predicts two pulses.
The intensity light curves show
similar pulse behavior but there is very little difference between the circular polarization
    predictions of one- and two-temperature models. Such discrepancies indicate that it is important to model some
aspect of two-temperature flow in indirect imaging procedures, like Stokes imaging, especially at the edges of extended
accretion regions, were the specific mass flow is low, and especially for ultraviolet data.
\end{abstract}


\section{Introduction}

Magnetic cataclysmic variables (mCVs), consisting of the classes known as polars and intermediate polars, 
  are composed of a Roche-lobe-filling M type main sequence star in orbit about a magnetic white dwarf 
  (see \citet{warner95} for a review). 
Mass is lost through the inner Lagrangian point, $L_{1}$, and flows toward the magnetosphere of the white dwarf 
  either predominately in a stream (polars) or after forming a truncated accretion disk 
  circulating around the white dwarf (intermediate polars).  
In either case, the ionized gas follows magnetic field lines to the surface of the white dwarf 
  after the gas reaches the magnetosphere where the magnetic pressure exceeds the gas ram pressure. 
Upon reaching the white dwarf surface the gas will be essentially at ``free fall'', 
  with highly supersonic velocities. 
The abrupt stop of the radial inflow near the surface of the white dwarf  
  leads to the formation of a shock, 
  which heats the inflowing material  \citep{fabian76,king79,lamb79,wu00}.   
The hot subsonic post-shock flow settles gradually onto the white dwarf,  
  and cools via emitting bremsstrahlung X-rays and optical/infra-red cyclotron radiation.

The hydrodynamic structure of the post-shock settling flow is determined 
  by radiative and particle energy processes, which are essentially characterized by   
  the bremsstrahlung cooling time $t_{\mbox{\small br}}$, 
  the cyclotron cooling time $t_{\mbox{\small cy}}$, 
  the electron-ion energy-exchange time  $t_{\mbox{\small ei}}$, 
  the electron-electron collisional time $t_{\mbox{\small ee}}$, and
  the ion-ion collisional time $t_{\mbox{\small ii}}$ \citep{king79,lamb79,imamura96,saxton05}. 
For weakly magnetic systems (with $B \sim 10^6~$G or weaker) with accretion luminosities 
$10^{31}-10^{33}$ erg cm$^{-2}$ s$^{-1}$, typical of mCVs, 
  $t_{\mbox{\small cy}} > t_{\mbox{\small br}} > t_{\mbox{\small ei}} > t_{\mbox{\small ee}}$.  
As the strength of the magnetic field increases to $B \: \raisebox{-0.2em}{$\stackrel{>}{\sim}$} \: 10^7$~G, 
  cyclotron cooling may dominate bremsstrahlung cooling, $t_{\mbox{\small cy}} < t_{\mbox{\small br}}$.   
For sufficiently strong magnetic fields and low specific accretion rates, $t_{\mbox{\small cy}}$
  is so short ($t_{\mbox{\small cy}} < t_{\mbox{\small ei}}$) that collisions between electrons and ions 
  cannot maintain an equal temperature between the two types of particles. 
The accretion flow is therefore in a two-fluid regime which requires a two-temperature (2T) hydrodynamic description. 
A strong magnetic field can also result in a situation where 
  cyclotron radiative loss is so rapid ($t_{\mbox{\small cy}} < t_{\mbox{\small ee}}$) 
  that electron-electron collisions are 
  not efficient enough to maintain a Maxwellian distribution.  
In the extreme situation where $t_{\mbox{\small cy}} < t_{\mbox{\small ii}}$, 
  the accretion flow is no longer hydrodynamic.   

Previous calculations of cyclotron radiation from the post-shock settling flow in mCVs have either assumed a uniform density and temperature (\citet{chanmugam81}; \citet{meggitt82};\\ \citet{barrett84};\\ \citet{wickramasinghe85};\\ \citet{wu88,wu89}) 
or a one-temperature (1T) structure (\citet{wu90,wu92}; \citet{potter02}). 
However, detailed 1D calculations of the hydrodynamic structure of a post-shock accretion column
   that self-consistently include cyclotron and bremsstrahlung cooling clearly 
   show that a 2T structure is to be expected in many physical situations relevant to mCVs 
   (\citet{imamura96}; \citet{woelk96}; \citet{saxton01}; \citet{wu03}; \citet{saxton05}). 

Here we have computed and compared the cyclotron radiation
from a cylindrical post-shock accretion column, with a uniform cylindrical radial structure, assuming both a 1T hydrodynamic
structure and a 2T hydrodynamic structure. The resulting cyclotron spectra for a grid of three white dwarf masses
(0.5, 0.7 and 1.0 $M_{\odot}$), three magnetic field strengths (10, 30 and 50 MG) and two mass flow rates
($10^{16}$ and $10^{14}$ g s$^{-1}$) were computed for various viewing inclination angles. 
For each case, using the computed viewing-angle dependent cyclotron spectra, 
Johnson bandpass \citep{johnson65,bessell90} filtered light curves over an orbital period were computed 
for a mCV with an orbital inclination of 45$^{\circ}$ and with the given accretion column at a co-latitude
of 30$^{\circ}$. 

This work is organized as follows. In \S \ref{methods} 
we outline the hydrodynamic formulation used to determine the density and temperature structure of the post-shock flows 
  and the radiative transfer through the ionized accreting gas. 
In \S \ref{results} we present the results of the polarized radiative transfer calculations, and 
  in \S \ref{discussion} we examine the differences between the spectro-polarization properties of the emission 
  from 1T and 2T accretion flows and discuss their implications. 
Concluding remarks are made in \S \ref{conclusion}.

\section{Method}\label{methods} 

Calculations were done for 36 cases: 3 values of magnetic field strength ($B_{7}$ = 1, 3 and 5) $\times$
3 values of white dwarf mass ($M_{\mbox{\tiny WD}} = 0.5$, 0.7 and 1.0 $M_{\odot}$) $\times$
2 values of mass flow rate ($\dot{M} = 10^{16}$ and $10^{14}$ g s$^{-1}$) $\times$ 2 values of column
hydrodynamic structure (1T and 2T). Here $B_{7}$ is the magnetic field strength at the accretion spot in units
of $10^{7}$ G. In all cases the area of the accretion spot, $A$, was taken to be constant at $2 \times 10^{14}$ cm$^{2}$.
This value for the area is based on the consideration that
\begin{equation}\label{e1}
A = 4 \pi R_{\mbox{\tiny WD}}^{2}\: \chi
\end{equation}
where $R_{\mbox{\tiny WD}}$ is the white dwarf radius and $\chi \sim 10^{-4}$ was chosen as being a typical value for the
fraction of white dwarf surface occupied by the accretion spot.
The white dwarf radius, in turn, is taken to be \citep{nauenberg72}:
\begin{equation}\label{e2}
\frac{R_{\mbox{\tiny WD}}}{R_{\odot}} = \frac{0.0225}{\mu_{\mbox{\tiny WD}}} 
\frac{\sqrt{1 - (M_{\mbox{\tiny WD}}/M_{3})^{4/3}}}{(M_{\mbox{\tiny WD}}/M_{3})^{1/3}}
\end{equation}
where $M_{3} = 5.816 M_{\odot}/\mu_{\mbox{\tiny WD}}^{2}$ is the Chandrasekhar mass
limit and $\mu_{\mbox{\tiny WD}}$ is the mean molecular weight of the white dwarf material
taken to be equal to 2.00. The chosen spot area, $A$, gave specific accretion rates,
$\dot{m} = \dot{M}/A$, of 50.0 g cm$^{-2}$s$^{-1}$ for $\dot{M} = 10^{16}$ and 0.5 g cm$^{-2}$s$^{-1}$ for $\dot{M} = 10^{14}$.  With this constraint
on the specific mass accretion rate, 1T and 2T hydrodynamic accretion column structures
were computed as described in \S \ref{hydro}. The computed electron temperature and density
in the column, as a function of height, was then determined for the center of each cube, of individual side length $\Delta h$, in
a computational grid-of-cubes. Each individual cube was modeled as having a uniform temperature and density at
the center values. Ray-tracing was then done
through the computational grid-of-cubes as described in \S \ref{ray}.
The radiation transfer calculation method for the polarized cyclotron radiation
from the accretion column emerging from the top of the grid-of-cubes for various column orientations
(viewing angles) is also described in \S \ref{ray}. The method for computing a 
broadband light curve over the course of a binary orbit is given in \S \ref{orbit}.

\subsection{Hydrodynamics}\label{hydro}

The hydrodynamic structure of the post-shock accretion column was computed 
  following the methods given by \citet{saxton05}. 
The hydrodynamic structure for the two-temperature flow 
  is specified by the conservation of mass, momentum, energy of the electrons and energy of the ions, respectively, 
  as the height above the white dwarf, $z$, changes according to the relationships explicitly given by
\begin{equation}
u \frac{d \rho}{d z} + \rho \frac{d u}{d z} = 0
\end{equation}
\begin{equation}
\frac{d P_{\mbox{\small e}}}{d z} + \frac{d P_{\mbox{\small i}}}{d z} + \rho u
\frac{d u}{d z} = 0
\end{equation}
\begin{equation}
u \frac{d P_{\mbox{\small e}}}{d z} - \gamma \frac{u P_{\mbox{\small e}}}{\rho}
\frac{d \rho}{d z} = (\gamma - 1)(\Gamma_{\mbox{\small ei}} - \Lambda)
\end{equation}
\begin{equation}
u \frac{d P_{\mbox{\small i}}}{d z} - \gamma \frac{u P_{\mbox{\small i}}}{\rho}
\frac{d \rho}{d z} = -(\gamma - 1) \Gamma_{\mbox{\small ei}}
\end{equation}
  where $u$ is the flow velocity, $\rho$ is the density, $P_{\mbox{\small e}}$ is the electron partial pressure, 
  $P_{\mbox{\small i}}$ is the ion partial pressure, $\Gamma_{\mbox{\small ei}}$ is the electron-ion energy exchange rate, 
  $\Lambda$ is the electron cooling rate and $\gamma$ is the adiabatic index. 
An ideal gas law for mono-atomic species was assumed for both electrons and ions so that $\gamma = 5/3$. 
The total pressure is $P = P_{\mbox{\small e}} +P_{\mbox{\small i}}$ and the number densities, $n_{j}$, 
  are given by $P_{j} = n_{j} k T_{j}$ where $k$ is the Boltzmann constant and $j =$ e or i. 
The cooling rate is the sum of the cooling due to bremsstrahlung and cyclotron emissions, 
   $\Lambda = [\Lambda_{\mbox{\small br}} + \Lambda_{\mbox{\small cy}}] \propto [(1/t_{\mbox{\small br}}) + (1/t_{\mbox{\small cy}})]$. 
In relative terms, the cyclotron cooling is most efficient in the hot, less dense plasma just below the shock 
  and the bremsstrahlung is more efficient at the base of the accretion column 
  where the the gas is dense. 
The electron-ion energy exchange rate $\Gamma_{\mbox{\small ei}} \propto 1/t_{\mbox{\small ei}}$ 
  is a function of the electron and ion number densities and temperatures.
One-temperature flow is similarly given by 
\begin{equation}
u \frac{d \rho}{d z} + \rho \frac{d u}{d z} = 0
\end{equation} 
\begin{equation}
\frac{d P}{d z} + \rho u
\frac{d u}{d z} = 0
\end{equation} 
\begin{equation}
u \frac{d P}{d z} - \gamma \frac{u P}{\rho}
\frac{d \rho}{d z} = -(\gamma - 1) \Lambda.
\end{equation}

The flow is assumed to have a constant density, pressure and temperature across the column area $A$. 
Note that variation of structure across the column can substantially affect the radiation produced \citep{wu90,wu92} 
  but here we assume the simpler structure so that the effects of the 1T and 2T structures, alone, 
  in the production of radiation may be discerned. 
In the formulation given by \citet{saxton05} (see also \citet{imamura96})  
   the pressure ratio of the electrons to ions at the shock, $\sigma_{\mbox{\small s}} = P_{\mbox{\small e,s}}/P_{\mbox{\small i,s}}$ 
   was taken to be a free parameter. Here we take $\sigma_{s} = \overline{Z}_{\odot} = 1.099$, 
   which corresponds to an assumption of solar metallicity.

\begin{figure*}
\begin{center}
\includegraphics[scale=0.85]{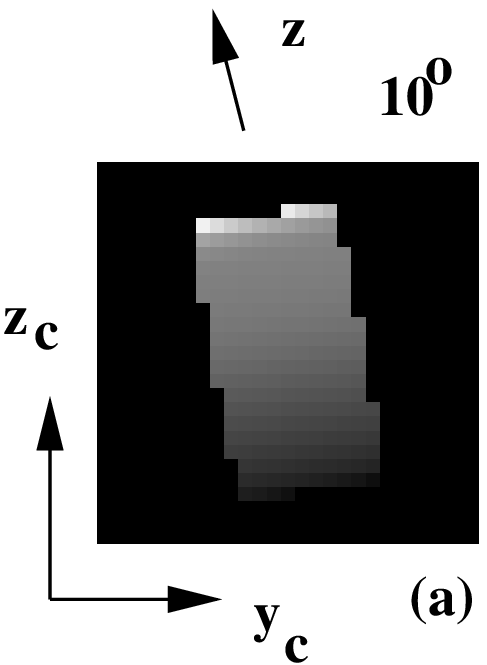}\hspace*{2em}
\includegraphics[scale=0.85]{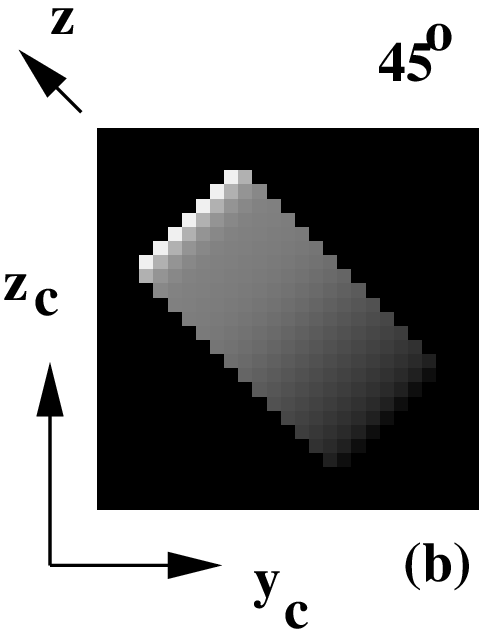}\hspace*{2em} 
\includegraphics[scale=0.85]{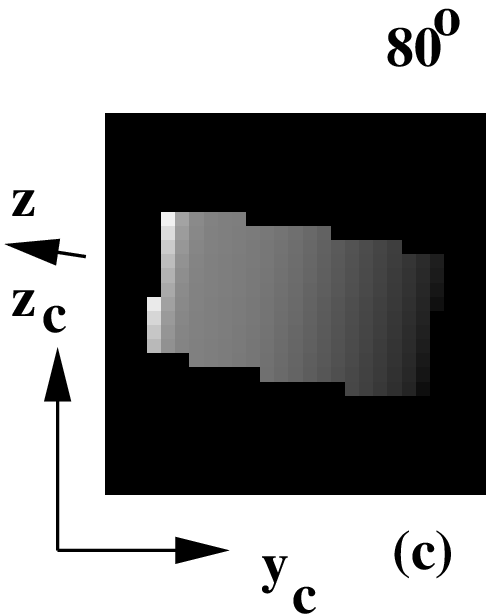}

\vspace*{2em}
\includegraphics[scale=0.85]{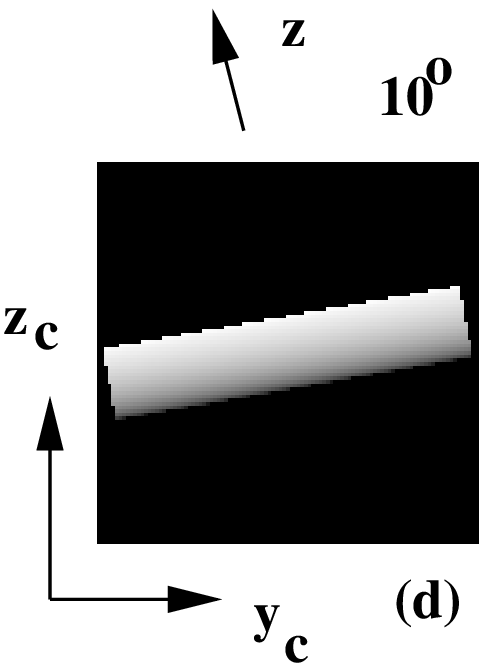}\hspace*{2em}
\includegraphics[scale=0.85]{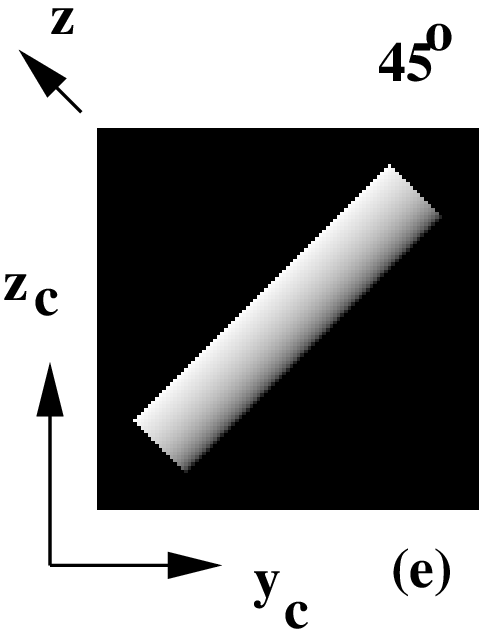}\hspace*{2em}
\includegraphics[scale=0.85]{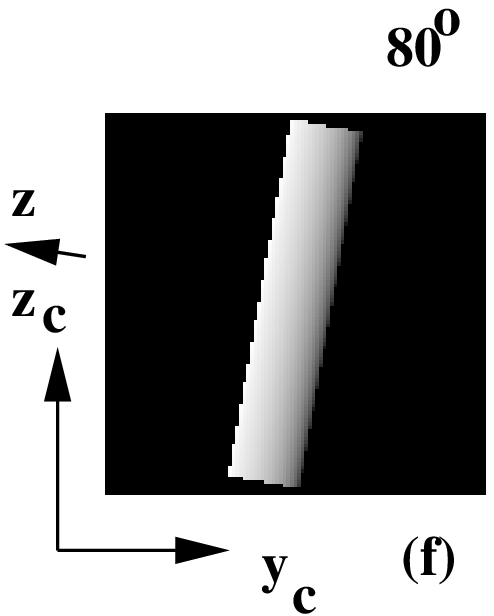}
\end{center}
\caption{The orientation of the post-shock accretion column in the ray-tracing
computational grid of cubes for a tall and short column for various viewing angles (indicated
above the images). The greyscale
within the columns is proportional to the computed electron temperature with white being hot.
(The greyscale is relative to the hottest temperature for each column; the temperatures of the two different columns
cannot be compared directly to each other simply by looking at the greyscale.)  The
side view, in section through the middle, of the columns are shown and the columns are circular in cross-section. Both columns
are from the $M_{\mbox{\tiny WD}} = 1.0 M_{\odot}$ cases. The tall column, shown in panels (a)-(c),
is from the 2T calculation with $B_{7} = 1.0$ and $\dot{M} = 10^{14}$ g s$^{-1}$ ($\dot{m} = 0.5$ g cm$^{-2}$s$^{-1}$). The short
column, shown in panels (d)-(f), is from the 2T calculation with $B_{7} = 1.0$ and $\dot{M} = 10^{16}$ g s$^{-1}$ ($\dot{m} = 50$ g cm$^{-2}$s$^{-1}$).
The height of the tall column is $30.2 \times 10^{6}$ cm and the height of the short
column is $3.2 \times 10^{6}$ cm; both columns have a spot area $A = 0.20 \times
10^{15}$ cm$^{2}$. Both columns are divided into 20 strata along the
$z$ direction with a resulting grid size $N_{t} \times N \times N = 13 \times 29 \times 29$ for the
tall column and $N_{t} \times N \times N = 101 \times 107 \times 107$ for the short column. Both grids have the
same number of temperature bins in the $z$ direction but the aspect ratio of the short column required a much larger
grid, and corresponding computation time, for the ray tracing radiative transfer calculations.
The surrounding black boxes in the images represent the ray-tracing computational grid of cubes (the individual cubes are the size of
pixels in these images) and ray tracing was
always done in the $z_{c}$ direction; the computational box was always viewed from above when summing the emission from
the individual cubes.  \label{figgrid}}
\end{figure*}

The results of the hydrodynamic calculations for the column profile 
  were transferred to a computational grid-of-cubes of size $N_{t} \times N \times N$ as the column was rotated in that grid. 
More exactly, the temperatures and densities of a radially uniform cylinder 
  were computed at the center of each cube in the grid and those  values were taken to be uniform over the individual cube. 
The distance between cube centers on the grid, 
  $\Delta h$, was uniform in the $x_{c}$, $y_{c}$ and $z_{c}$ grid directions and was determined 
  by dividing the computed column (shock) height, $h_{s}$, by a fixed number of strata 
  so that the spatial resolution in the $z$ direction of the column was fixed. 
The number of strata was set at 20 for all cases except in six (unimportant, it turned out) cases 
  where it resulted in a grid size so large that subsequent computations were impractically long. 
Those six cases were the high mass flow rate ($\dot{M} = 10^{16}$ g s$^{-1}$), $M_{\mbox{\tiny WD}} = 0.5 M_{\odot}$ cases 
  where 5 strata were used and where it turned out that 
  there were no differences between the 1T and 2T cases. 
With the number of strata fixed, the computational grid-of-cubes size was determined as follows. 
The extent of the grid in the $y_{c}$ and $z_{c}$ directions was taken as $\lceil c \rceil$ 
  (meaning round $c$ up to the nearest full multiple of $\Delta h$) 
  where $c = \sqrt{h_{s}^{2} + d^{2}}$ and $d = 2 \sqrt{A/\pi}$ where $A$ is the column spot area of $2 \times 10^{14}$ cm$^{2}$. 
The thickness of the grid in the $x_{c}$ direction was $\lceil d \rceil$. 
With the grid sized this way, the accretion column could be rotated completely within the grid and the ray-tracing 
  through the column was always done vertically through the grid, in the $z_{c}$ direction, 
  with the computed radiation emerging from the top $x_{c}y_{c}$ surface of the computational grid-of-cubes as
  shown in Fig.~\ref{figgrid}. 
The resulting grid size, in number of grid cubes, was then $N_{t} \times N \times N$.
With the shorter columns this size could be quite large, on the order of $300 \times 300 \times 300$. 

\subsection{Polarized radiative transfer}\label{ray}

After the values of electron temperature and number density, $T_{e}(z)$ and $n_{e}(z)$, 
  were determined for the $z$ (height above the white dwarf surface) value at the center of each cube in the computational grid, 
  the optical depth parameter ${\cal S}$ for each cube was computed according to
\begin{equation}
{\cal S}(z) = 2.01 \times 10^{5} \left( \frac{\Delta h}{10^{5} \mbox{cm}} \right)
\left( \frac{n_{e}(z)}{10^{16} \mbox{cm}^{-3}} \right) \left(\frac{3 \times 10^{7} \mbox{G}}{B} \right)
\end{equation}
   where $B$ is the magnetic field strength\\ \citep{wickramasinghe85}. 
Finally rays were traced from the bottom to the top of each $(x_{c},y_{c})$ column of  the computational grid 
  using a finite difference approximation (i.e. $dz_{c} = \Delta h$) of
\begin{equation}
\frac{d}{dz_{c}} \left[ \!\!\! \begin{array}{c} I_{\nu} \\ Q_{\nu} \\ U_{\nu} \\ V_{\nu} \end{array} \!\!\! \right] \!\! = \!\!
\left[ \!\!\! \begin{array}{c} \epsilon_{I,\nu} \\ \epsilon_{Q,\nu} \\ \epsilon_{U,\nu} \\ \epsilon_{V,\nu} \end{array} \!\!\! \right] +
\left[ \!\! \begin{array}{cccc} -\kappa_{\nu} & -q_{\nu} & 0 & -v_{\nu} \\
                           -q_{\nu} & -\kappa_{\nu} & -f_{\nu} & 0 \\
                           0 & f_{\nu} & -\kappa_{\nu} & -h_{\nu} \\
                           -v_{\nu} & 0 & h_{\nu} & -\kappa_{\nu} \end{array} \!\! \right] \!\!
\left[ \!\!\! \begin{array}{c} I_{\nu} \\ Q_{\nu} \\ U_{\nu} \\ V_{\nu} \end{array} \!\!\! \right]
\end{equation} 
   \citep{pacholczyk75}, 
   where $I_{\nu}$, $Q_{\nu}$, $U_{\nu}$ and $V_{\nu}$ are the (frequency dependent) Stokes intensities 
   (see e.g. \citet{rybicki79}),
  $\kappa_{\nu}$, $q_{\nu}$, 0 and $v_{\nu}$ are the respective corresponding absorptivities, 
  $f_{\nu}$  and $h_{\nu}$ are the Faraday mixing coefficients 
  and $\epsilon_{I,\nu}$, $\epsilon_{Q,\nu}$, $\epsilon_{U,\nu}$ and $\epsilon_{V,\nu}$ are the corresponding emissivities. 
The absorptivities and emissivities are related by Kirchhoff's law as the accretion column's electrons 
  are in local thermodynamic equilibrium and the absorptivities and emissivities 
  are (essentially) functions of ${\cal S}$, $T_{e}$, $\nu$ and $\theta$ \citep{meggitt82}, 
  where $\theta$ is the angle between the magnetic field direction ($z$ axis) and the viewing direction ($z_{c}$ axis). 
The Stokes intensities emerging from the top of each column of cubes were then summed over the columns 
   to give the four total intensities emerging from the top of the grid-of-cubes.

From the Stokes intensities, the linear polarization $L_{\nu} = \sqrt{Q_{\nu}^{2} + U_{\nu}^{2}}/I_{\nu}$ 
  and circular polarization $C_{\nu} = V_{\nu}/I_{\nu}$ were computed for 18 angles between 0$^{\circ}$ and 90$^{\circ}$ 
  and for 25 frequencies, expressed in cyclotron harmonic number $H=\omega/\omega_{c}$, 
  satisfying $2 \leq H \leq 12$. Here $\omega_{c} = eB/m_{e}c$ is the cyclotron frequency 
  where $e$ and $m_{e}$ are the electron charge and mass, respectively, $B$ is the magnetic field strength and 
  $c$ is the speed of light. 

The ray tracing calculations were done on a Beowulf cluster of 20 computers with each node in the cluster
   working separately on one of the 30 cases with 20 cylinder strata. 
Each node consisted of a 1 GHz Pentium III processor with 512 MB of RAM. 
The smaller grids required under a day to complete while the larger grids took several weeks of CPU time.
The 6 large problems where 5 cylinder stata were used were parallelized to compute one viewing angle per node 
  to reduce computing time and even then it took about 4 weeks to complete the computations. 
The code was optimized to minimize the use of computer memory, 
  so that the entire computational cube's data did not have to be stored, but no directed effort 
  was given to optimize the speed of the computations; 
  an improvement in the interpolation of the column temperatures and densities 
  to the cube centers would likely increase the computational speed significantly.

\subsection{Orbital polarization light curves}\label{orbit}

To better assess the difference in observational predictions between a 1T and a 2T accretion column, 
  the intensity, linear and circular polarization of the emission from the column as seen from Earth over a binary orbital period was computed. 
For this the angle, $\theta$, between the viewing direction and the magnetic field at each angular position of the white dwarf 
  relative to the mass donor star, was needed. 
Taking the magnetic field to be normal to the white dwarf surface at the accretion column, the viewing inclination angle, $\theta$, 
   is given by
\begin{equation}
\cos \theta = \sin \iota \cos \alpha \sin \phi + \cos \iota \cos \phi
\end{equation}
   where $\iota$ is the inclination of the binary orbit and $\phi$ is the co-latitude of the accretion spot on the white dwarf surface. 
The orbital position is $\alpha$ with $\alpha = 0$ representing inferior conjunction of the mass donor star 
  (mass donor star between the distant observer and the white dwarf) 
  when the accretion spot is located directly below the white dwarf rotation axis as seen by the observer. 
We assume that the rotation of the white dwarf is synchronized with the orbital period as this is one of the defining features 
  for CVs classified as polars \citep{warner95}.

The angle of linear polarization, $\chi_{T}$, was also computed over a binary orbital period. 
The angle $\chi_{T} = \chi_{\nu} + \delta$ where $\chi_{\nu}$ is the linear polarization angle 
   relative to the computational grid-of-cubes and  $\delta$ is the angle that the grid-of-cubes needs to be rotated by, 
   about the $z_{c}$ axis, to line up with the projection of the accretion column on the sky. 
Taking the orbital axis as defining zero position angle we have
\begin{equation}
\tan 2 \chi_{\nu} = U_{\nu}/Q_{\nu}
\end{equation}
and
\begin{equation}
\tan (\delta - \pi/2) = \frac{\sin \iota \: \cos \phi - \cos \iota \: \cos \alpha \: \sin \phi}{\sin \alpha \: \sin \phi}.
\end{equation}
A nominal configuration of $\phi = 30^{\circ}$ and $\iota = 45^{\circ}$ was chosen 
   to illustrate a relatively wide range of viewing angles without the accretion spot disappearing behind the limb of white dwarf.

The cyclotron harmonic range between 2 and 12 corresponds to a wavelength range of 5350 to 892 \AA\ 
  for $B_{7} = 1$, 1785 to 298 \AA\ for $B_{7} = 3$ and, 1070 to 178 \AA\ for $B_{7} = 5$. 
With the exception of the higher frequencies in the $B_{7} = 5$ case, these wavelength ranges
  are covered by the LKJIRVBU filters of the Johnson system \citep{johnson65,bessell90}. 
The response of those filters, in cases where the 1T and 2T solutions were different, 
  was multiplied against the computed spectra to give Stokes vector bandpass intensities for each point 
  in the orbital light curve.

\section{Results}\label{results}

In all cases, substantial linear polarization was evident generally only at high viewing angles, where the accretion column is viewed side-on, so it was at high viewing angles where the largest differences between the linearly polarized emission from 1T and 2T columns would be seen.
Cyclotron harmonic spike-features were visible in the linear polarization spectra at low harmonic numbers,  
   where the emission is marginally optically thick at the spectral harmonic peaks 
   and is optically thin in between the spectral harmonic peaks.  
The linear polarization spectra became smooth at higher harmonic numbers and the cyclotron harmonic spike-features disappear,  
   since the emission becomes optically thin at shorter wavelengths. 
At the high accretion rate of $\dot{M} = 10^{16}$ g s$^{-1}$,  
  there was no obvious difference in the linear polarization spectra of the 1T and 2T models. 
This high accretion rate is sufficient for the
   electron-ion collisions to maintain approximately equal ion and electron temperatures, 
   in spite of radiative loss.  
Circular polarization was most prominent when viewed along the accretion column (magnetic field) at low viewing angles, so it was at low viewing angles where the largest differences between the circularly polarized emission from 1T and 2T columns would be seen.    
The relative difference between the predicted circularly polarized radiation by the 1T and 2T hydrodynamic formulations 
  was negligible, 
  as the accretion flow is essentially 1T given the efficient energy exchange between the electrons and the ions.   
At the lower mass flow rate of $\dot{M} = 10^{14}$ g s$^{-1}$, the situation is very different.  
Electron-ion collisions become less efficient at lower densities.   
For a sufficiently strong magnetic field, 
   electrons will lose energy rapidly via cyclotron radiation 
   but their collisions with ions are unable to maintain an equal temperature between the two particle species. 
For $B_{7} = 3$ and 5, the flow is essentially 2T, 
  and we find  
  very different predicted polarization for the 1T and 2T formulations.  
  
To investigate the observational difference in the predictions of a 1T and 2T accretion column for orbital light curves, 
  the Johnson filter with the highest frequency pass-band that contained the computed emission, where the 1T and 2T emission differences were the greatest, was chosen 
  for the $B_{7} = 1$ and 3 cases. For $B_{7} = 1$ the filter choice corresponds to the infra-red J band, 
  for $B_{7} = 3$ the choice corresponds to the U band.  
For the $B_{7} = 5$ case, the light curve at the frequency corresponding to the cyclotron harmonic 9.92 
  (the 20th of the 25 frequencies computed), in the ultraviolet beyond the U filter, was computed. 
These choices emphasize the differences between the 1T and 2T model predictions at the higher cyclotron harmonics. 
Since polars are not generally observed in the ultraviolet, V band orbital light curves for the $B_{7} = 3$ and 5 cases, 
  where the flow is essentially a 2T flow, were also computed.
 
\begin{figure*}
\begin{center}

\vspace*{-8em}
\begin{picture}(30,100)(0,0)\put(-5,-35){\scriptsize \bf \boldmath B$_{7} = 1$ \unboldmath}
\put(-5,-45){\scriptsize \bf J Filter}\end{picture}
\includegraphics[angle=-90,scale=0.16]{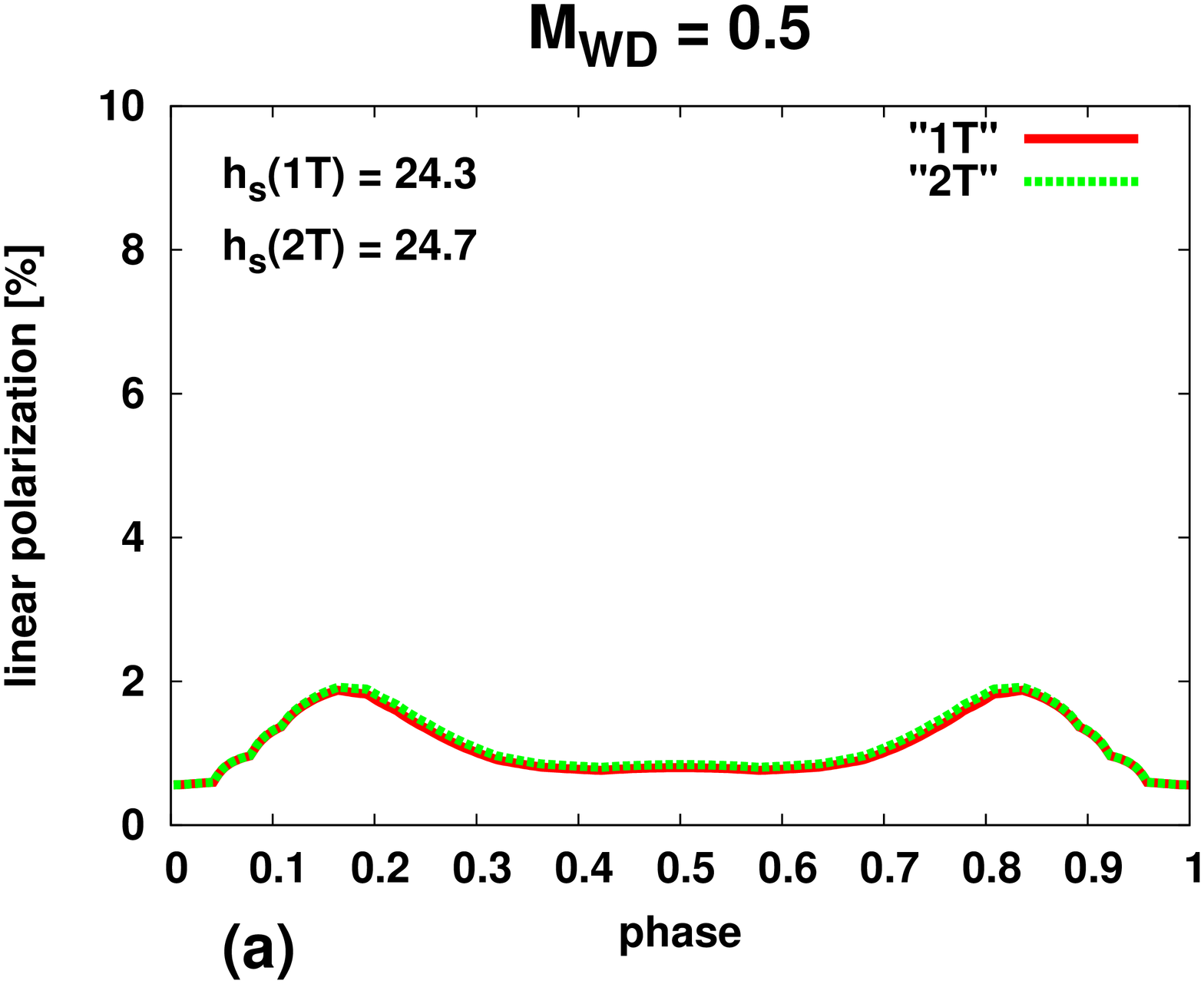}
\includegraphics[angle=-90,scale=0.16]{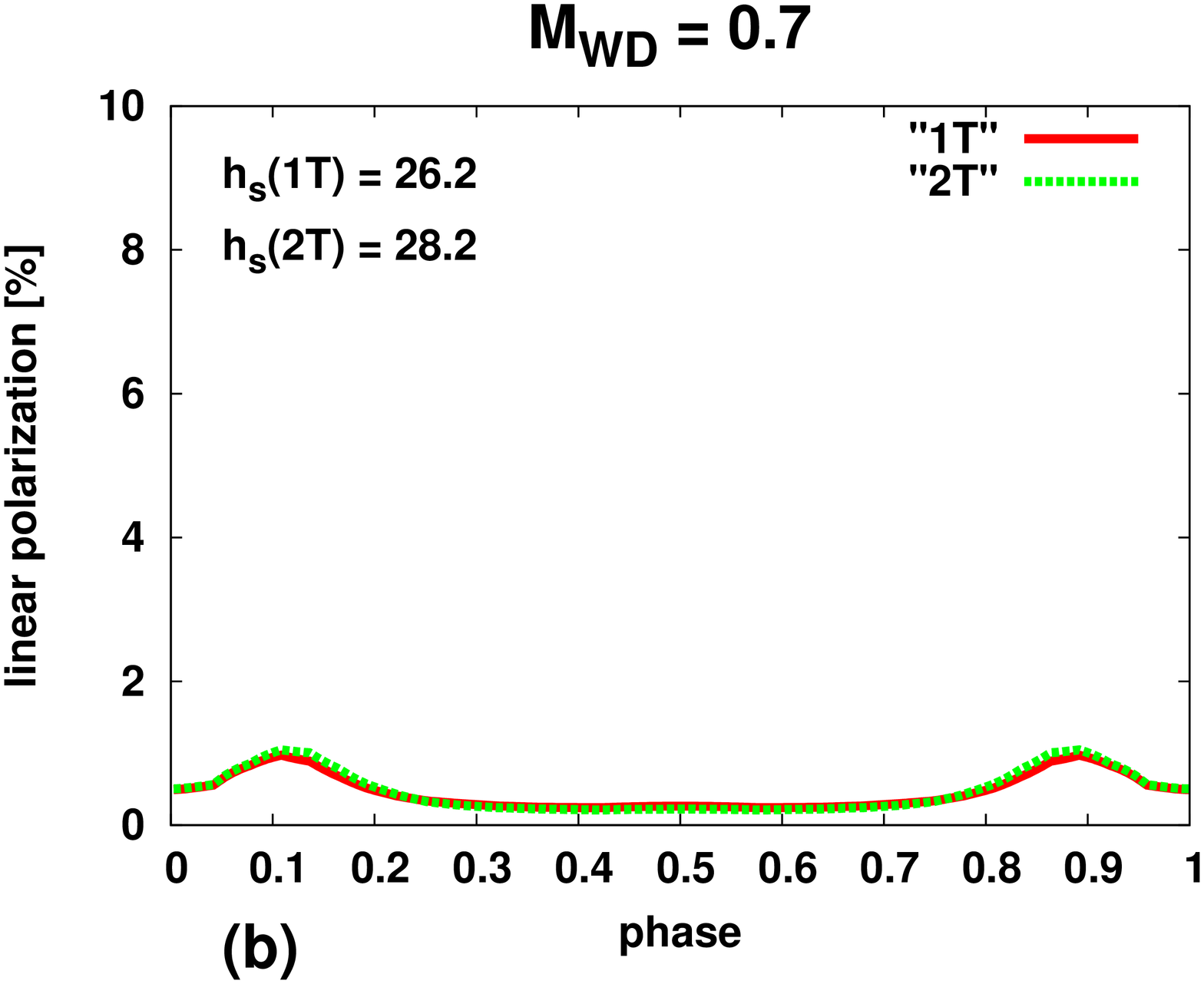}
\includegraphics[angle=-90,scale=0.16]{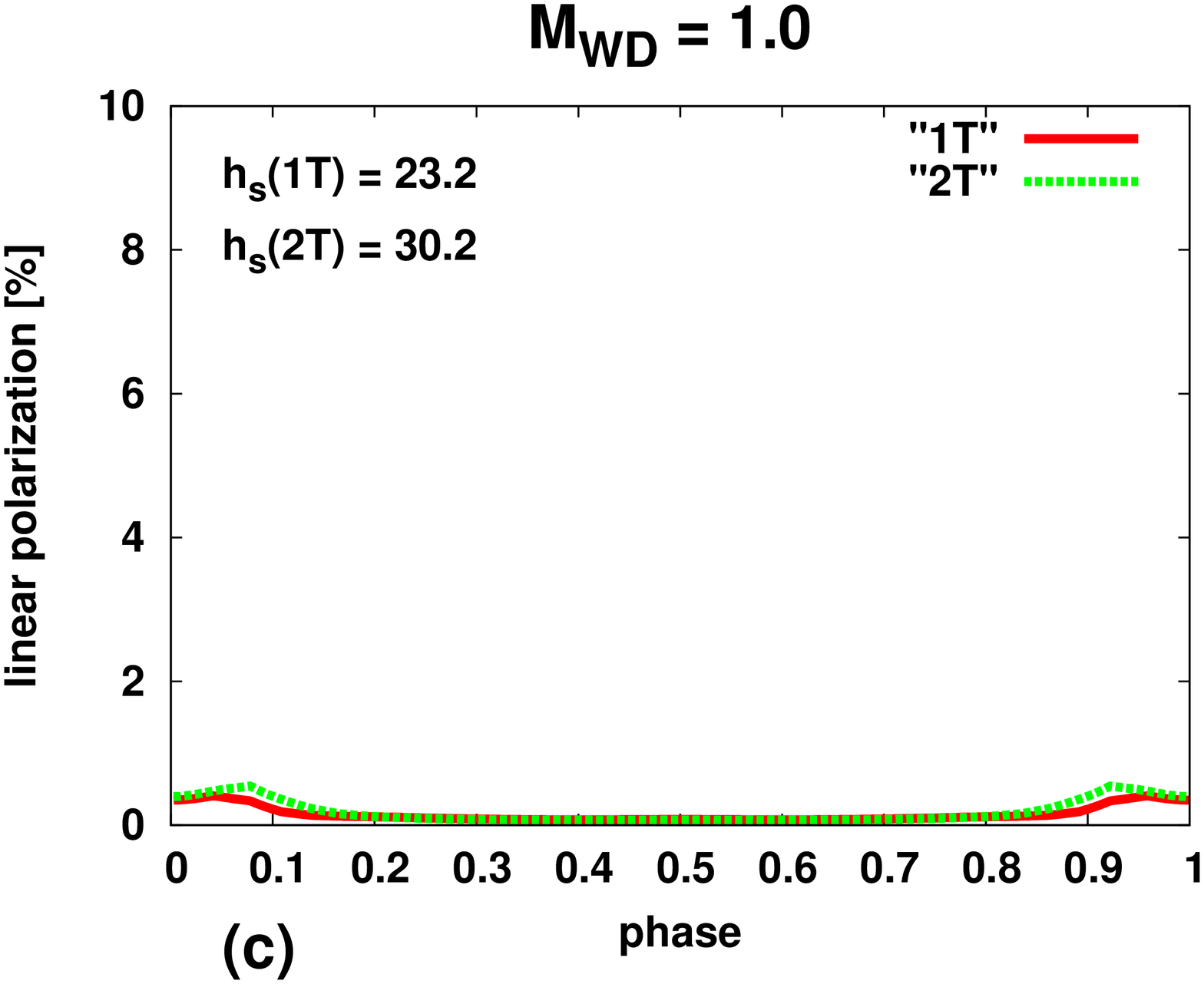}
\begin{picture}(25,100)(0,0)\put(0,-40){\scriptsize \bf \boldmath $\dot{M} = 10^{14}$ \unboldmath}
\end{picture}

\vspace*{-8em}
\begin{picture}(30,100)(0,0)\put(-5,-35){\scriptsize \bf \boldmath B$_{7} = 3$ \unboldmath}
\put(-5,-45){\scriptsize \bf U Filter}\end{picture}
\includegraphics[angle=-90,scale=0.16]{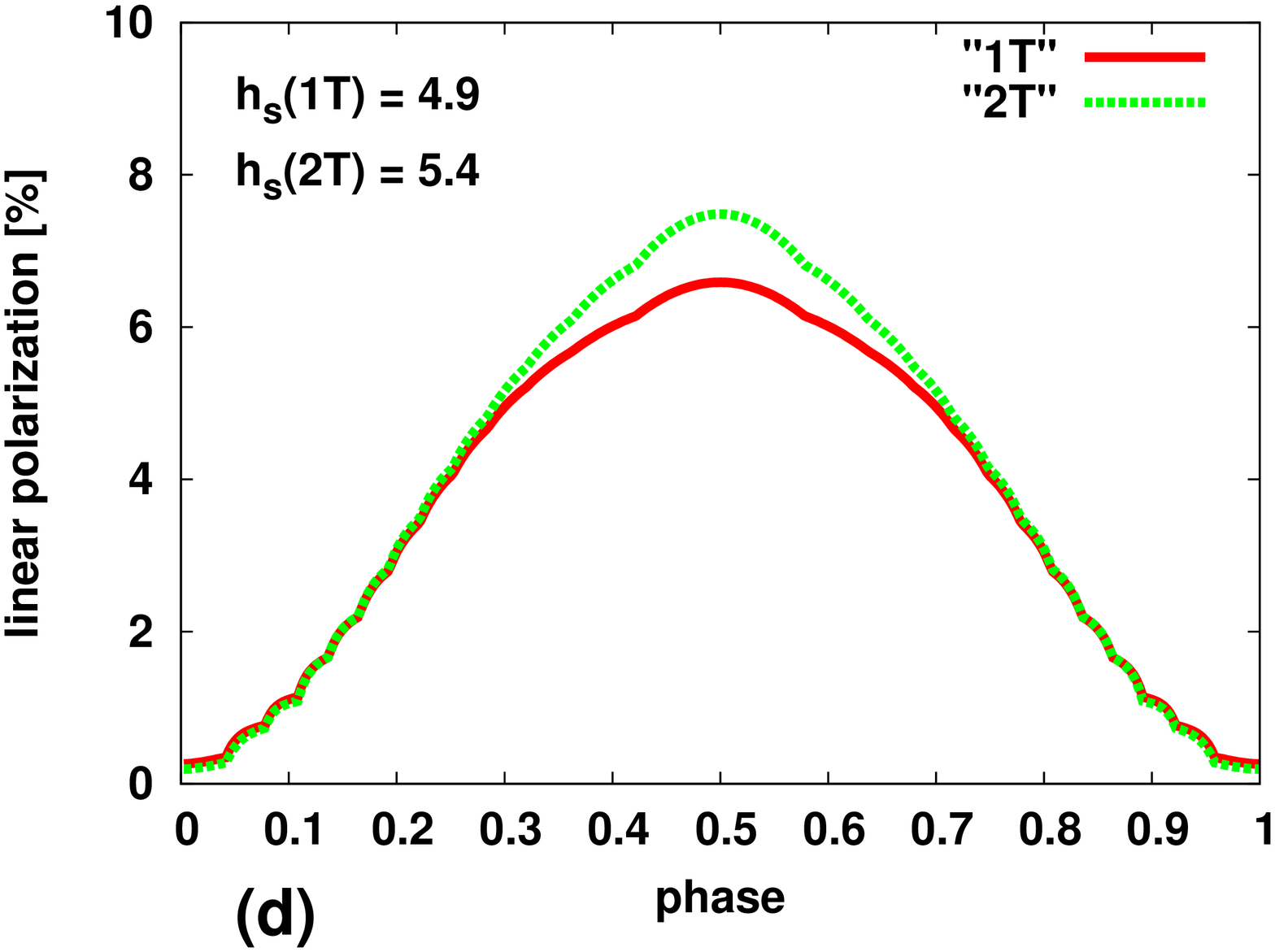}
\includegraphics[angle=-90,scale=0.16]{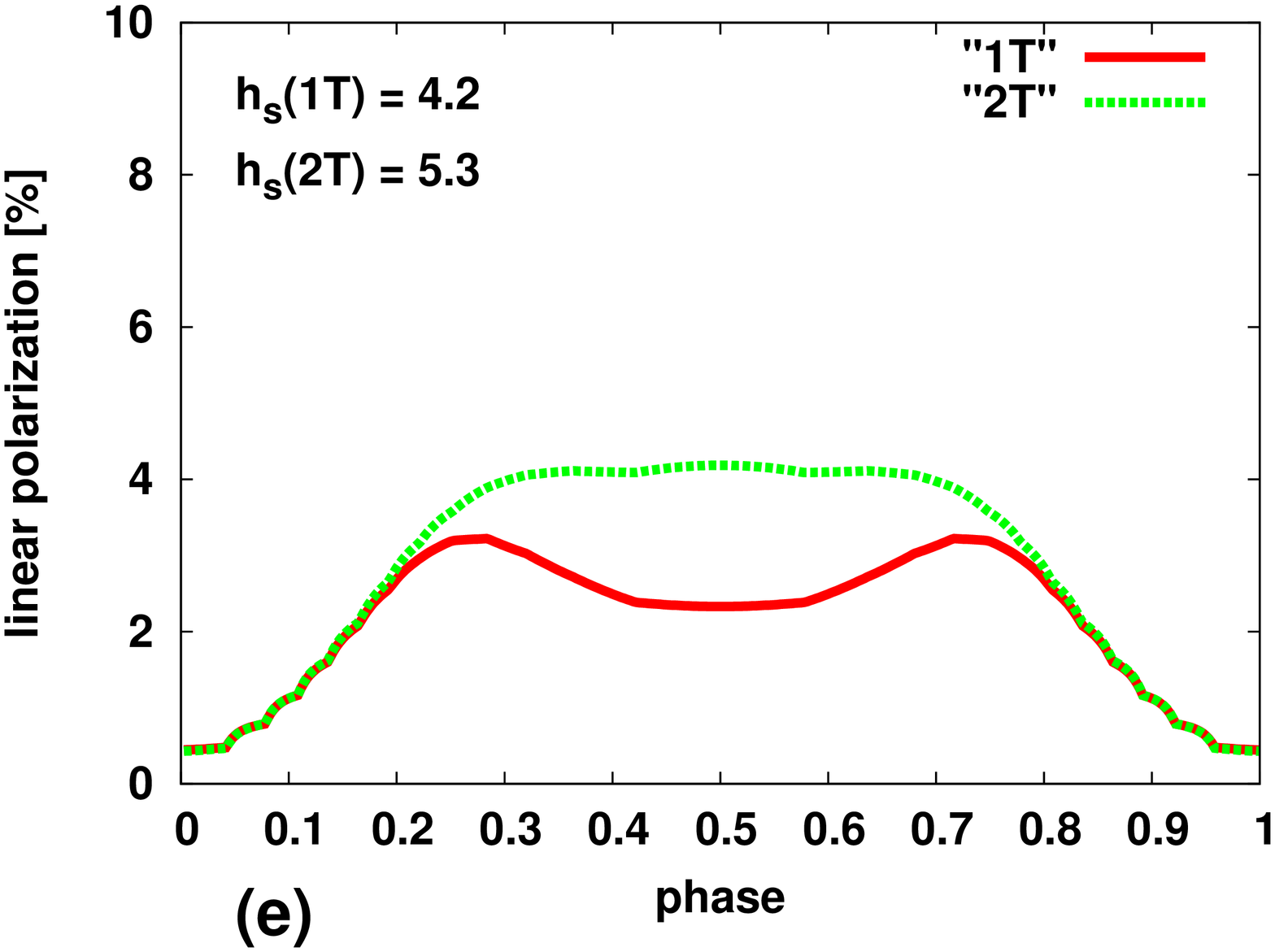}
\includegraphics[angle=-90,scale=0.16]{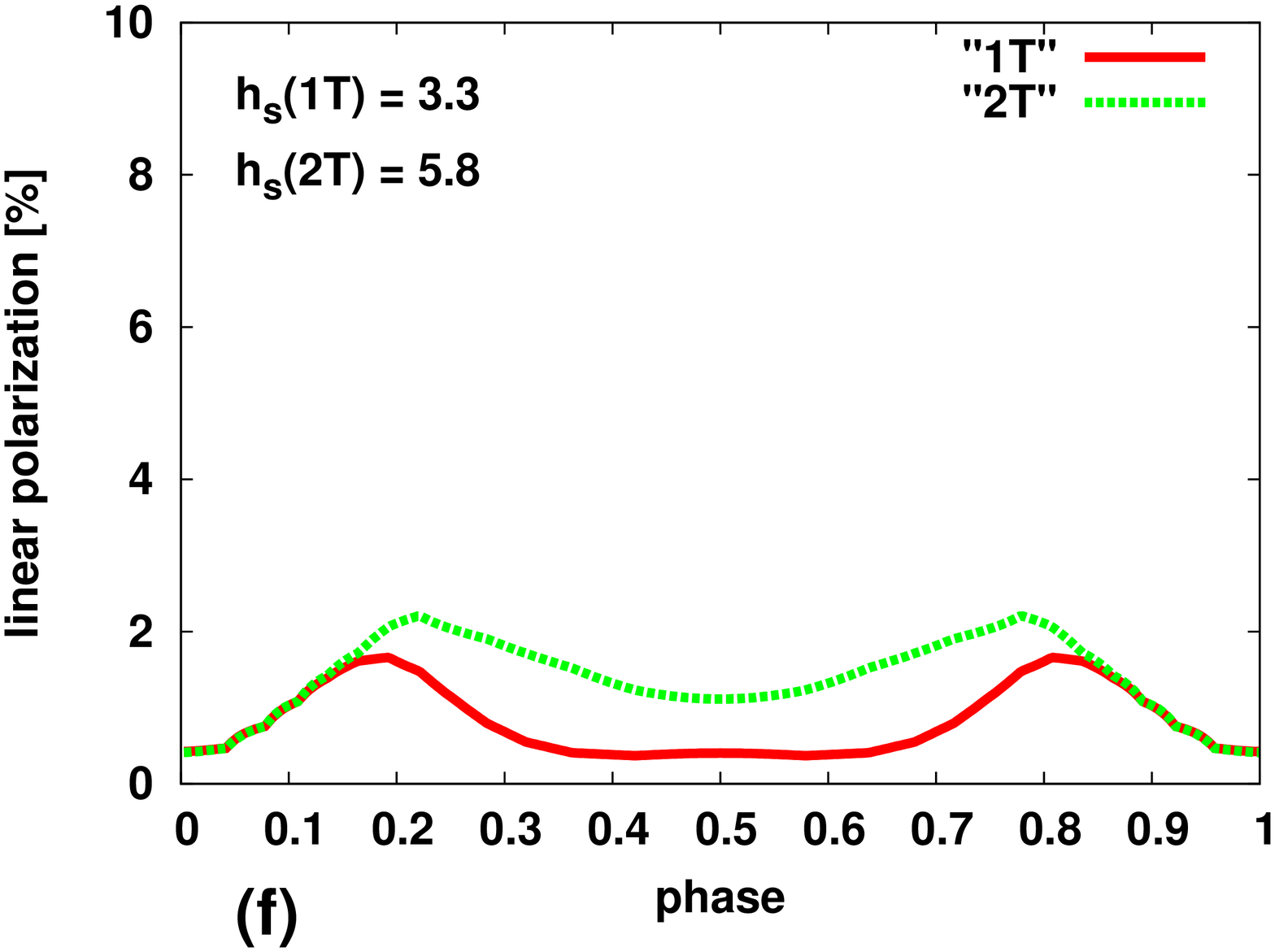}
\begin{picture}(25,100)(0,0)\put(0,-40){\scriptsize \bf \boldmath $\dot{M} = 10^{14}$ \unboldmath}
\end{picture}

\vspace*{-8em}
\begin{picture}(30,100)(0,0)\put(-5,-35){\scriptsize \bf \boldmath B$_{7} = 5$ \unboldmath}
\put(-5,-45){\scriptsize \bf H = 9.92}\end{picture}
\includegraphics[angle=-90,scale=0.16]{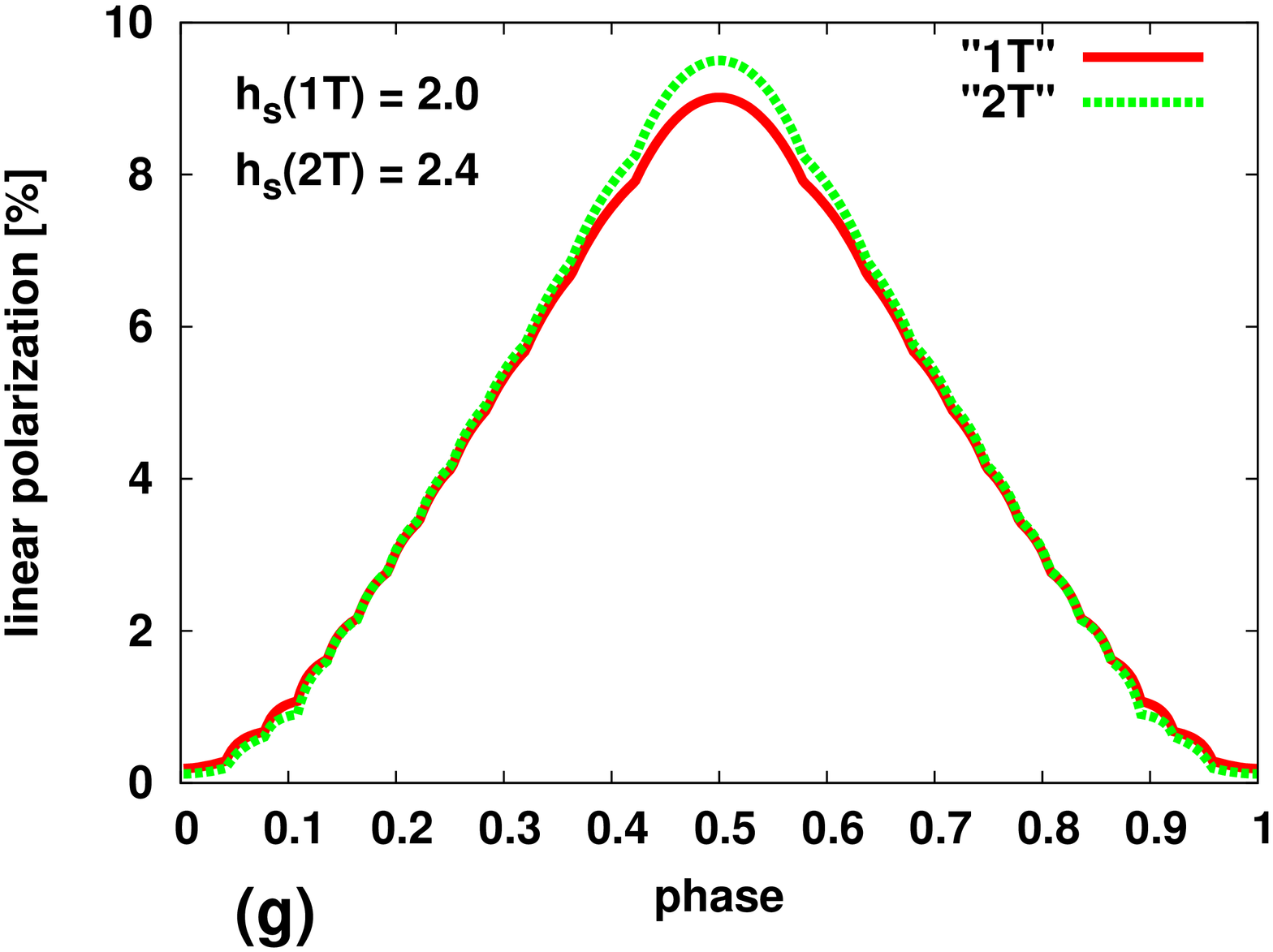}
\includegraphics[angle=-90,scale=0.16]{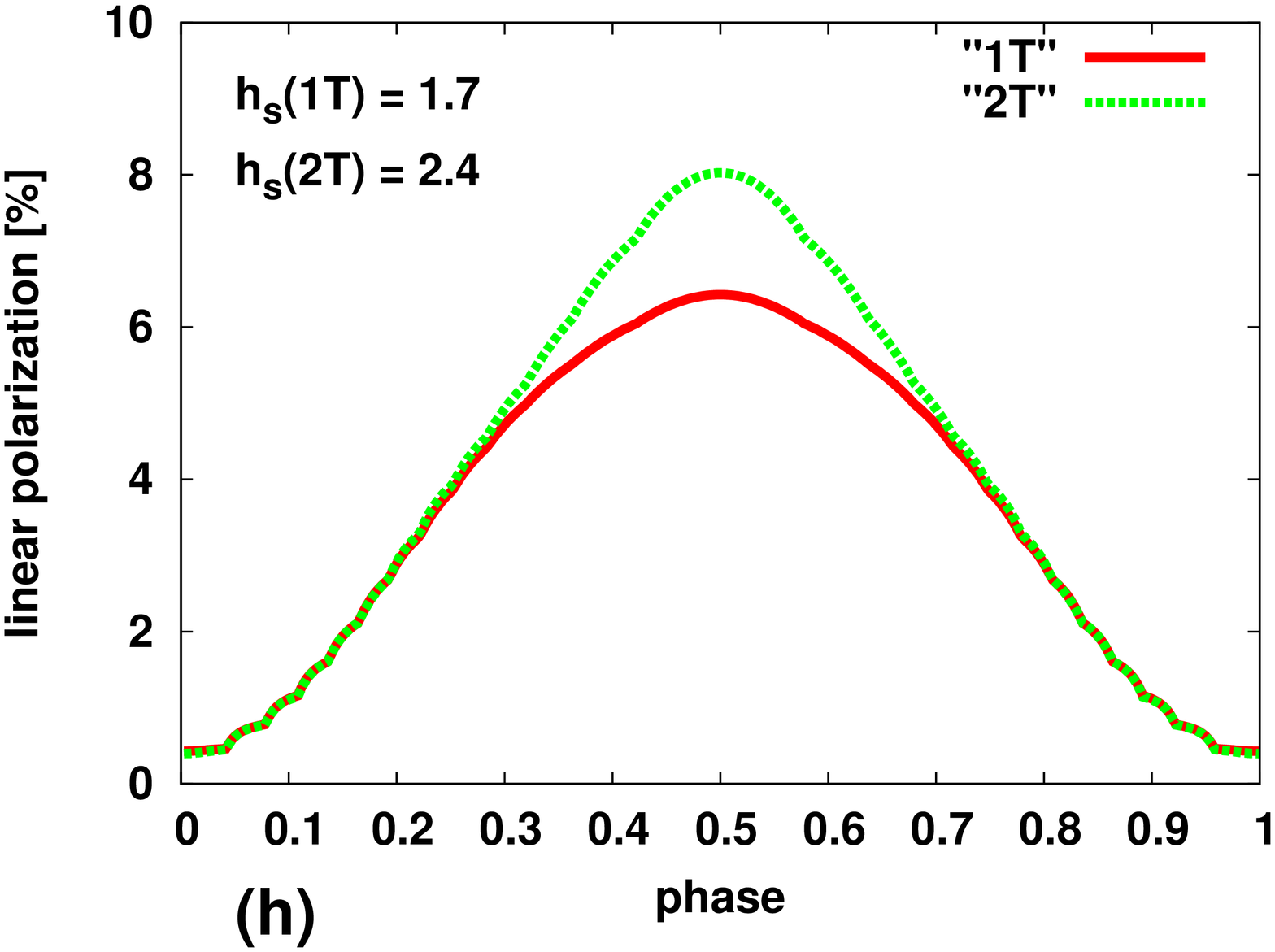}
\includegraphics[angle=-90,scale=0.16]{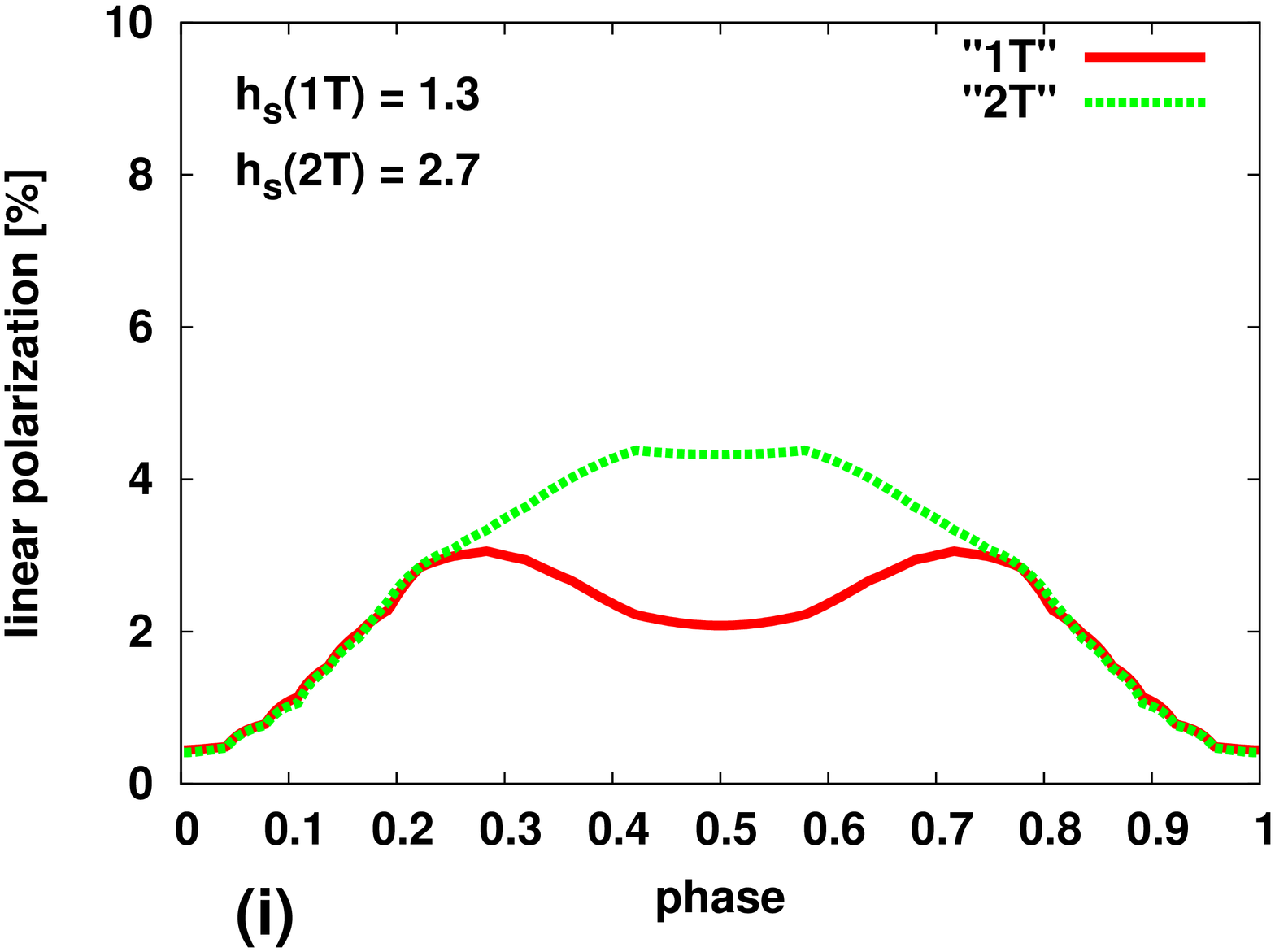}
\begin{picture}(25,100)(0,0)\put(0,-40){\scriptsize \bf \boldmath $\dot{M} = 10^{14}$ \unboldmath}
\end{picture}

\vspace*{-8em}
\begin{picture}(30,100)(0,0)\put(-5,-35){\scriptsize \bf \boldmath B$_{7} = 3$ \unboldmath}
\put(-5,-45){\scriptsize \bf V Filter}\end{picture}
\includegraphics[angle=-90,scale=0.16]{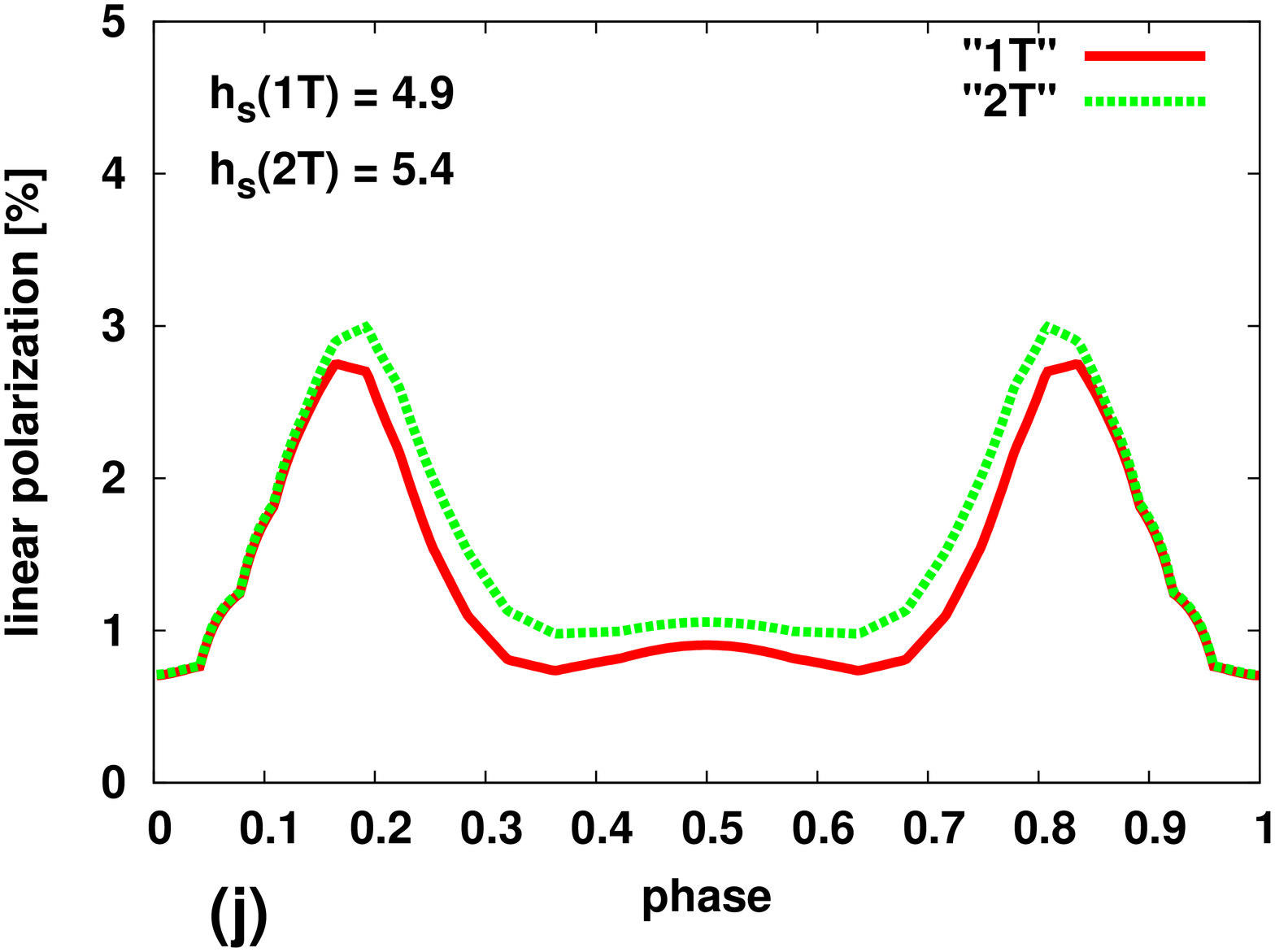}
\includegraphics[angle=-90,scale=0.16]{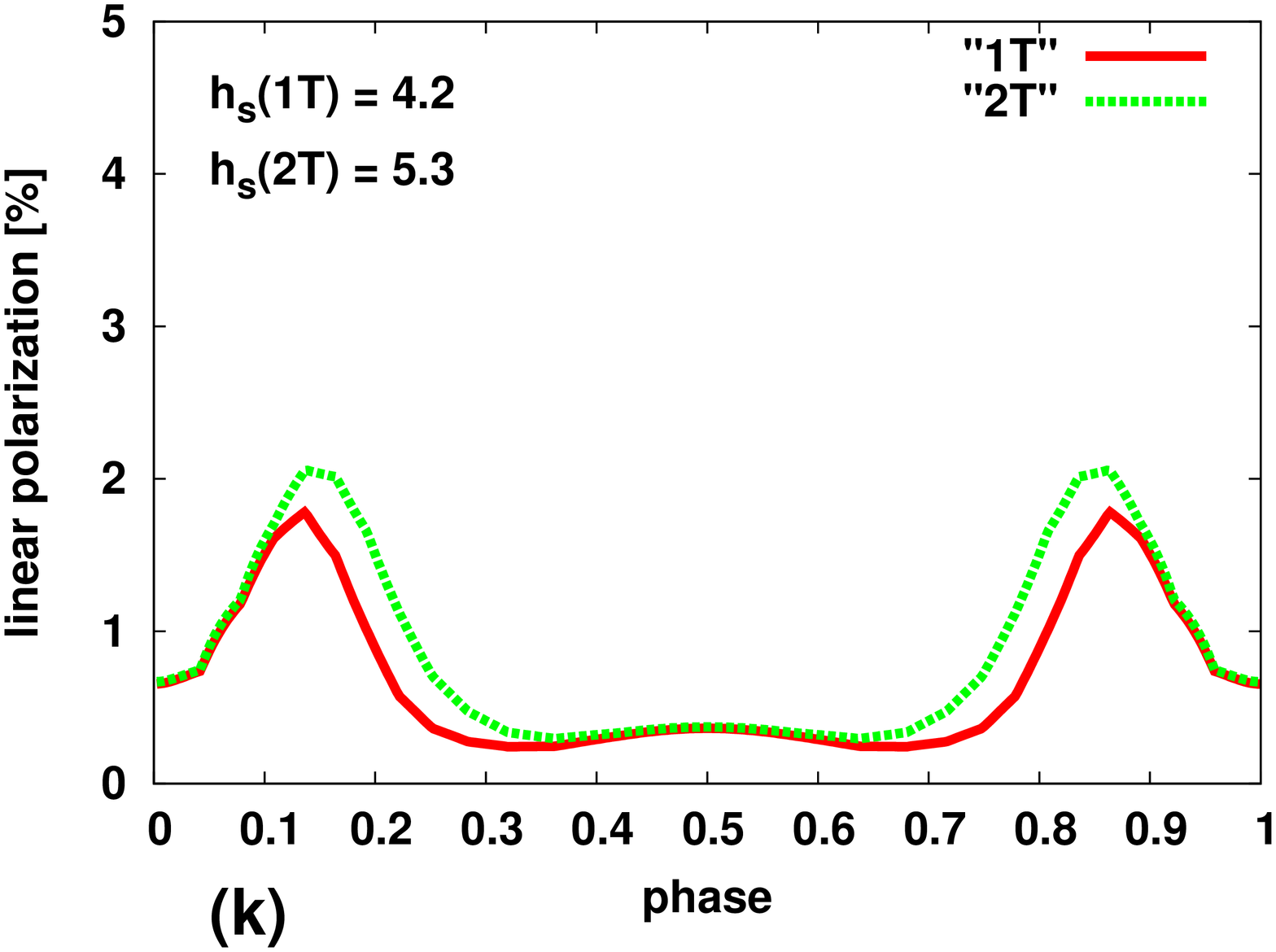}
\includegraphics[angle=-90,scale=0.16]{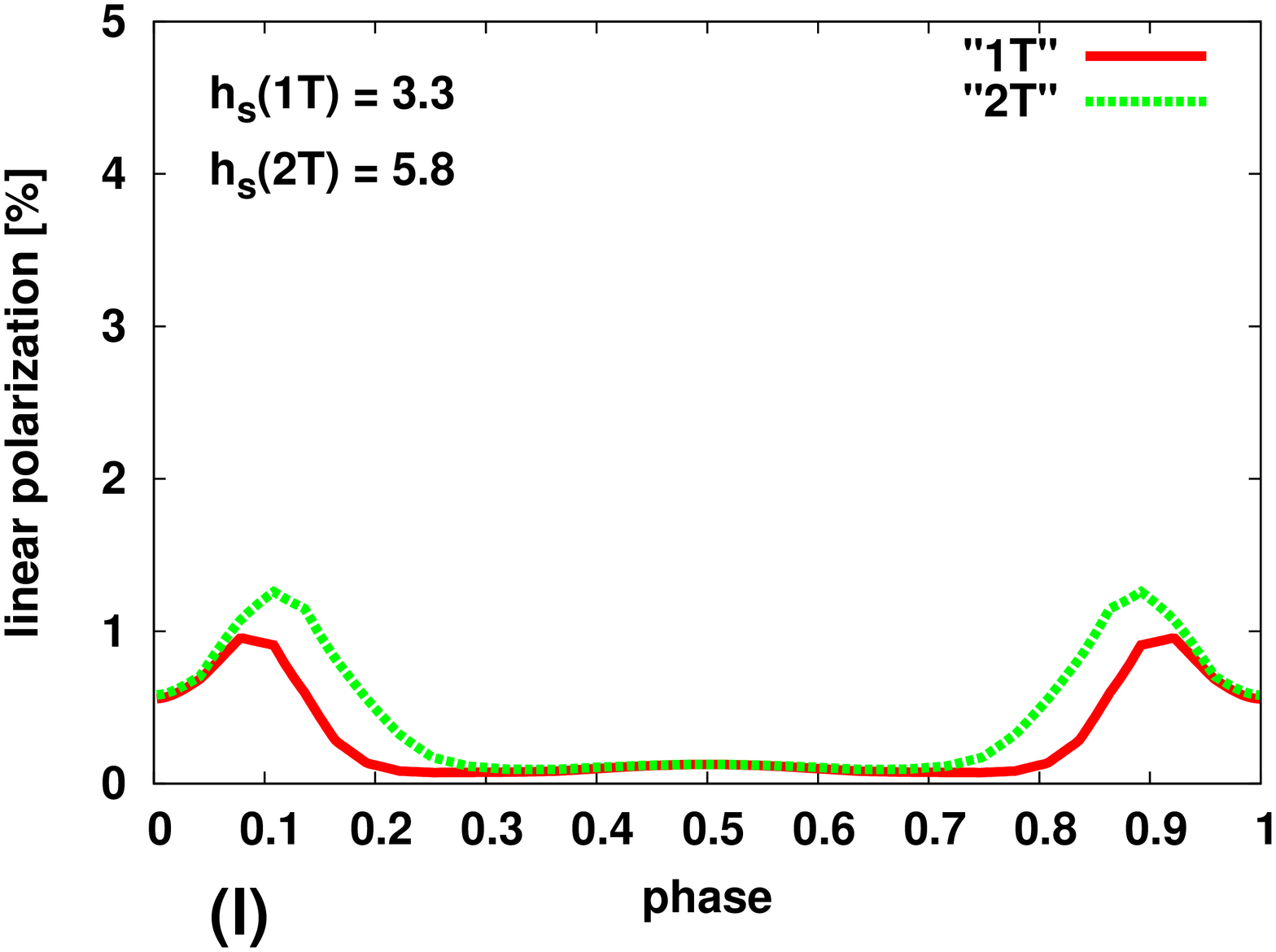}
\begin{picture}(25,100)(0,0)\put(0,-40){\scriptsize \bf \boldmath $\dot{M} = 10^{14}$ \unboldmath}
\end{picture}

\vspace*{-8em}
\begin{picture}(30,100)(0,0)\put(-5,-35){\scriptsize \bf \boldmath B$_{7} = 5$ \unboldmath}
\put(-5,-45){\scriptsize \bf V Filter}\end{picture}
\includegraphics[angle=-90,scale=0.16]{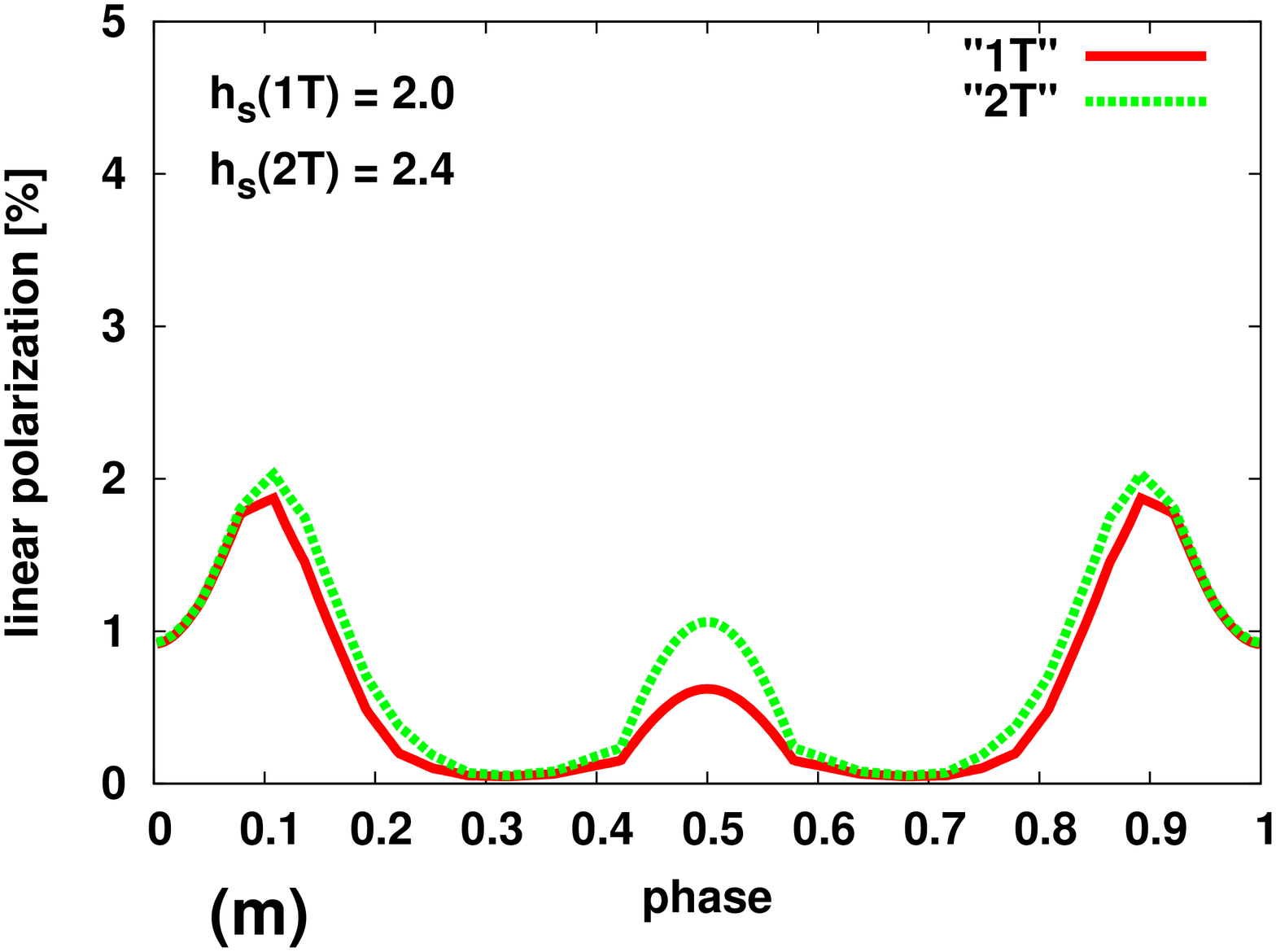}
\includegraphics[angle=-90,scale=0.16]{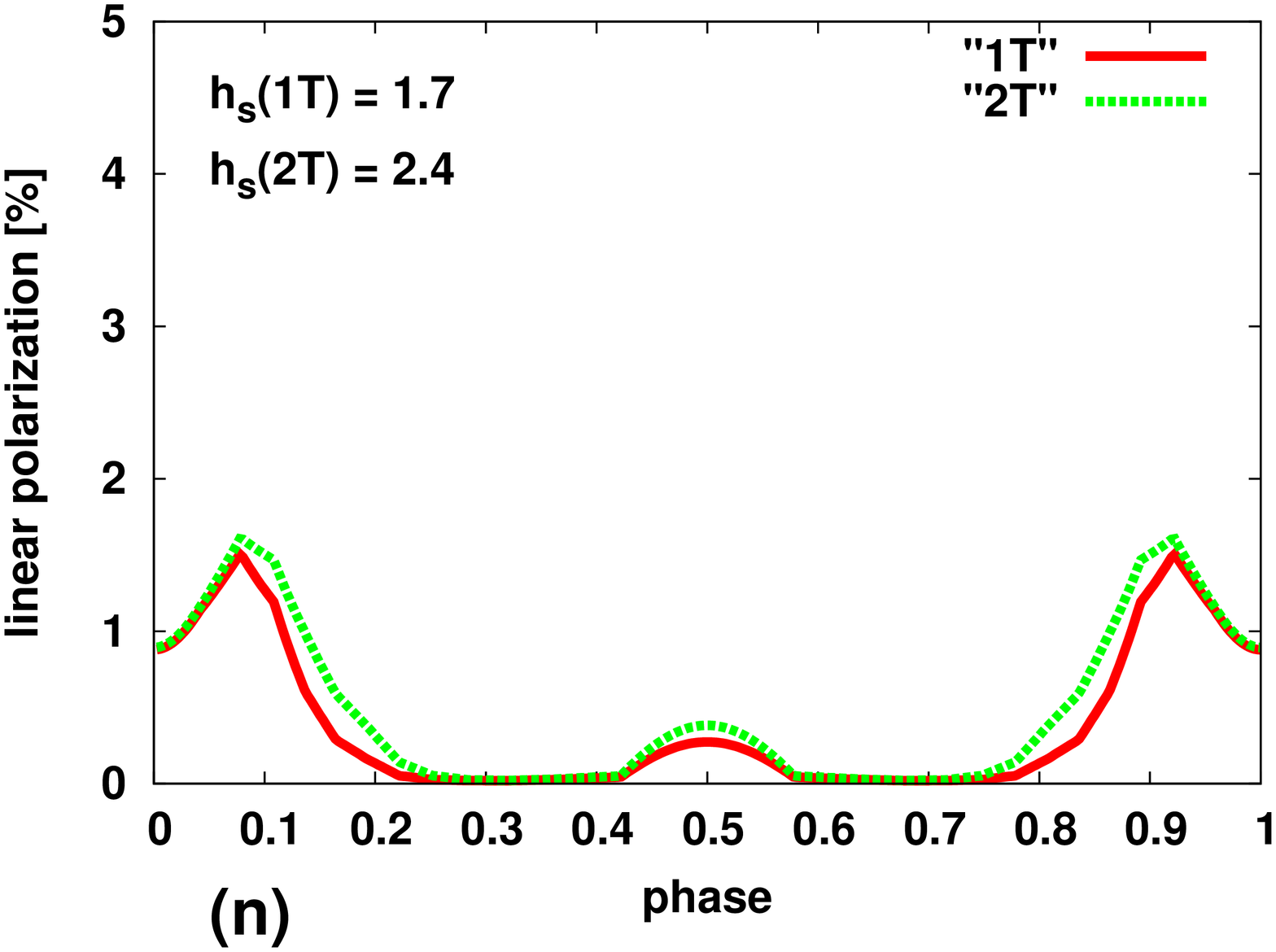}
\includegraphics[angle=-90,scale=0.16]{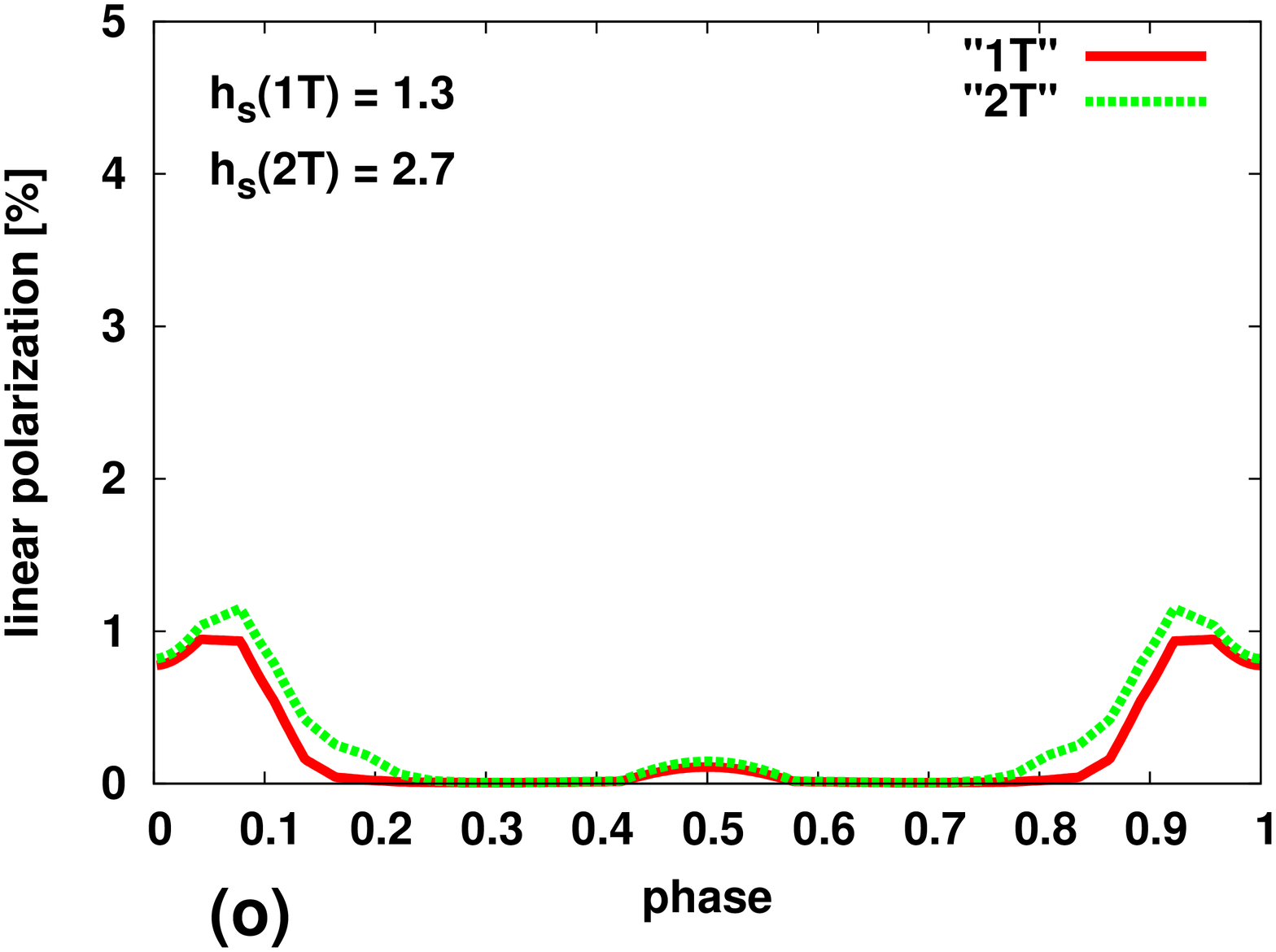}
\begin{picture}(25,100)(0,0)\put(0,-40){\scriptsize \bf \boldmath $\dot{M} = 10^{14}$ \unboldmath}
\end{picture}
\end{center}
\caption{Linear polarization orbital light curves comparing predictions for the
one-temperature model ({\tt "1T"}, red line) with the two-temperature ({\tt "2T"}, green line) model.
Only the low mass flow rate cases, $\dot{M} = 10^{14}$ g s$^{-1}$ ($\dot{m} = 0.5$ g cm$^{-2}$s$^{-1}$), are shown as there are
no differences in the two model predictions at the higher mass flow rate. Light curves through bandpass filters
at the higher harmonics are shown in the first nine panels (a) to (i). The cases in which the 2T model produced different
radiation output predictions from the 1T model are shown again in the last six panels (j) to (o) for the visual
V filter. \label{linorb}}
\end{figure*}

\begin{figure*}
\begin{center}

\vspace*{-8em}
\begin{picture}(30,100)(0,0)\put(-5,-35){\scriptsize \bf \boldmath B$_{7} = 1$ \unboldmath}
\put(-5,-45){\scriptsize \bf J Filter}\end{picture}
\includegraphics[angle=-90,scale=0.16]{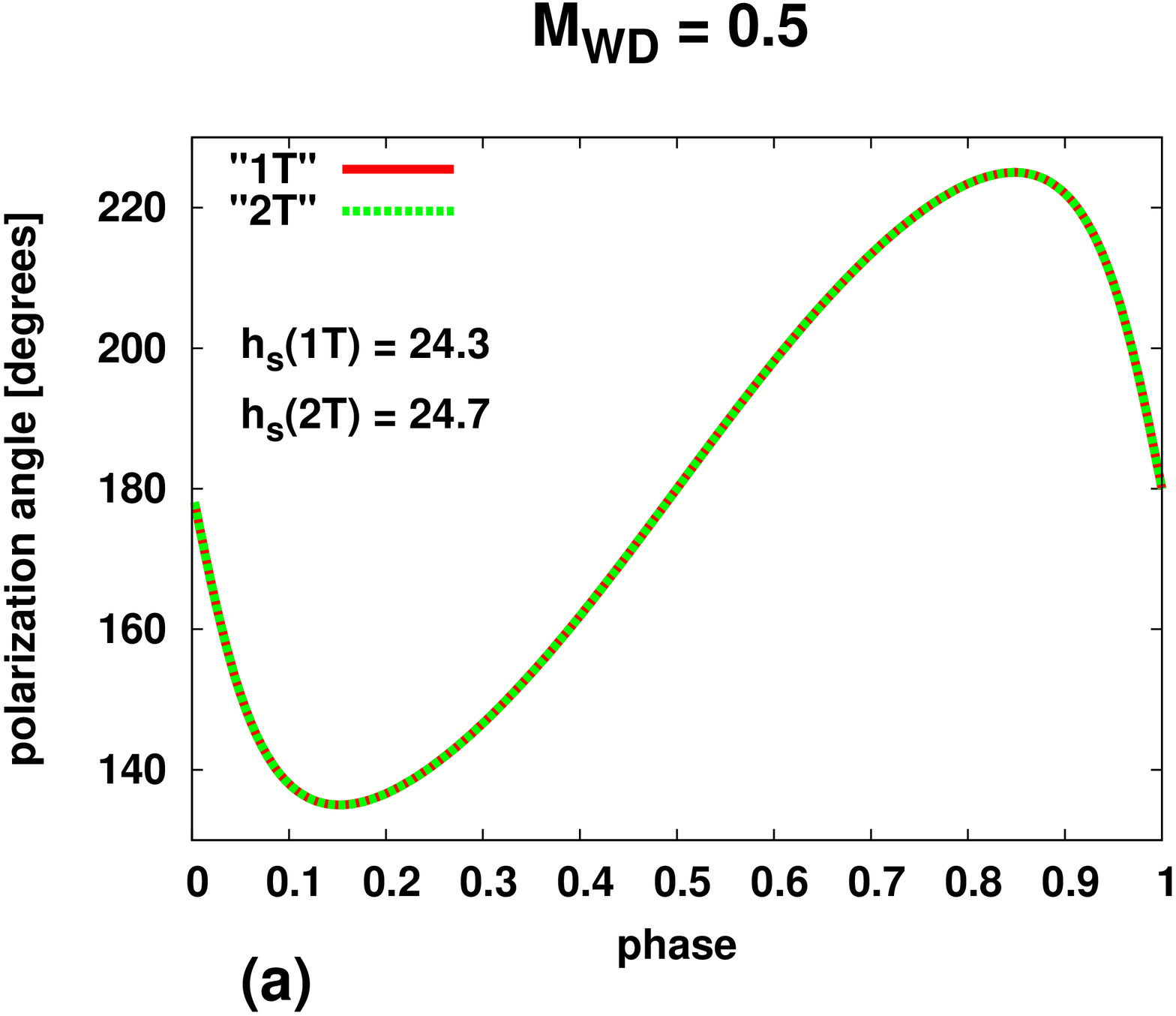}
\includegraphics[angle=-90,scale=0.16]{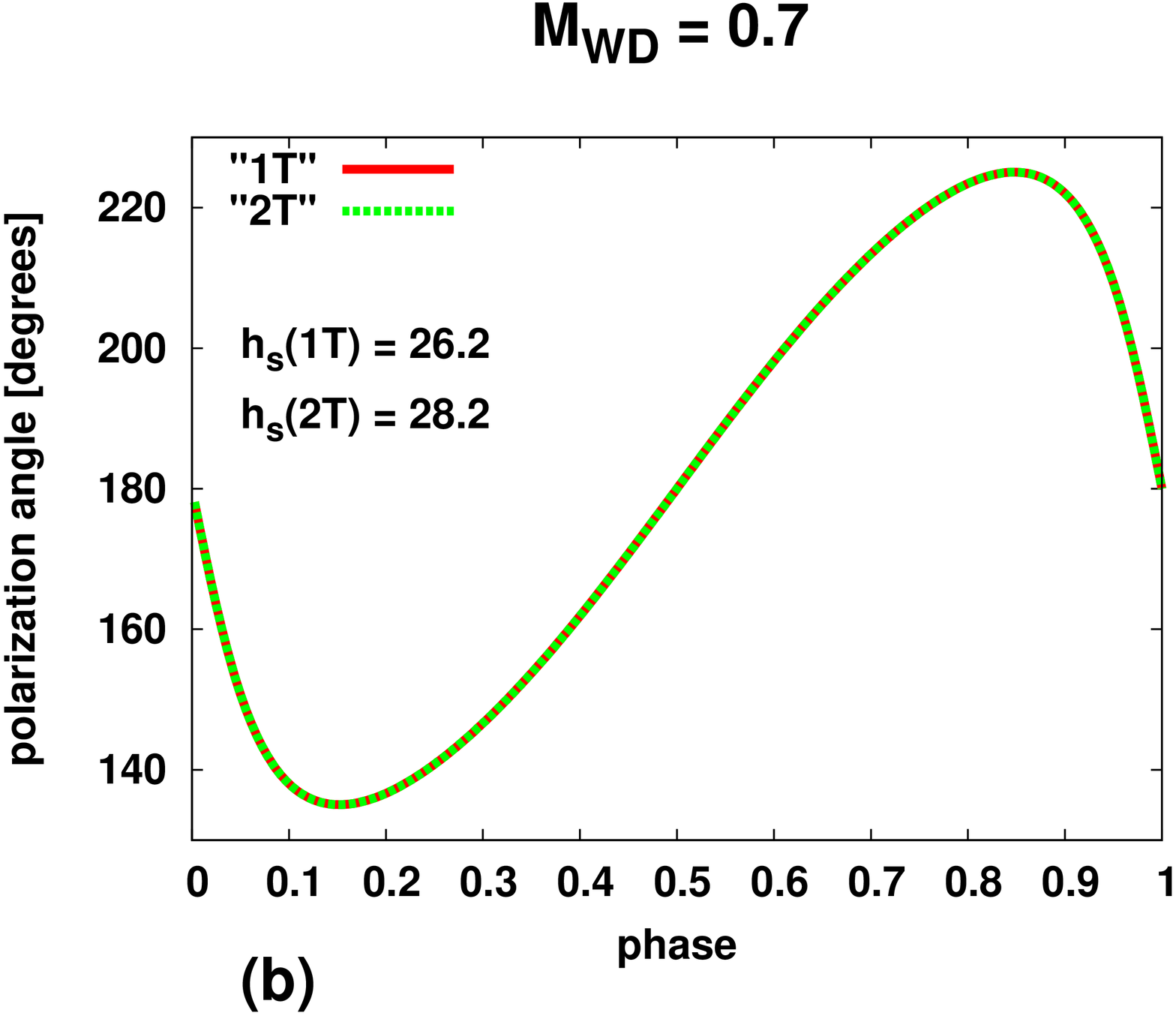}
\includegraphics[angle=-90,scale=0.16]{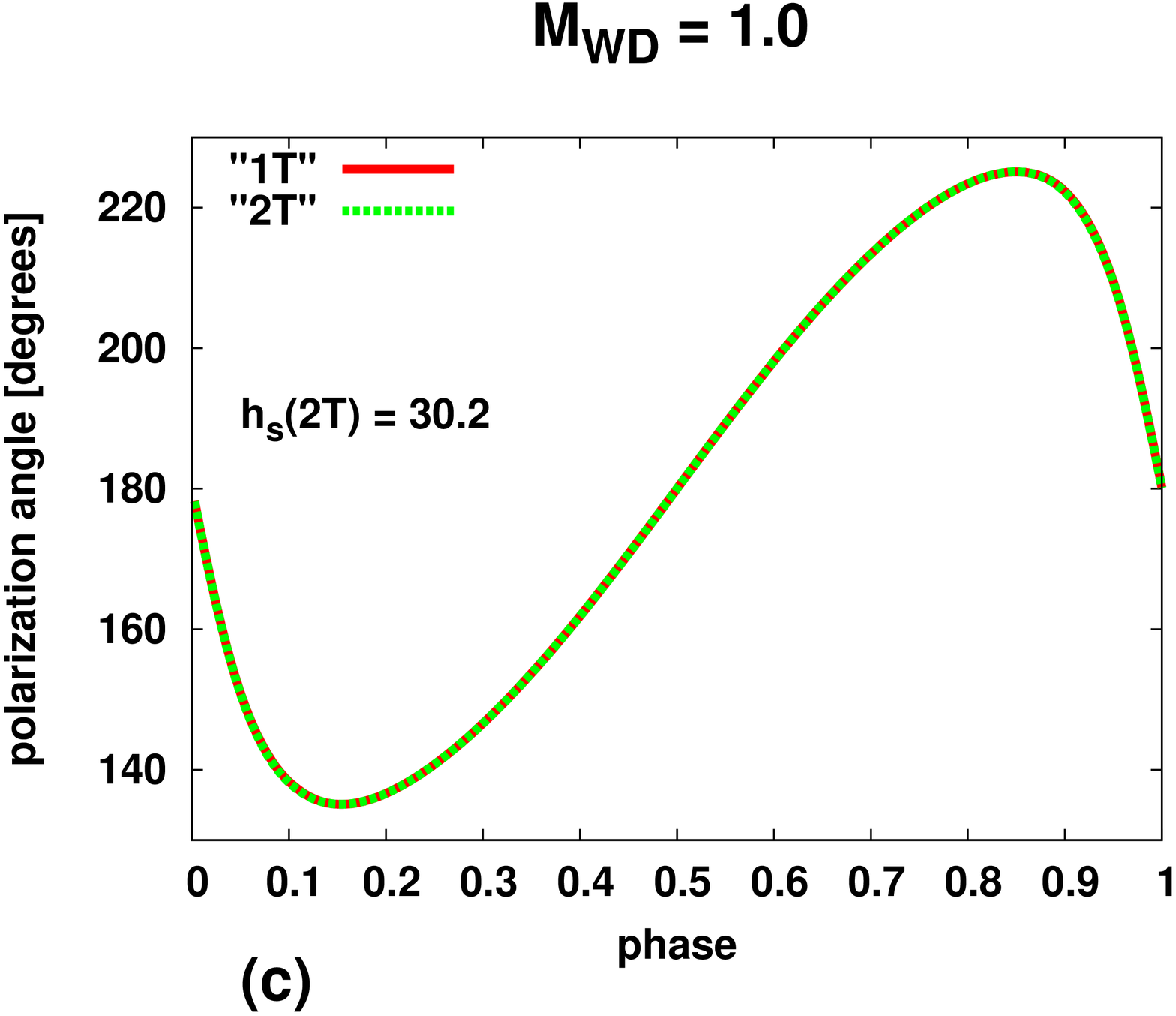}
\begin{picture}(25,100)(0,0)\put(0,-40){\scriptsize \bf \boldmath $\dot{M} = 10^{14}$ \unboldmath}
\end{picture}

\vspace*{-8em}
\begin{picture}(30,100)(0,0)\put(-5,-35){\scriptsize \bf \boldmath B$_{7} = 3$ \unboldmath}
\put(-5,-45){\scriptsize \bf U Filter}\end{picture}
\includegraphics[angle=-90,scale=0.16]{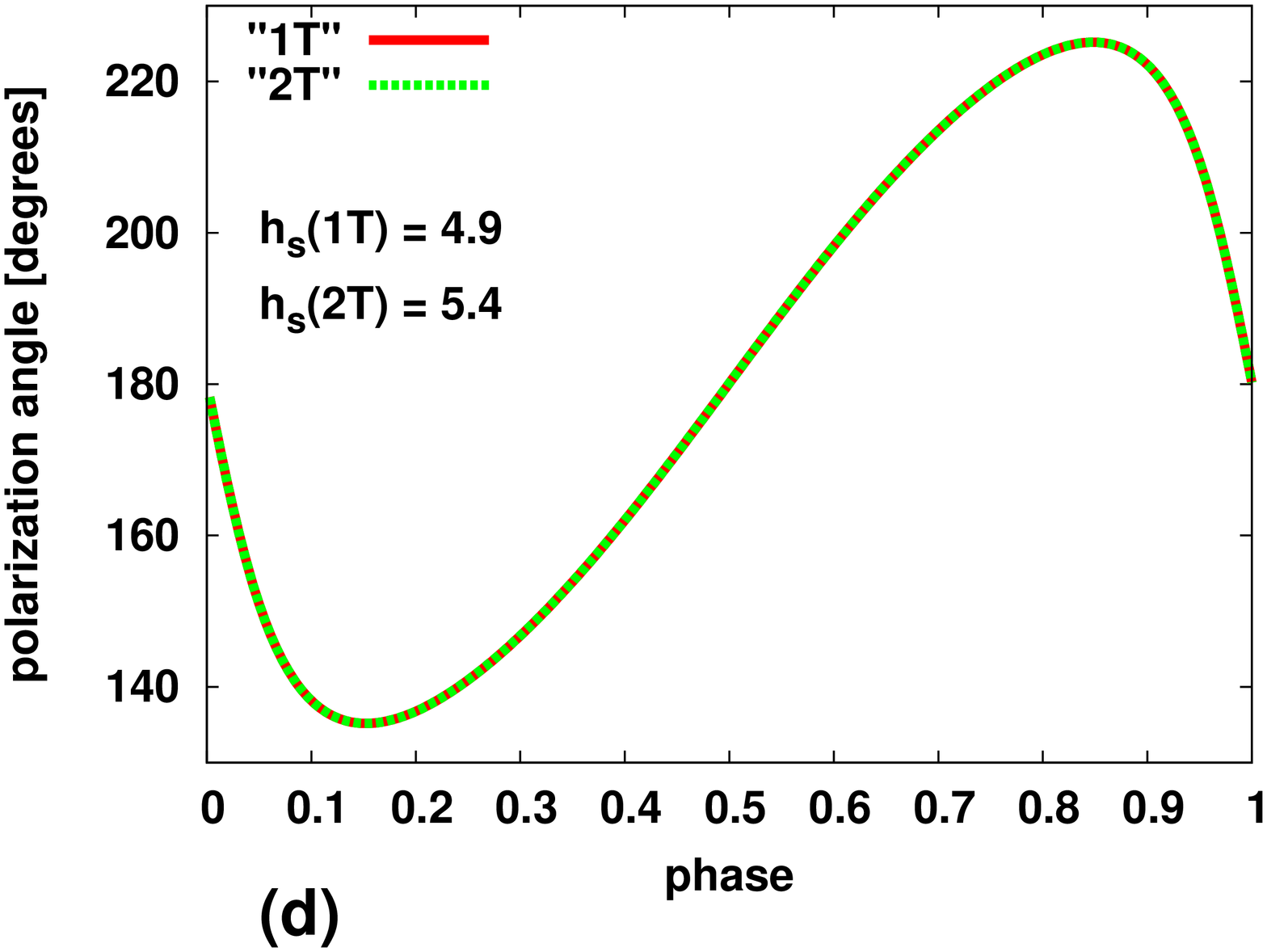}
\includegraphics[angle=-90,scale=0.16]{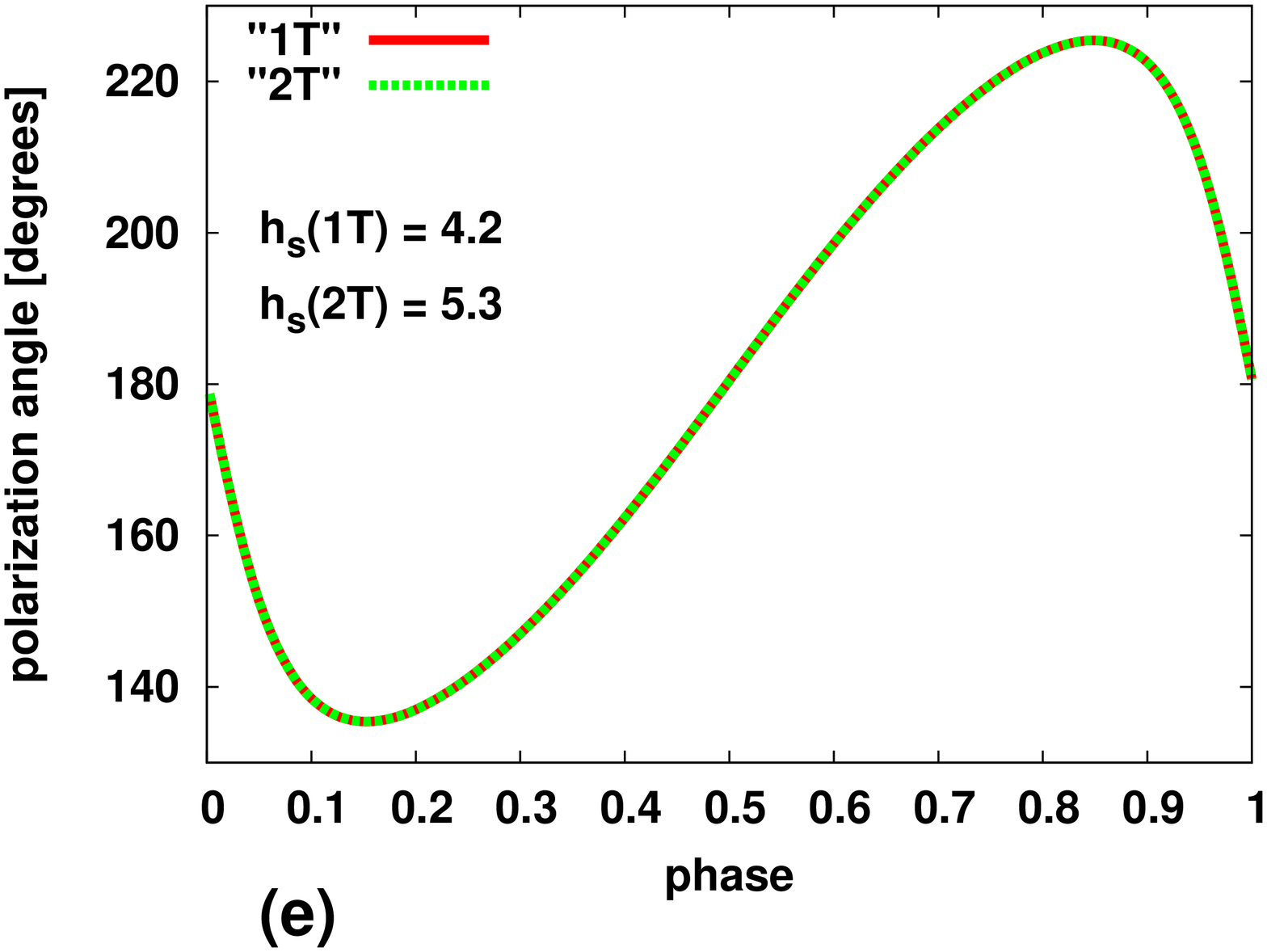}
\includegraphics[angle=-90,scale=0.16]{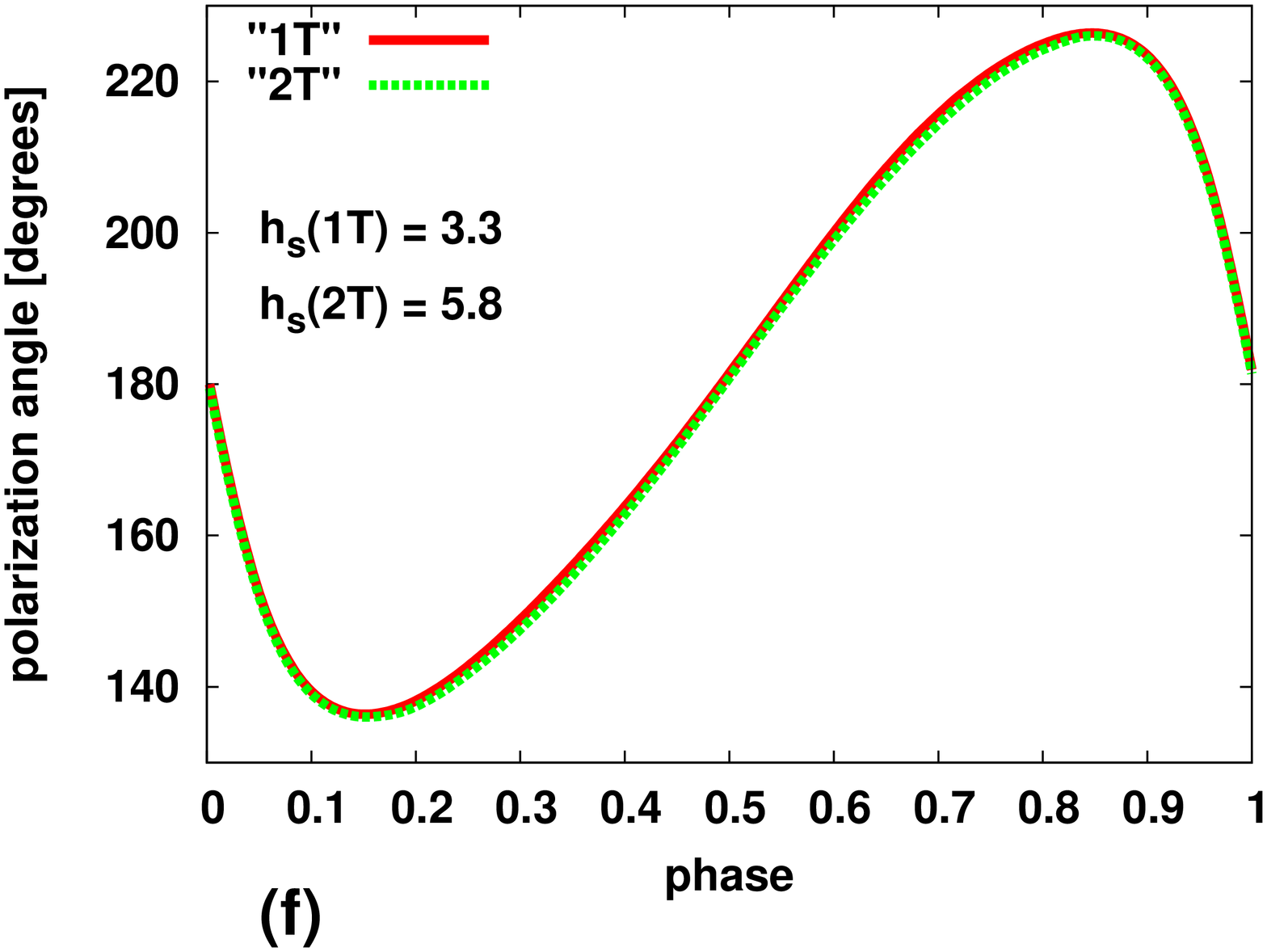}
\begin{picture}(25,100)(0,0)\put(0,-40){\scriptsize \bf \boldmath $\dot{M} = 10^{14}$ \unboldmath}
\end{picture}

\vspace*{-8em}
\begin{picture}(30,100)(0,0)\put(-5,-35){\scriptsize \bf \boldmath B$_{7} = 5$ \unboldmath}
\put(-5,-45){\scriptsize \bf H = 9.92}\end{picture}
\includegraphics[angle=-90,scale=0.16]{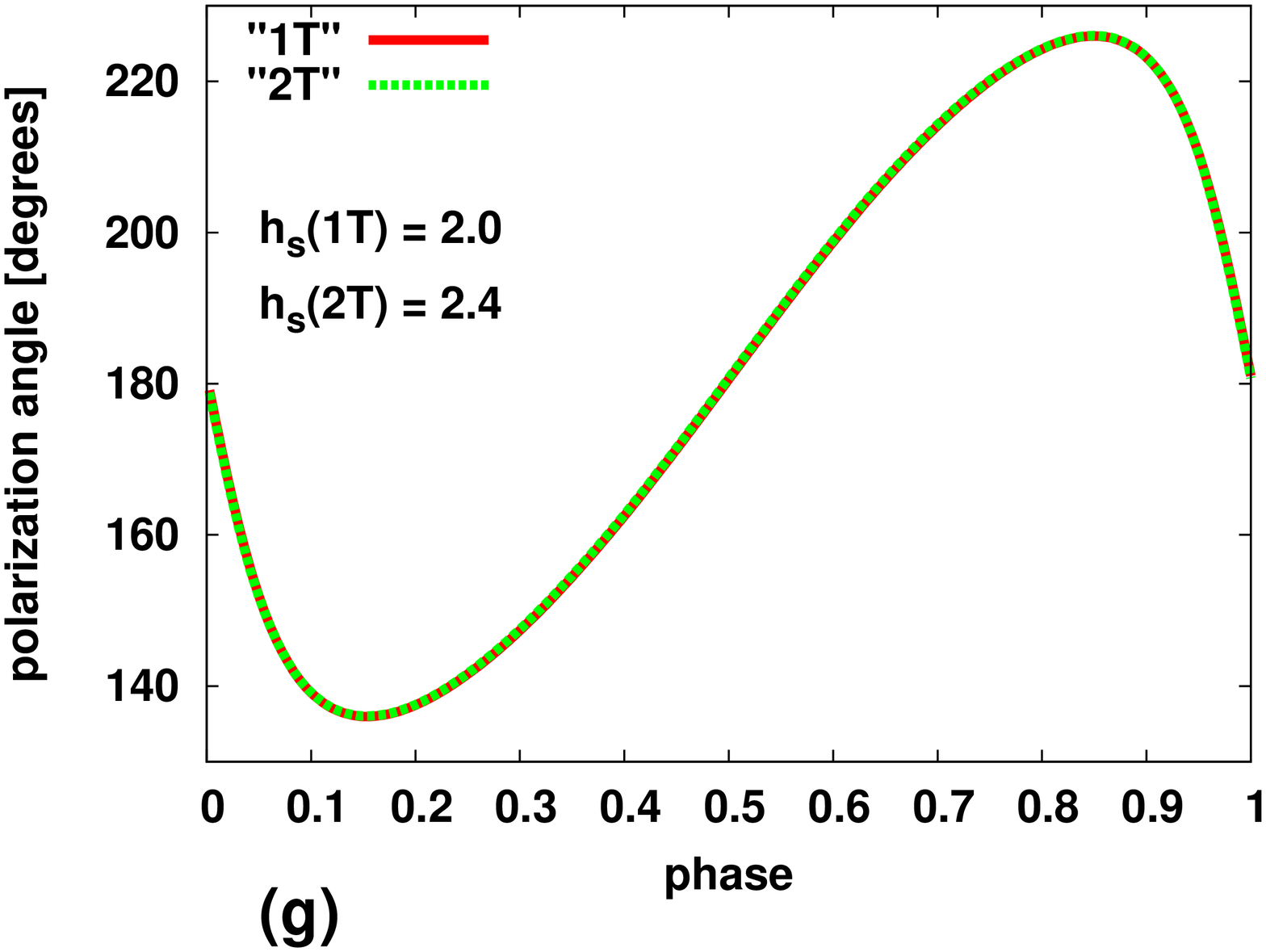}
\includegraphics[angle=-90,scale=0.16]{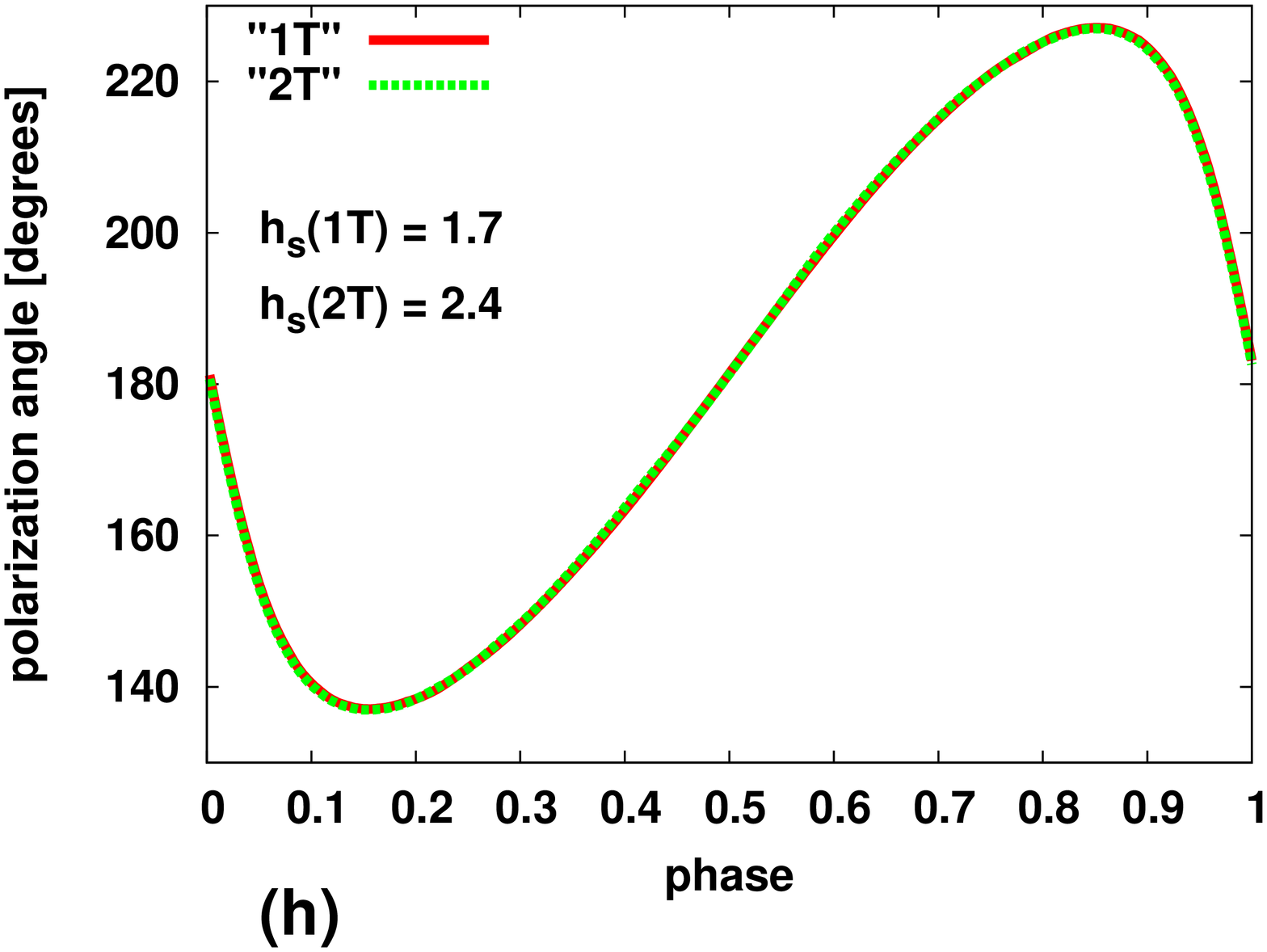}
\includegraphics[angle=-90,scale=0.16]{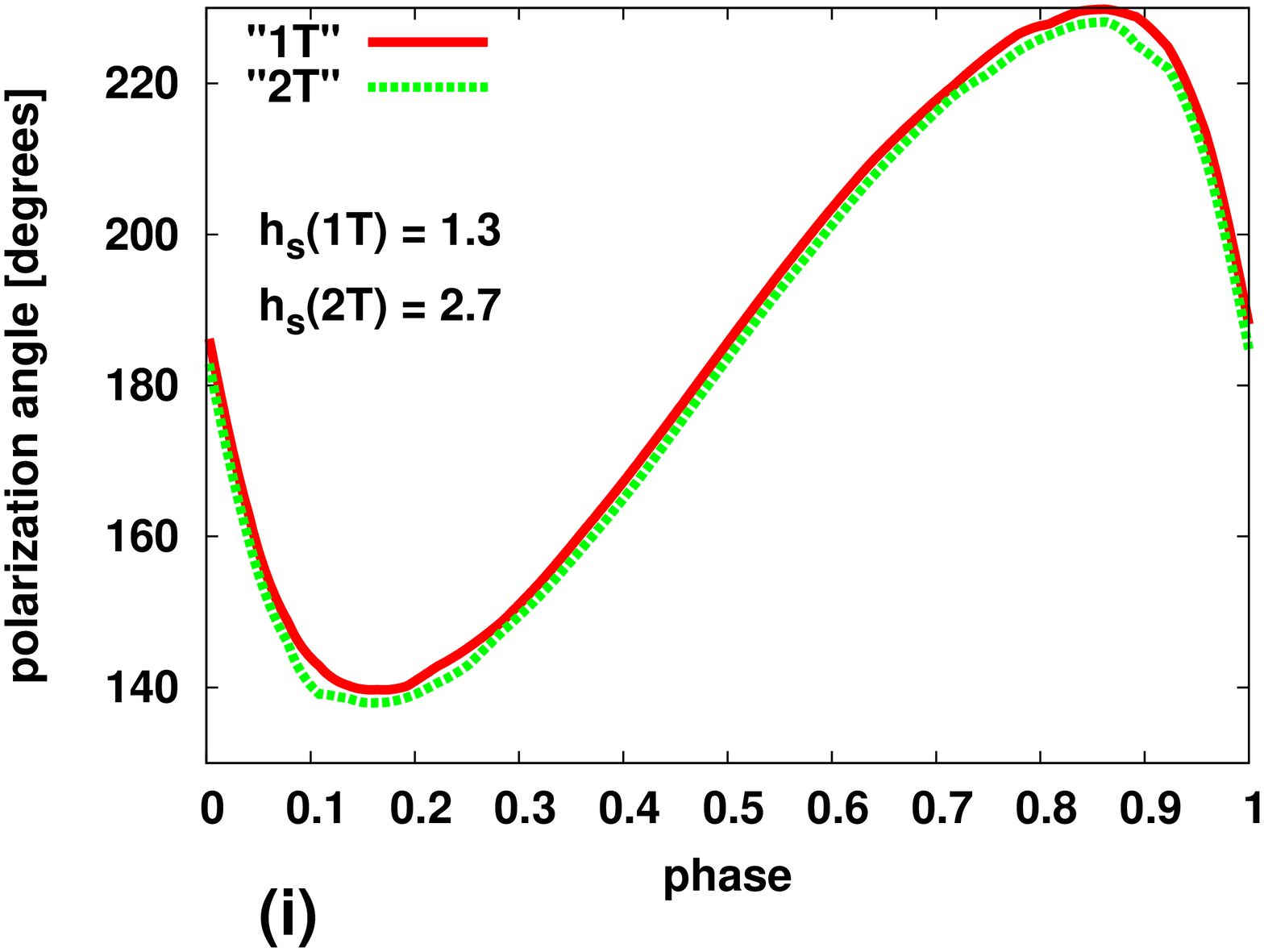}
\begin{picture}(25,100)(0,0)\put(0,-40){\scriptsize \bf \boldmath $\dot{M} = 10^{14}$ \unboldmath}
\end{picture}

\vspace*{-8em}
\begin{picture}(30,100)(0,0)\put(-5,-35){\scriptsize \bf \boldmath B$_{7} = 3$ \unboldmath}
\put(-5,-45){\scriptsize \bf V Filter}\end{picture}
\includegraphics[angle=-90,scale=0.16]{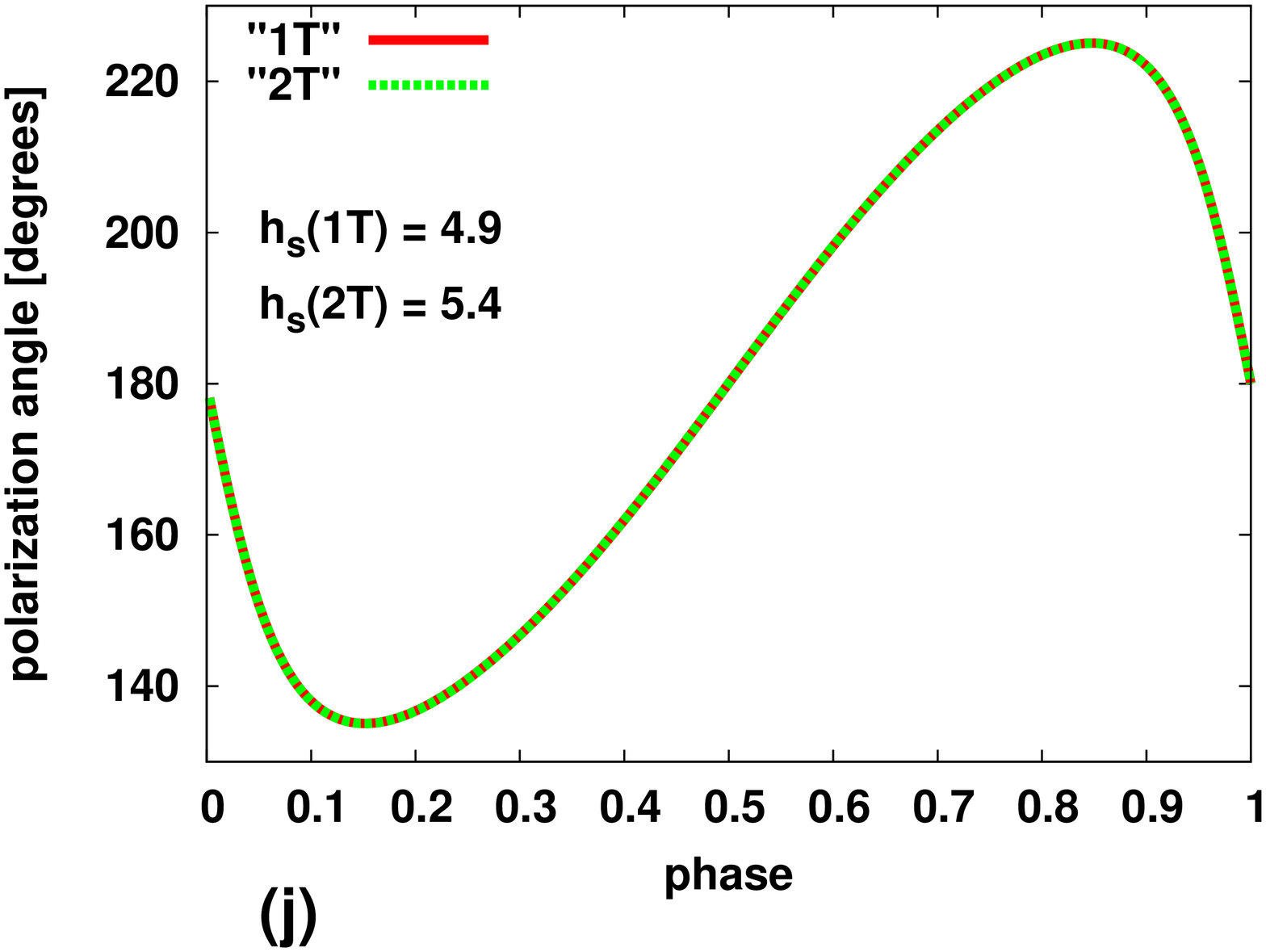}
\includegraphics[angle=-90,scale=0.16]{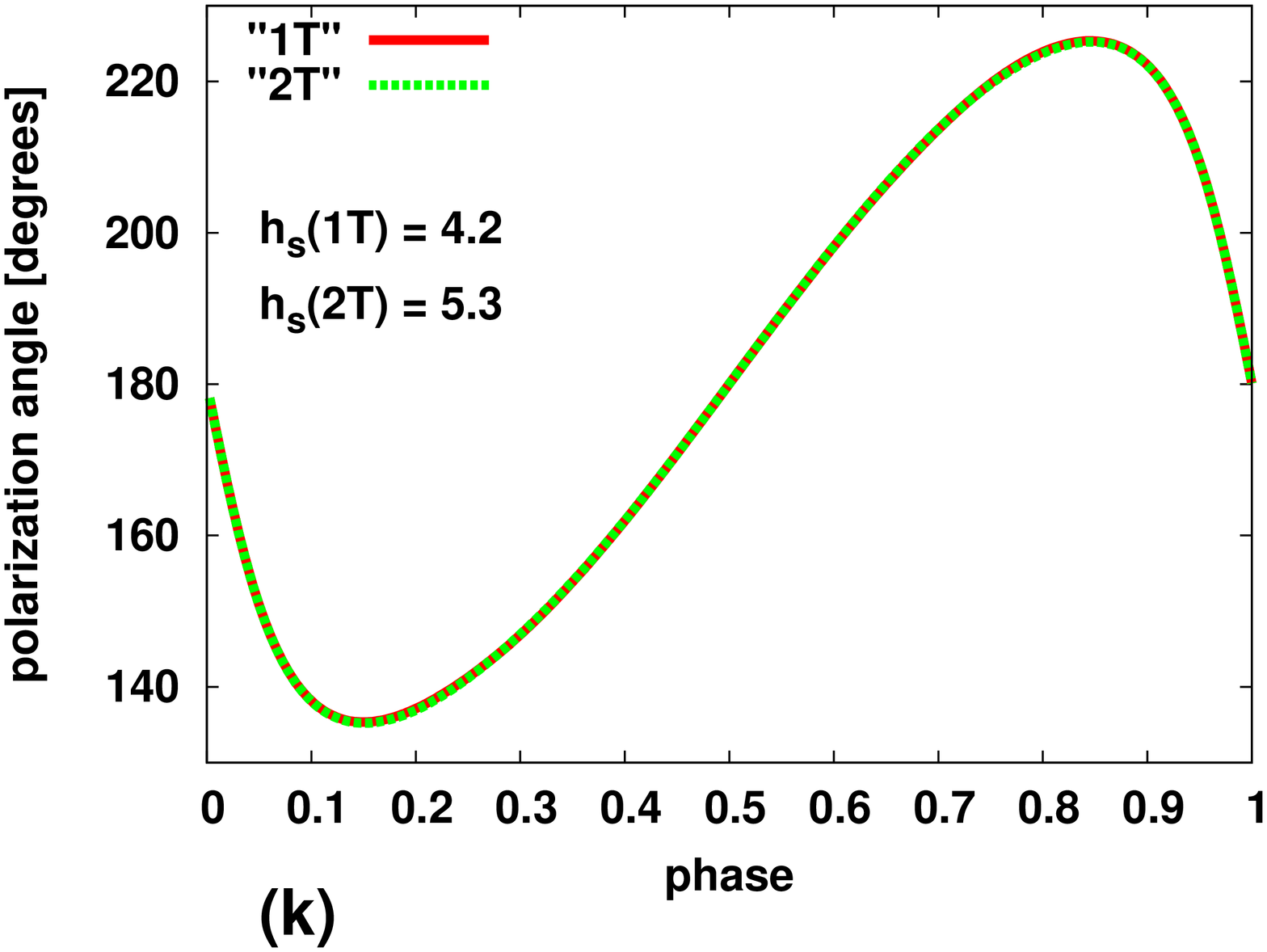}
\includegraphics[angle=-90,scale=0.16]{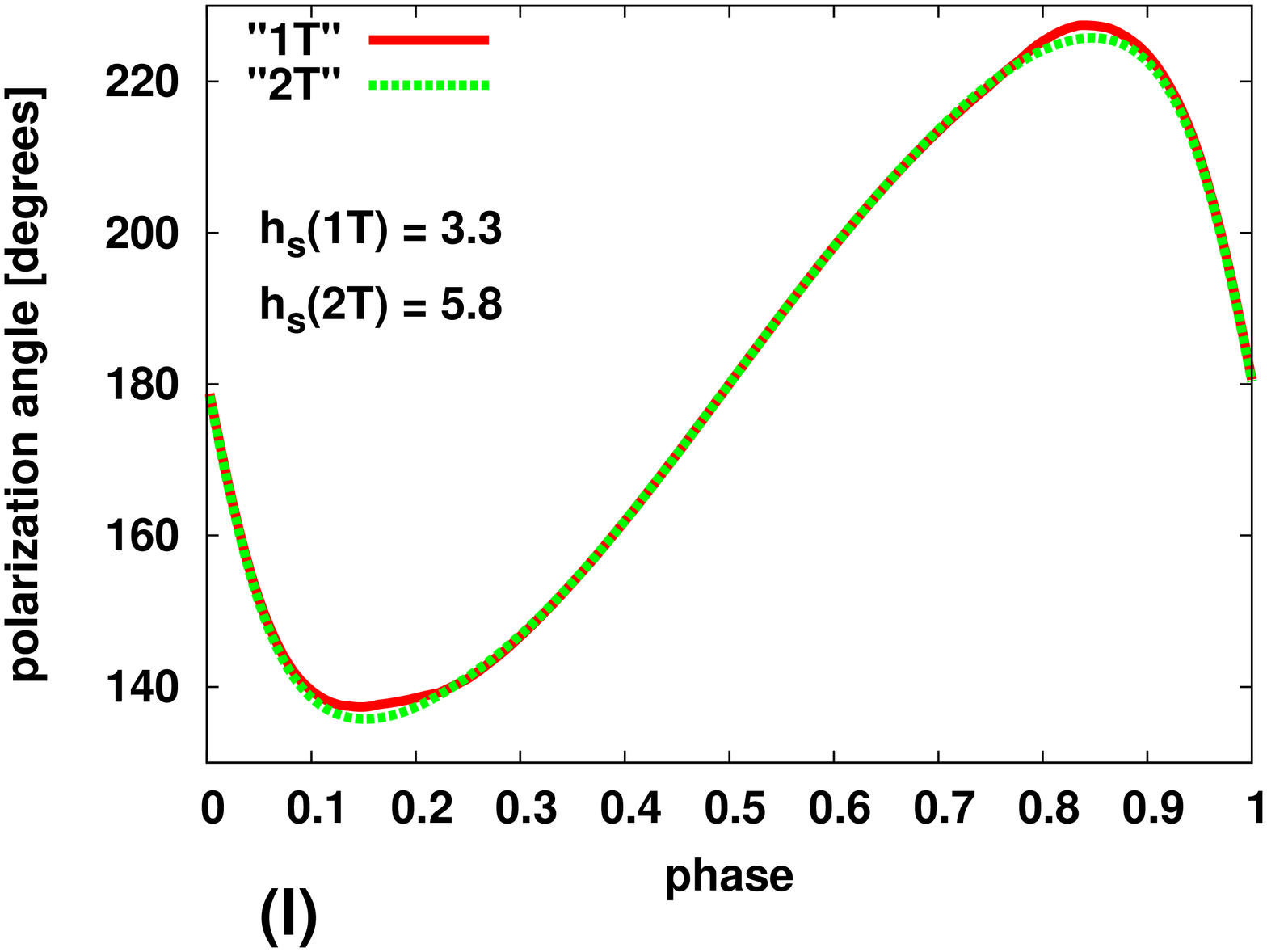}
\begin{picture}(25,100)(0,0)\put(0,-40){\scriptsize \bf \boldmath $\dot{M} = 10^{14}$ \unboldmath}
\end{picture}

\vspace*{-8em}
\begin{picture}(30,100)(0,0)\put(-5,-35){\scriptsize \bf \boldmath B$_{7} = 5$ \unboldmath}
\put(-5,-45){\scriptsize \bf V Filter}\end{picture}
\includegraphics[angle=-90,scale=0.16]{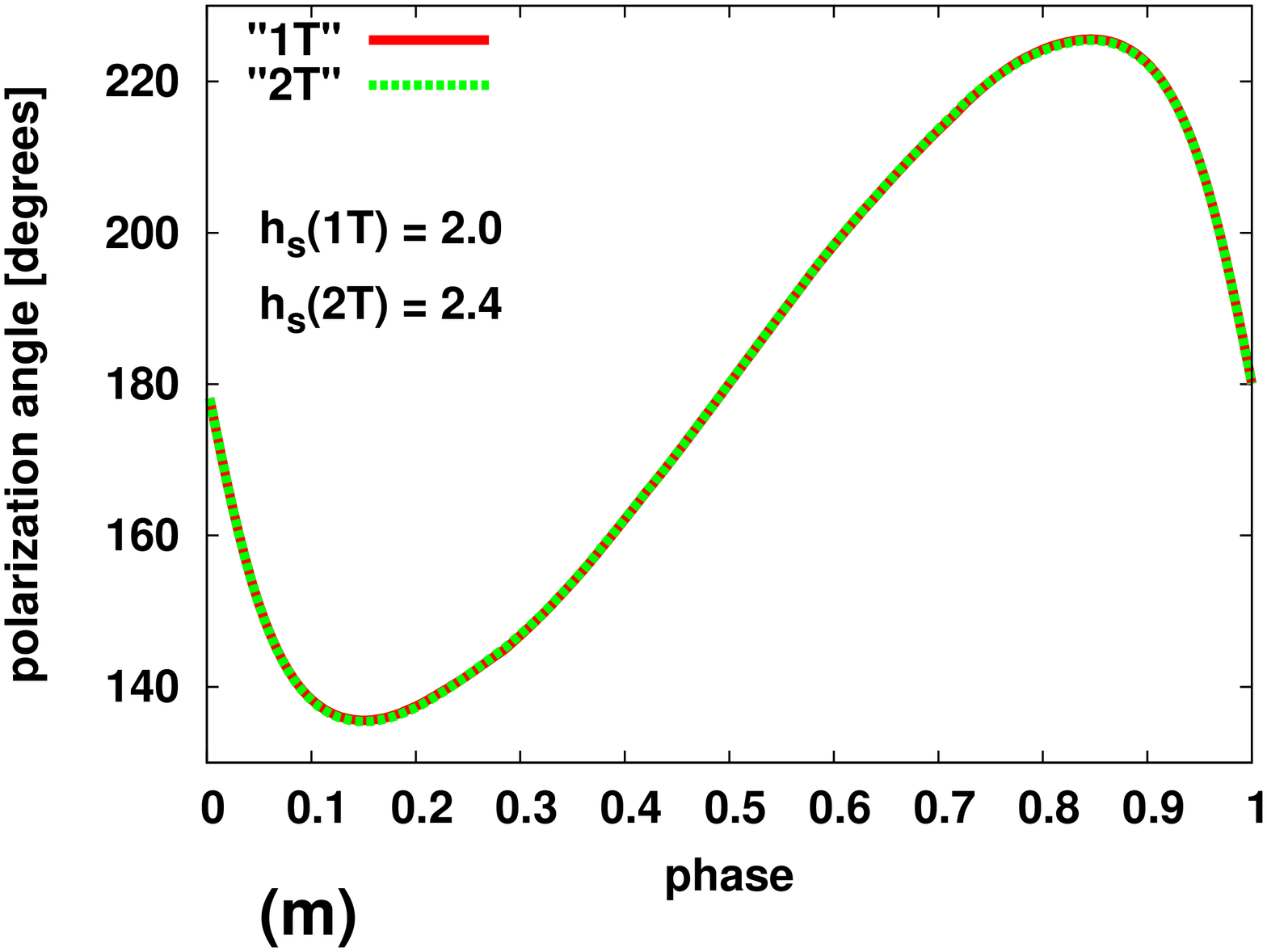}
\includegraphics[angle=-90,scale=0.16]{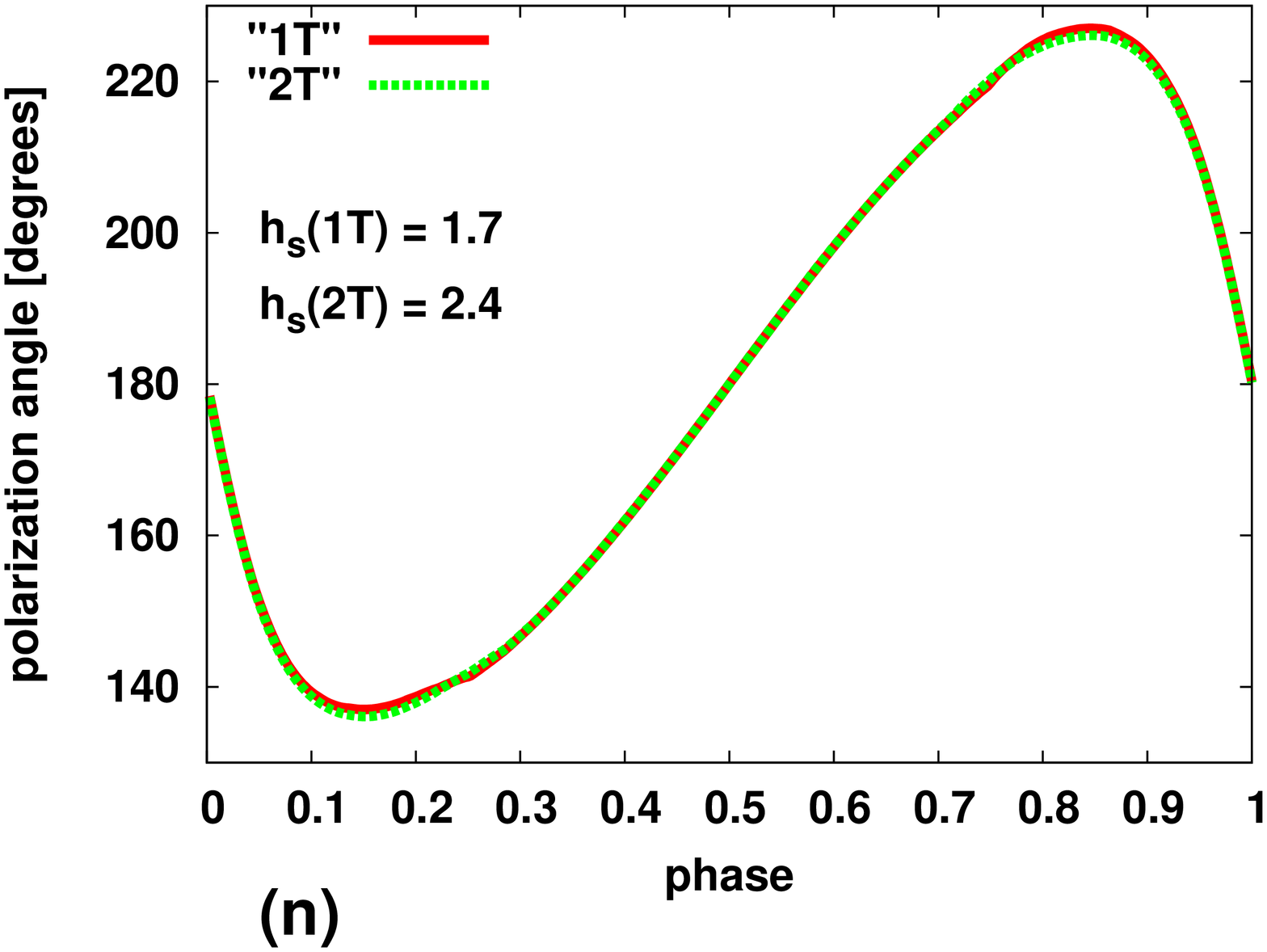}
\includegraphics[angle=-90,scale=0.16]{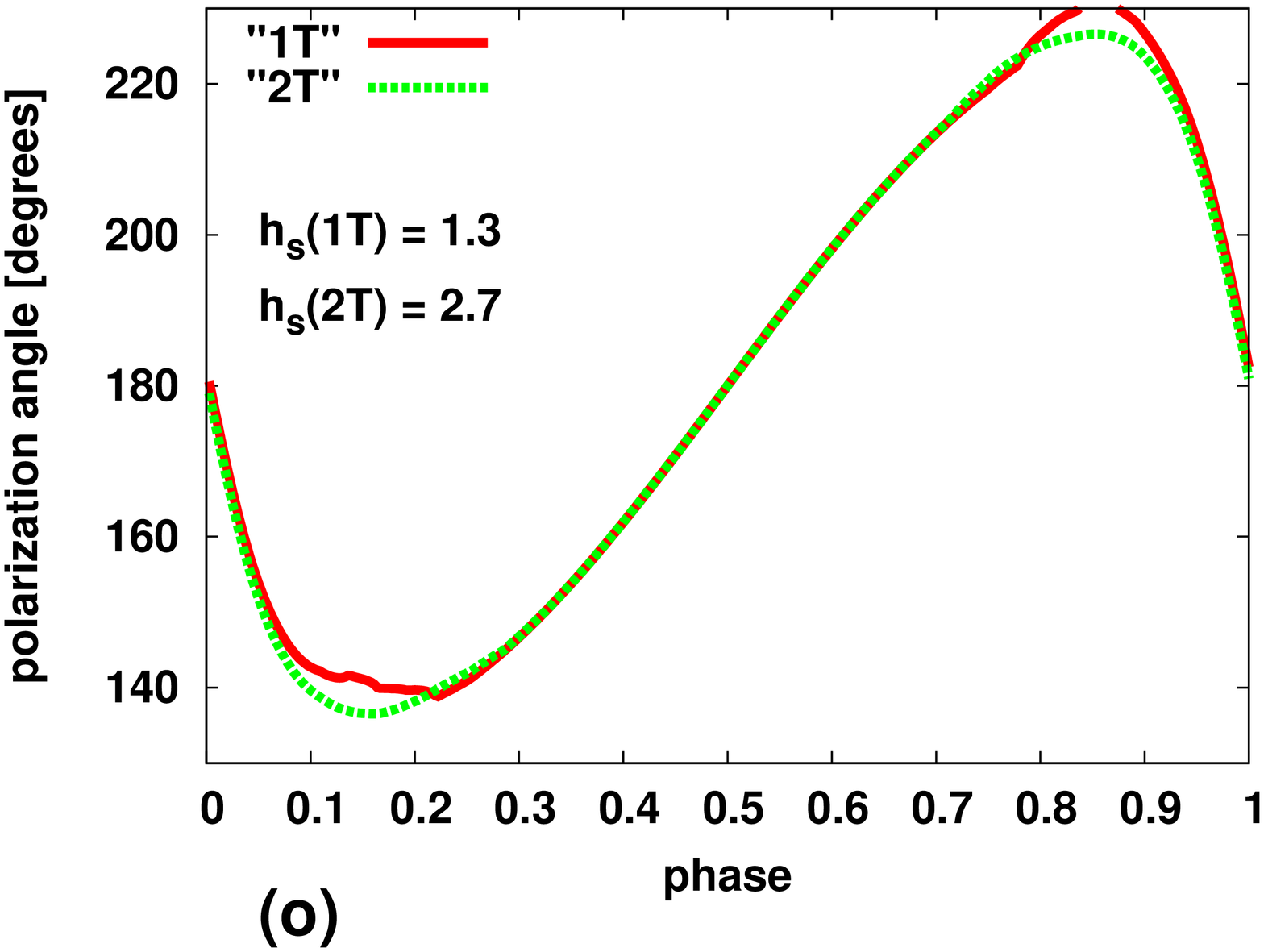}
\begin{picture}(25,100)(0,0)\put(0,-40){\scriptsize \bf \boldmath $\dot{M} = 10^{14}$ \unboldmath}
\end{picture}
\end{center}
\caption{Linear polarization angle orbital light curves comparing predictions for the
one-temperature model ({\tt "1T"}, red line) with the two-temperature ({\tt "2T"}, green line) model.
The figure is laid out in the same way as Fig.~\ref{linorb}. \label{angorb}}
\end{figure*}

\begin{figure*}
\begin{center}

\vspace*{-8em}
\begin{picture}(30,100)(0,0)\put(-5,-35){\scriptsize \bf \boldmath B$_{7} = 1$ \unboldmath}
\put(-5,-45){\scriptsize \bf J Filter}\end{picture}
\includegraphics[angle=-90,scale=0.16]{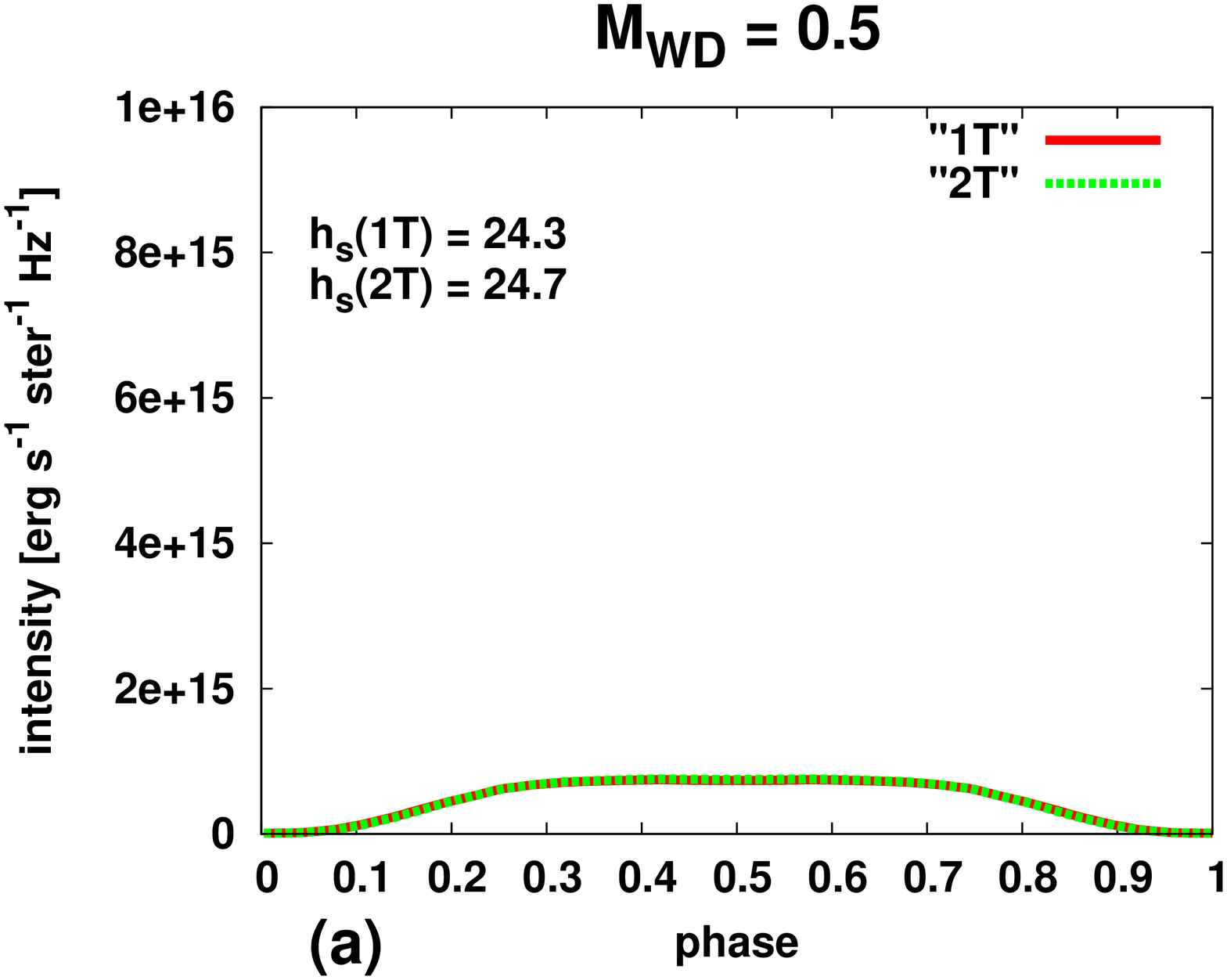}
\includegraphics[angle=-90,scale=0.16]{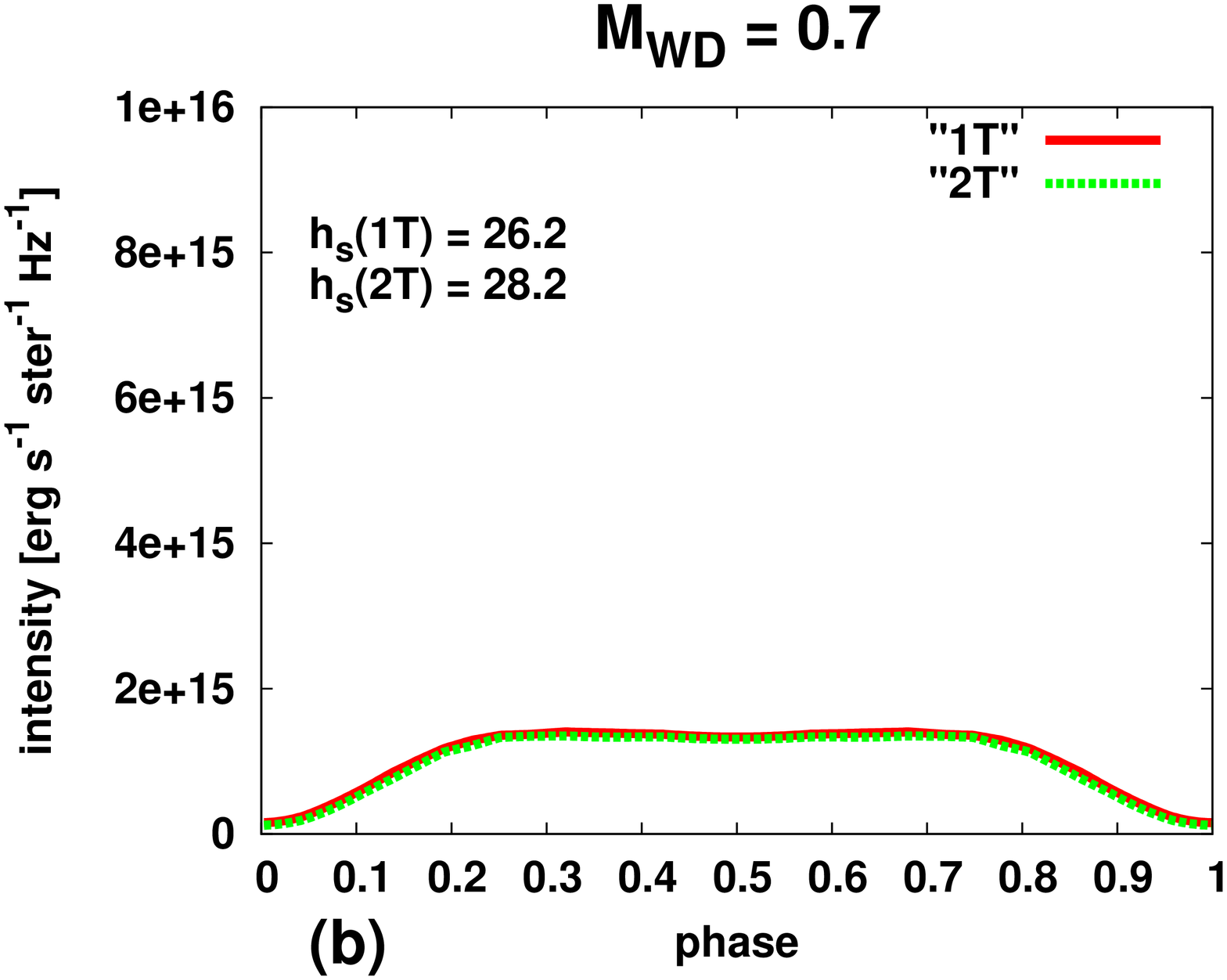}
\includegraphics[angle=-90,scale=0.16]{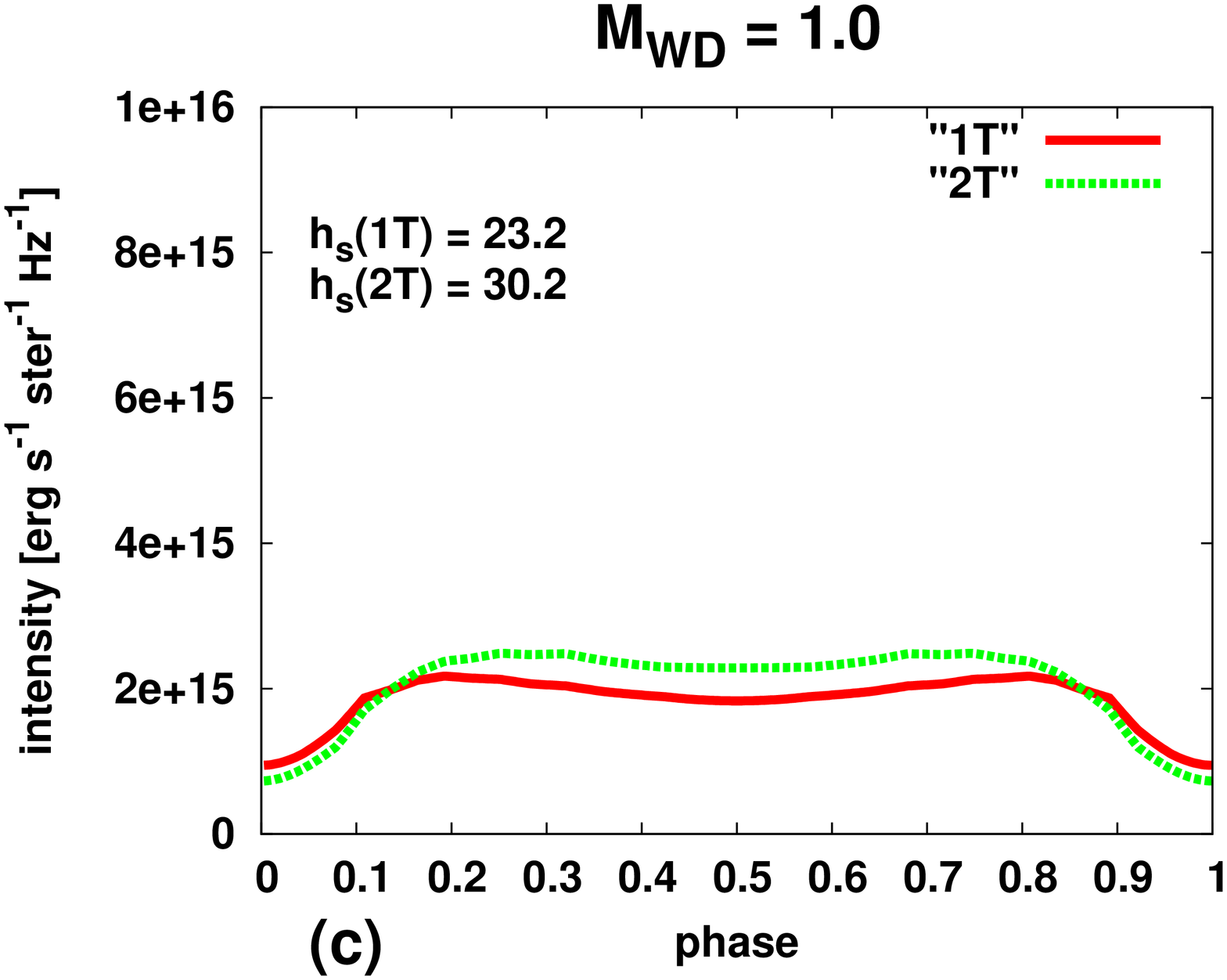}
\begin{picture}(25,100)(0,0)\put(0,-40){\scriptsize \bf \boldmath $\dot{M} = 10^{14}$ \unboldmath}
\end{picture}

\vspace*{-8em}
\begin{picture}(30,100)(0,0)\put(-5,-35){\scriptsize \bf \boldmath B$_{7} = 3$ \unboldmath}
\put(-5,-45){\scriptsize \bf U Filter}\end{picture}
\includegraphics[angle=-90,scale=0.16]{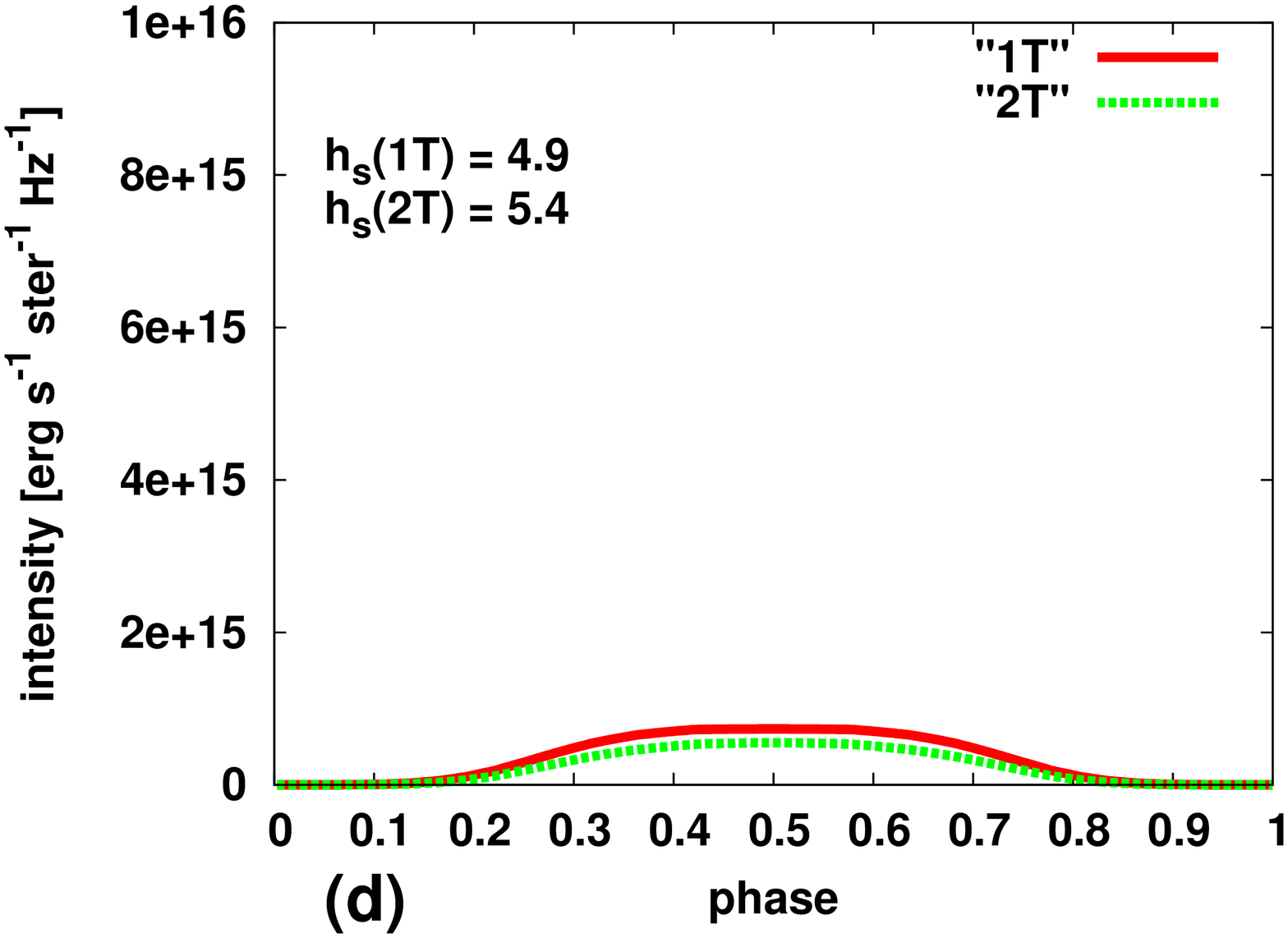}
\includegraphics[angle=-90,scale=0.16]{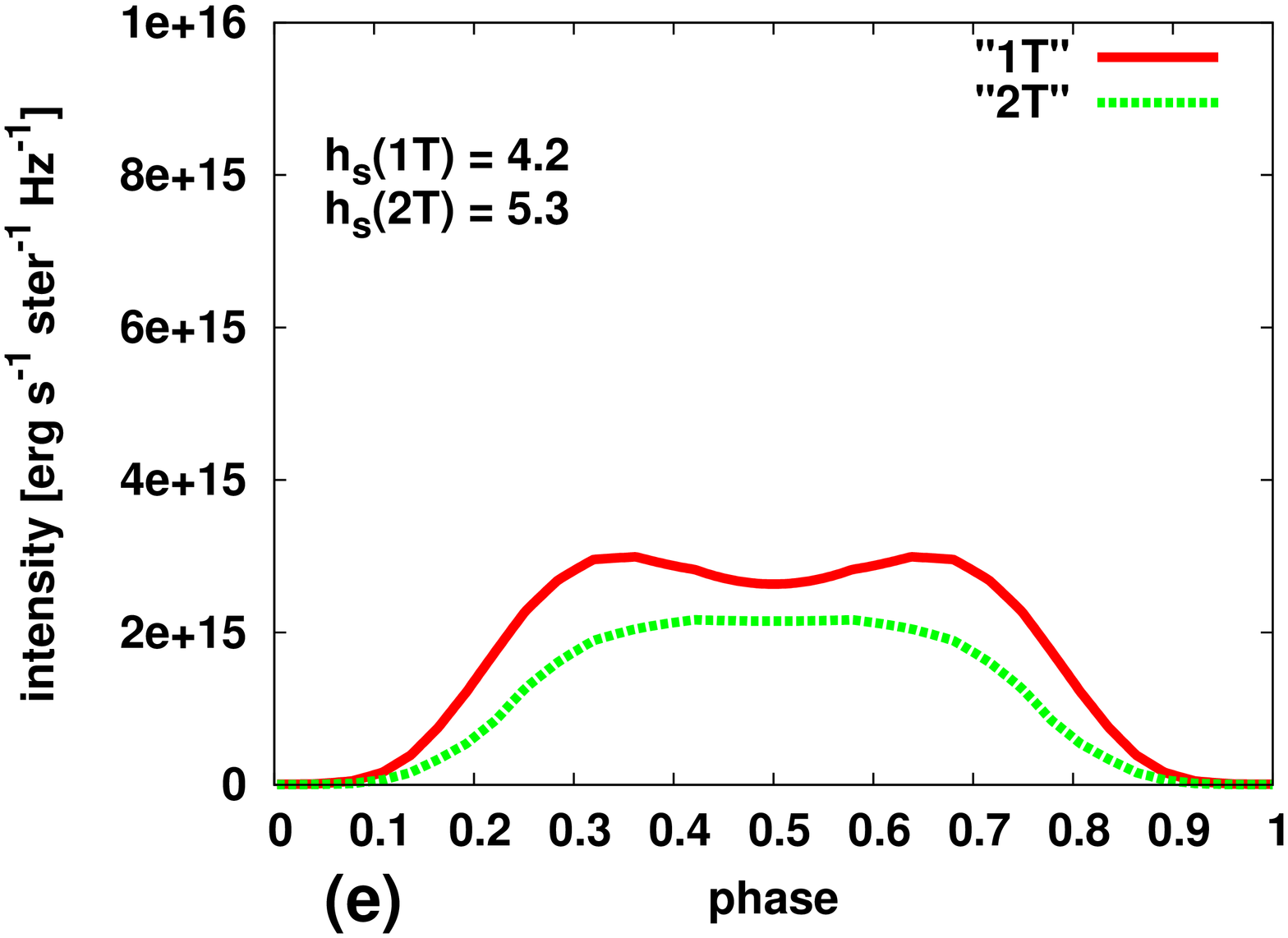}
\includegraphics[angle=-90,scale=0.16]{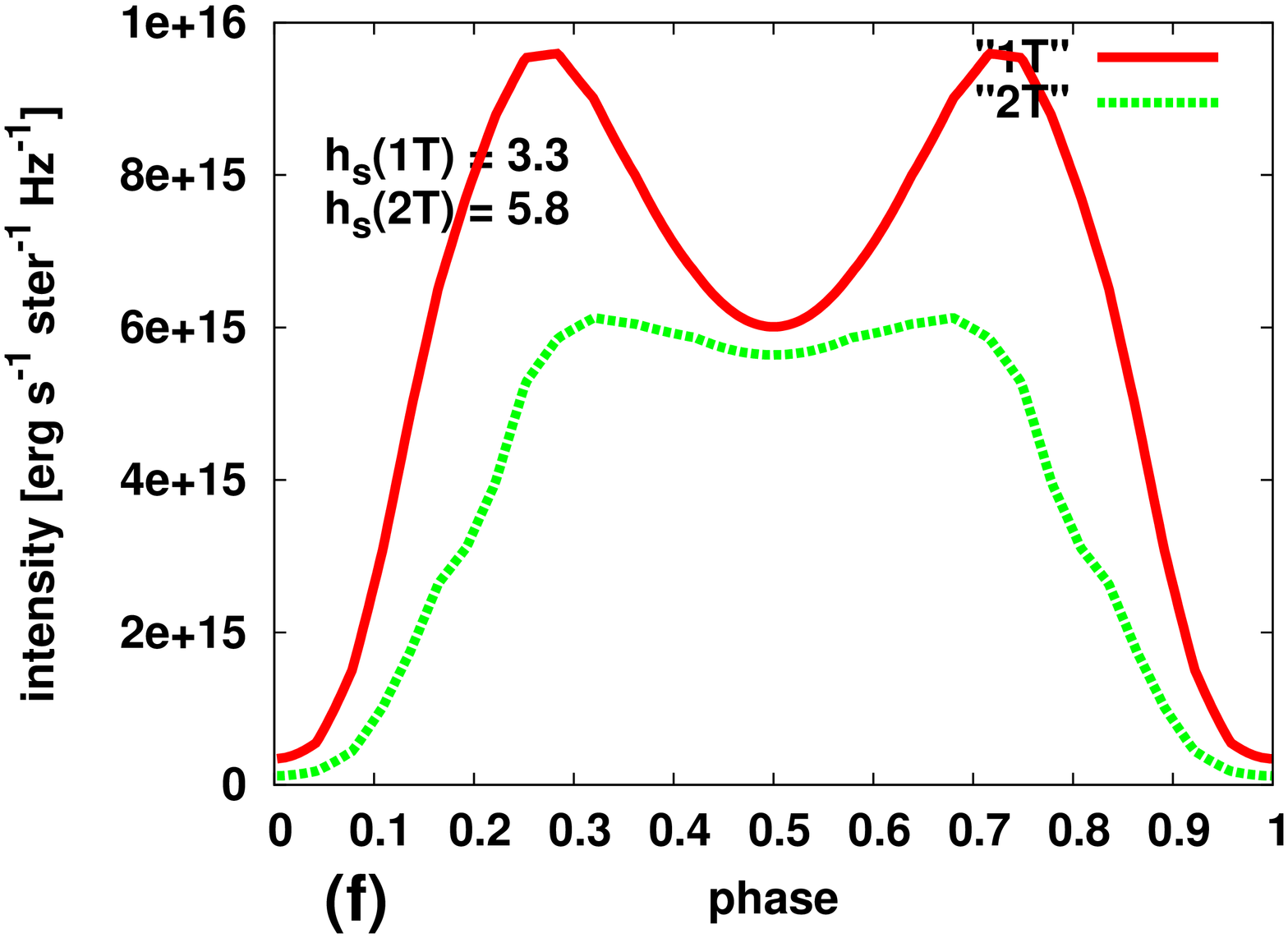}
\begin{picture}(25,100)(0,0)\put(0,-40){\scriptsize \bf \boldmath $\dot{M} = 10^{14}$ \unboldmath}
\end{picture}

\vspace*{-8em}
\begin{picture}(30,100)(0,0)\put(-5,-35){\scriptsize \bf \boldmath B$_{7} = 5$ \unboldmath}
\put(-5,-45){\scriptsize \bf H = 9.92}\end{picture}
\includegraphics[angle=-90,scale=0.16]{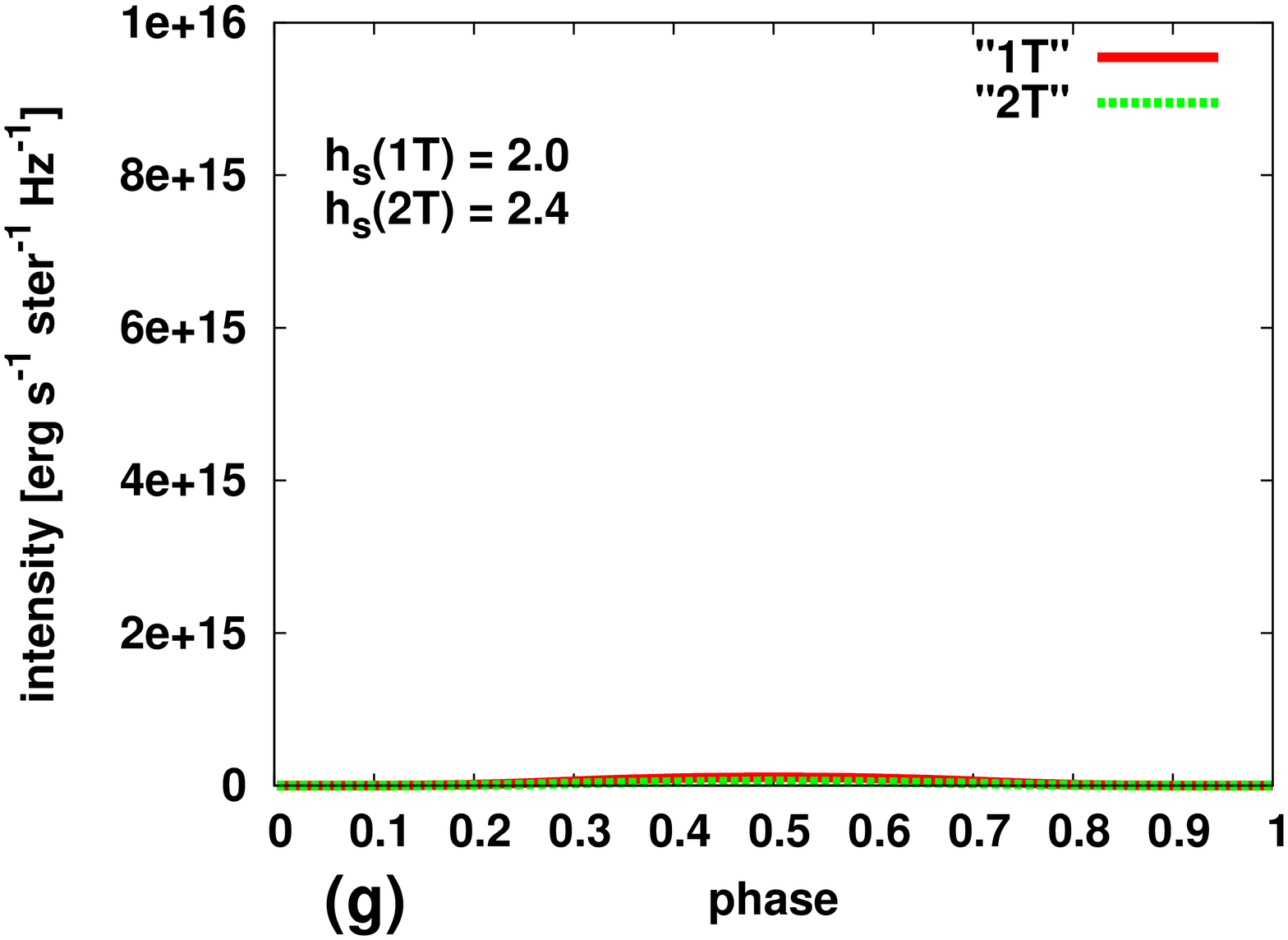}
\includegraphics[angle=-90,scale=0.16]{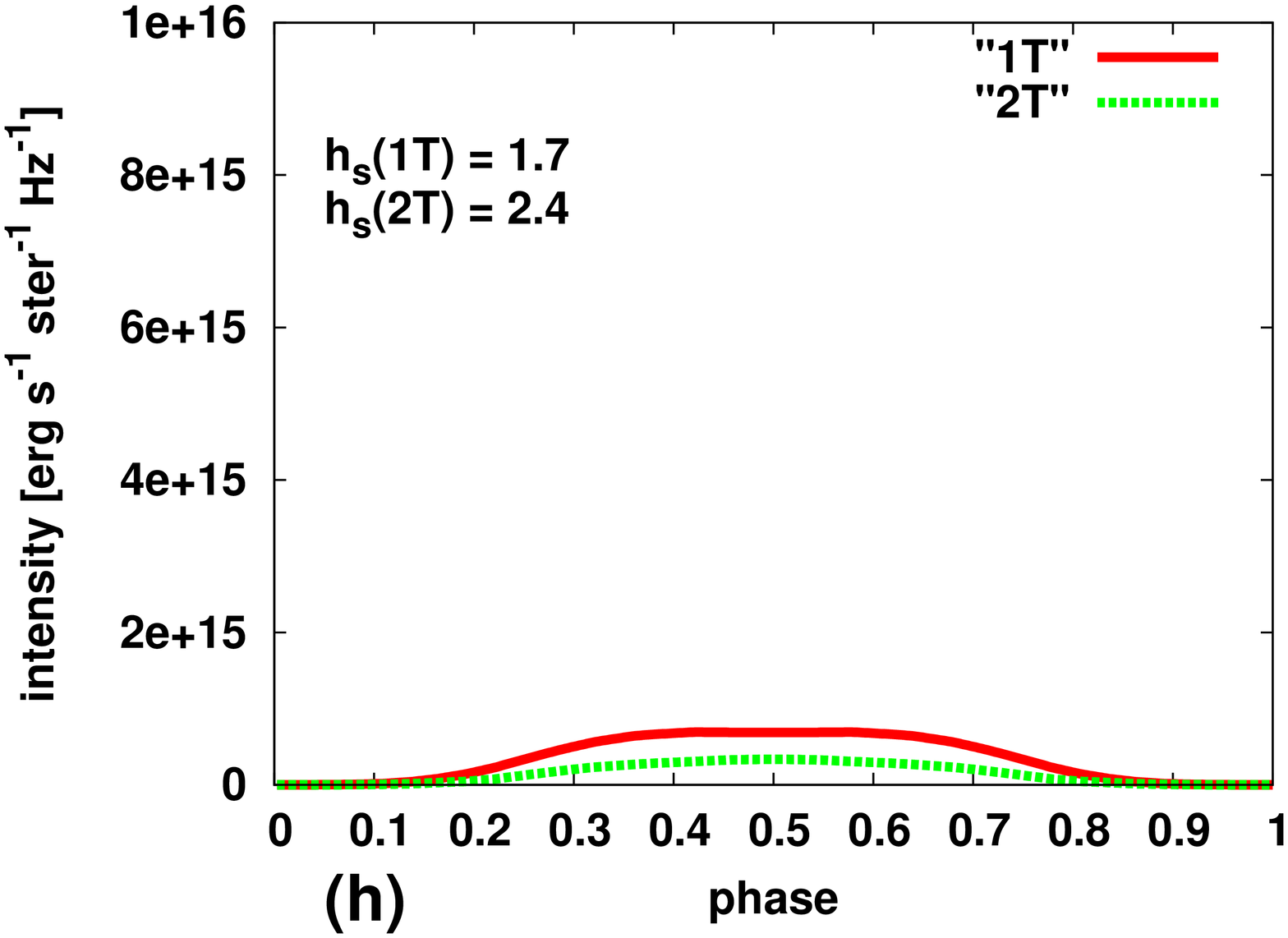}
\includegraphics[angle=-90,scale=0.16]{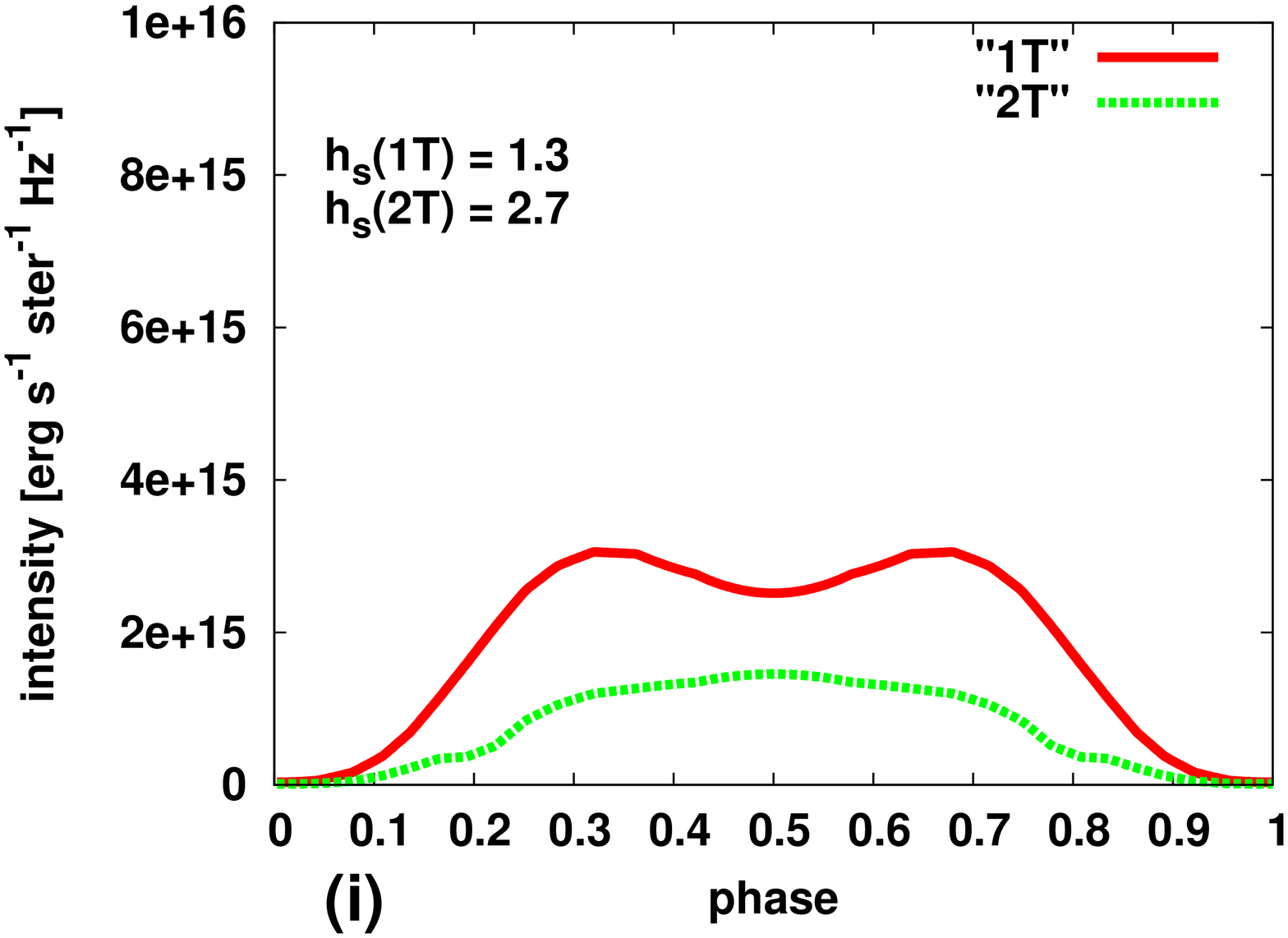}
\begin{picture}(25,100)(0,0)\put(0,-40){\scriptsize \bf \boldmath $\dot{M} = 10^{14}$ \unboldmath}
\end{picture}

\vspace*{-8em}
\begin{picture}(30,100)(0,0)\put(-5,-35){\scriptsize \bf \boldmath B$_{7} = 3$ \unboldmath}
\put(-5,-45){\scriptsize \bf V Filter}\end{picture}
\includegraphics[angle=-90,scale=0.16]{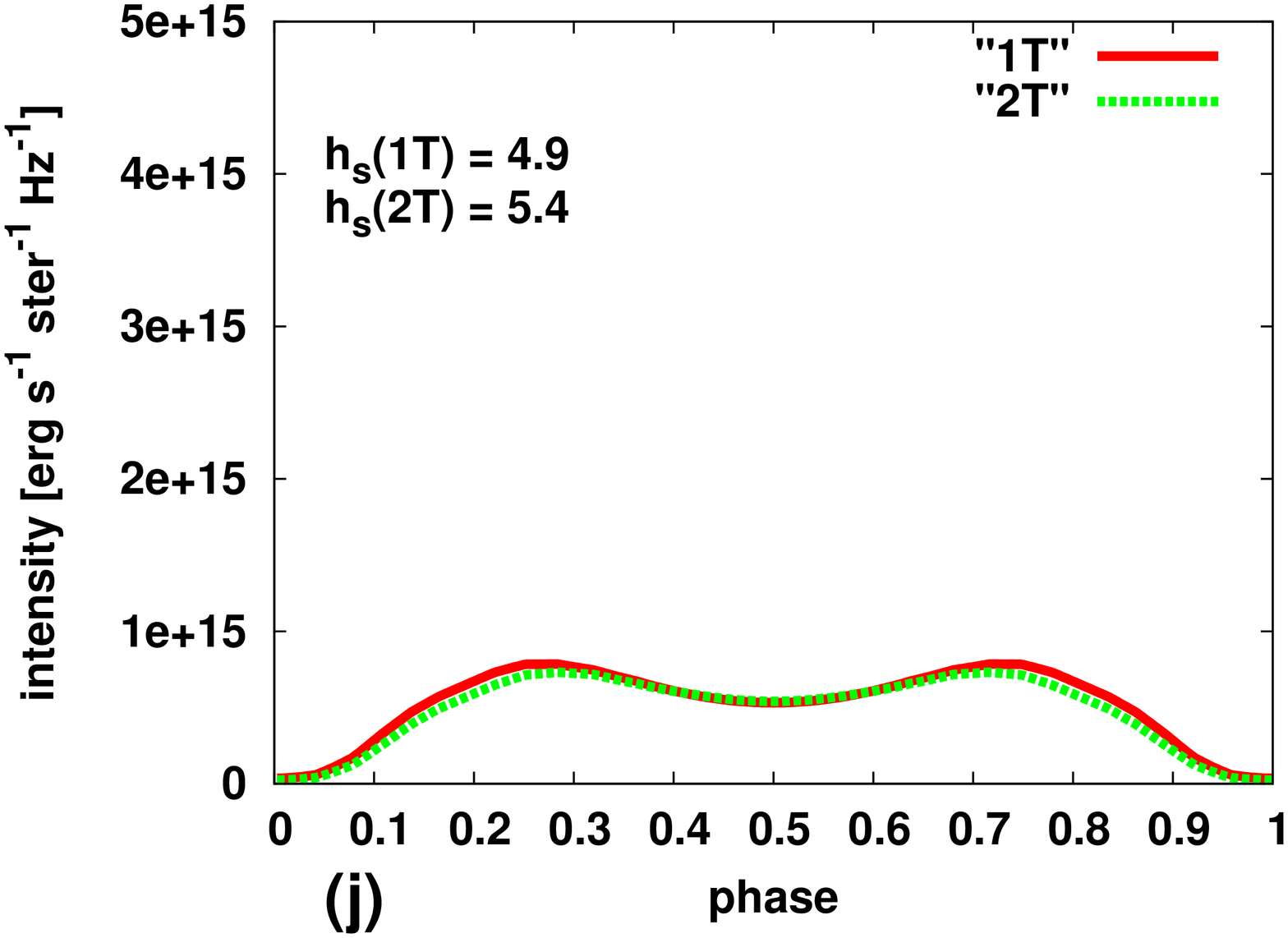}
\includegraphics[angle=-90,scale=0.16]{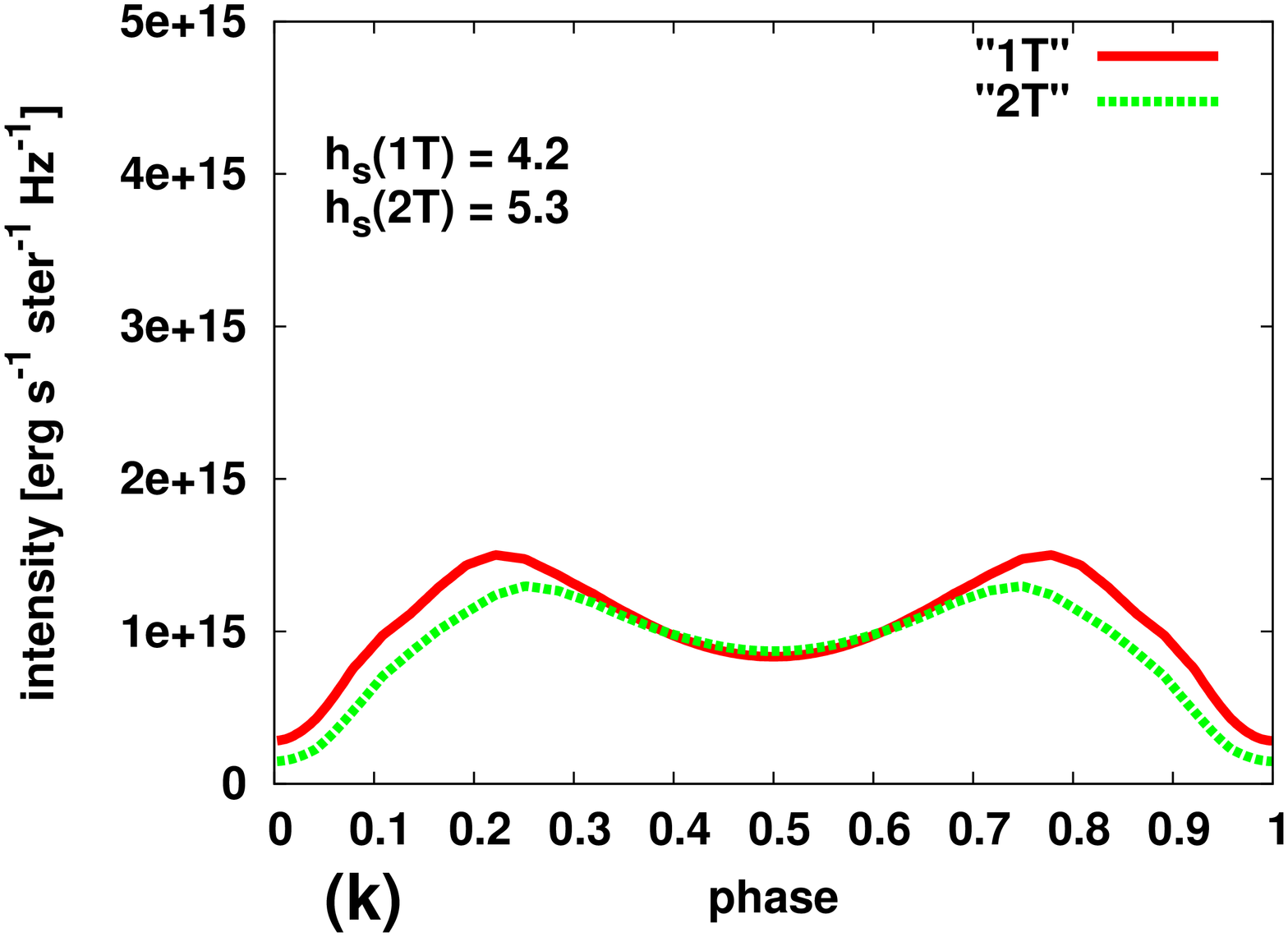}
\includegraphics[angle=-90,scale=0.16]{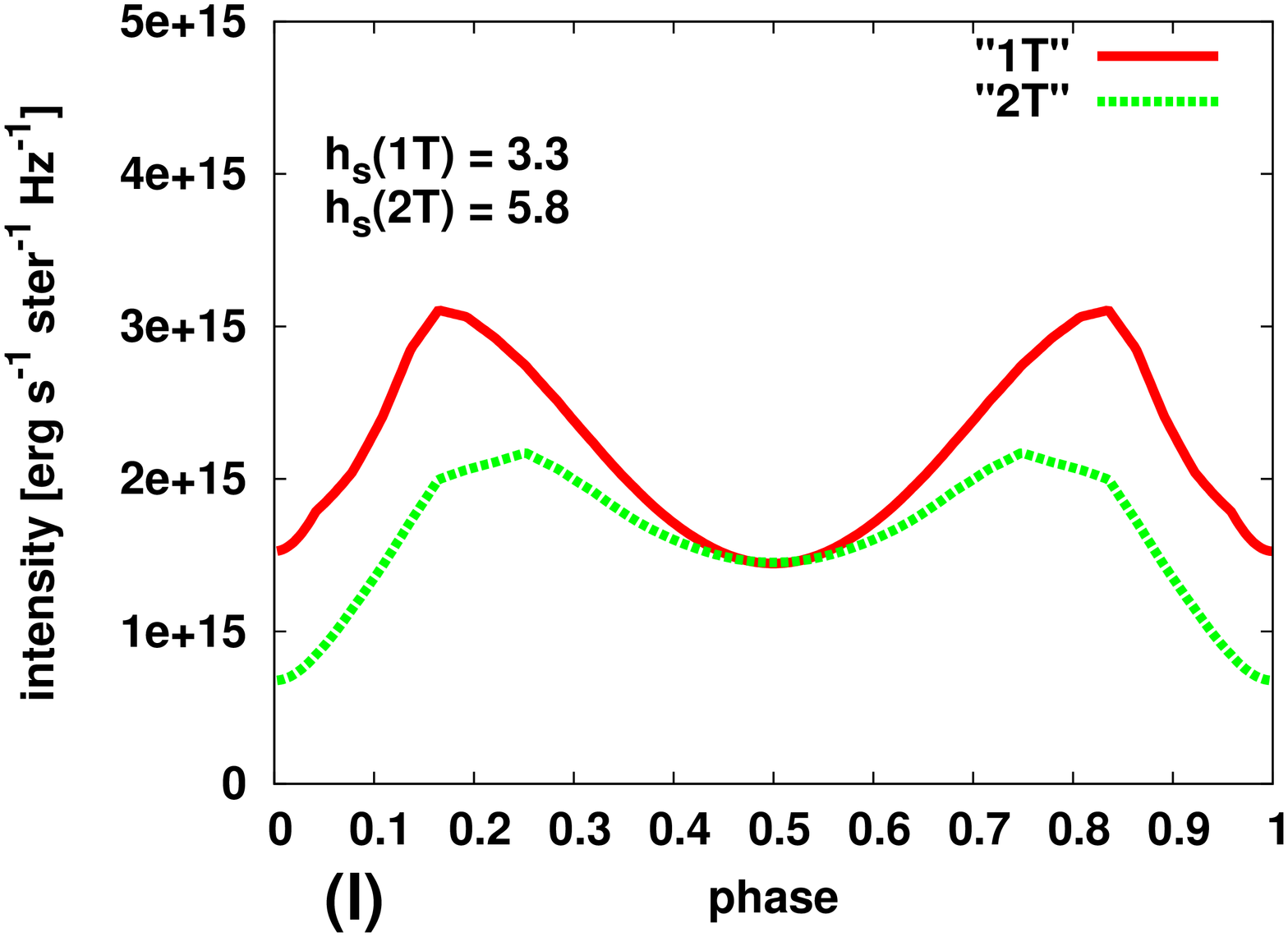}
\begin{picture}(25,100)(0,0)\put(0,-40){\scriptsize \bf \boldmath $\dot{M} = 10^{14}$ \unboldmath}
\end{picture}

\vspace*{-8em}
\begin{picture}(30,100)(0,0)\put(-5,-35){\scriptsize \bf \boldmath B$_{7} = 5$ \unboldmath}
\put(-5,-45){\scriptsize \bf V Filter}\end{picture}
\includegraphics[angle=-90,scale=0.16]{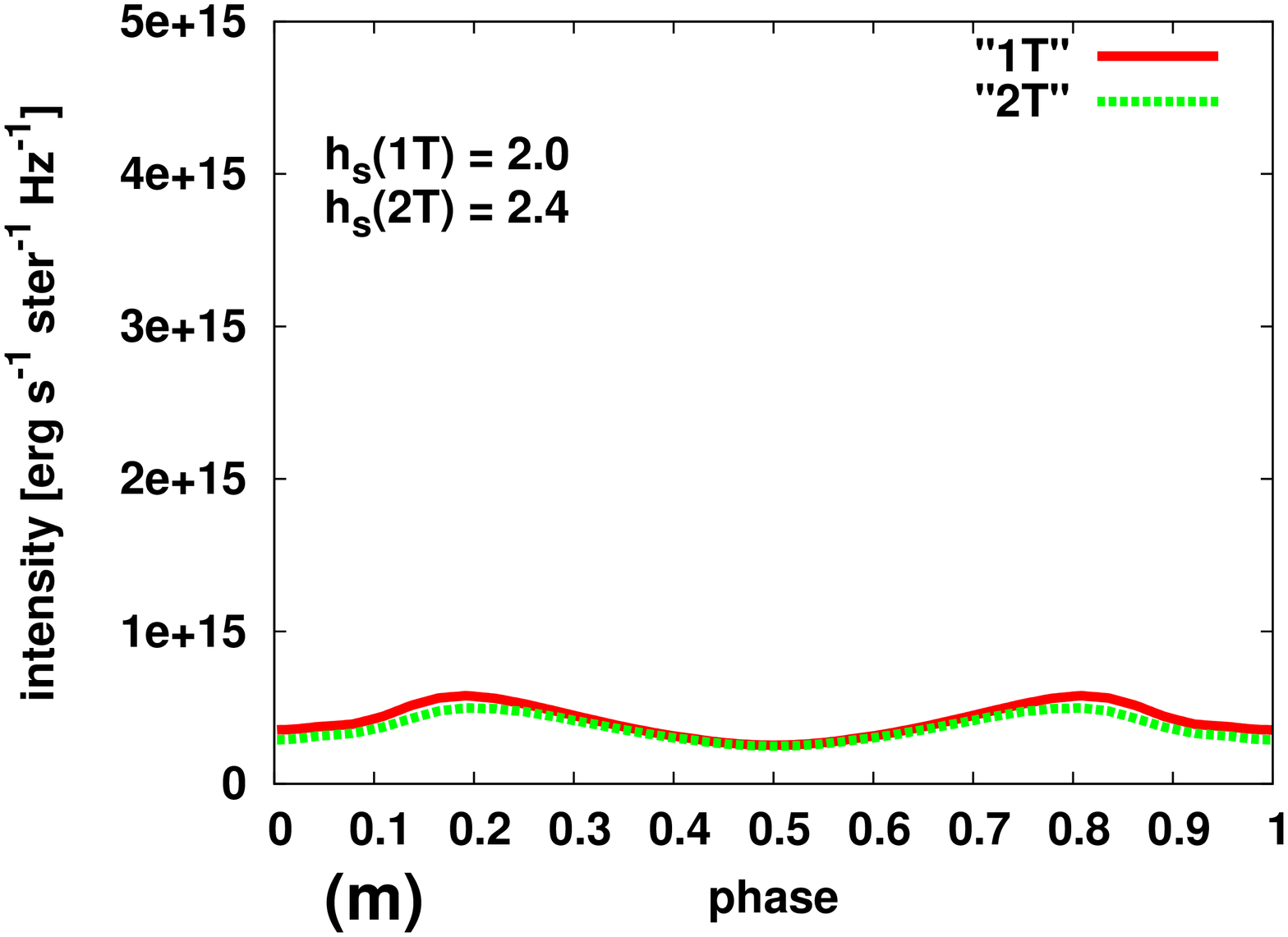}
\includegraphics[angle=-90,scale=0.16]{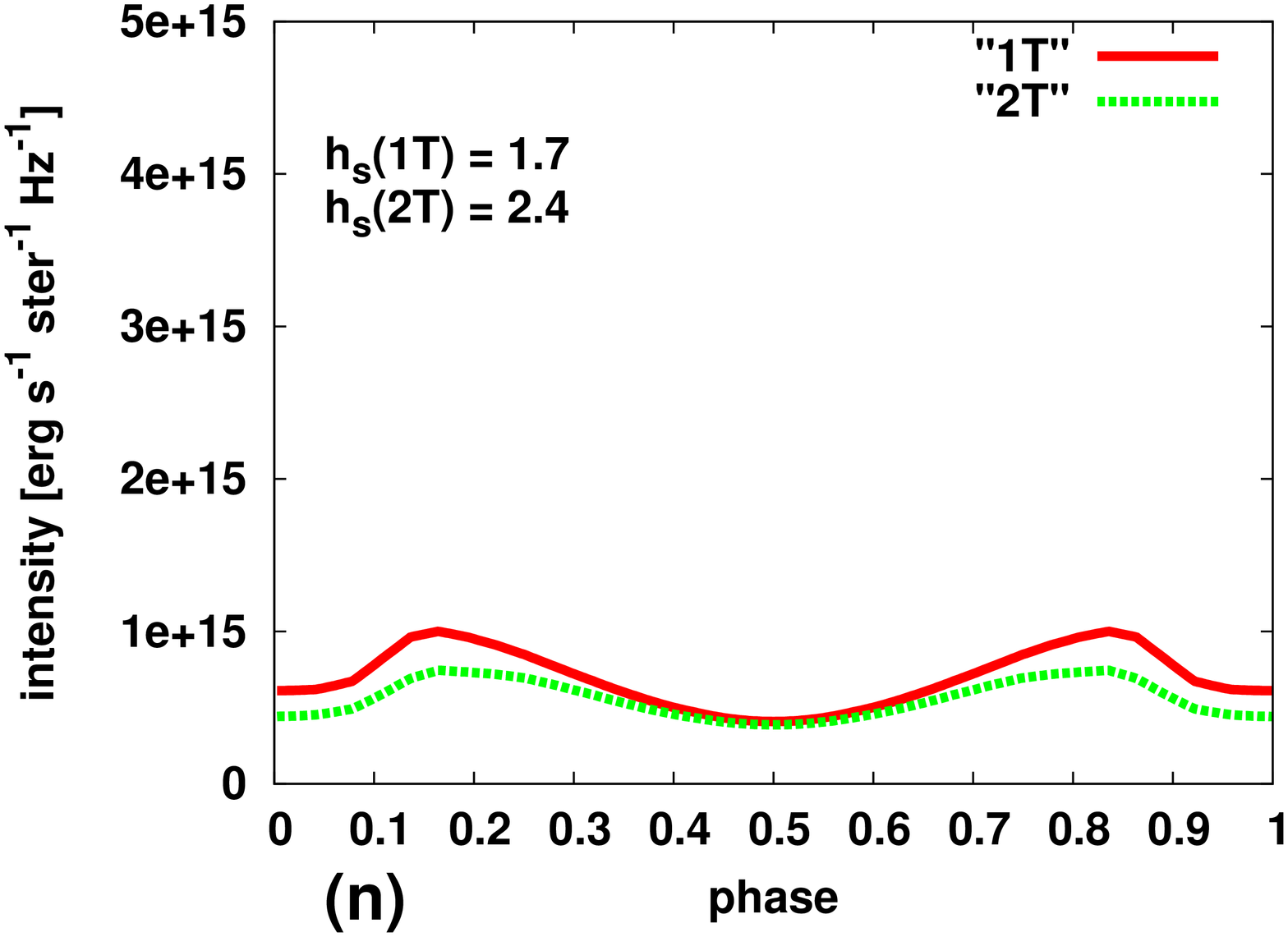}
\includegraphics[angle=-90,scale=0.16]{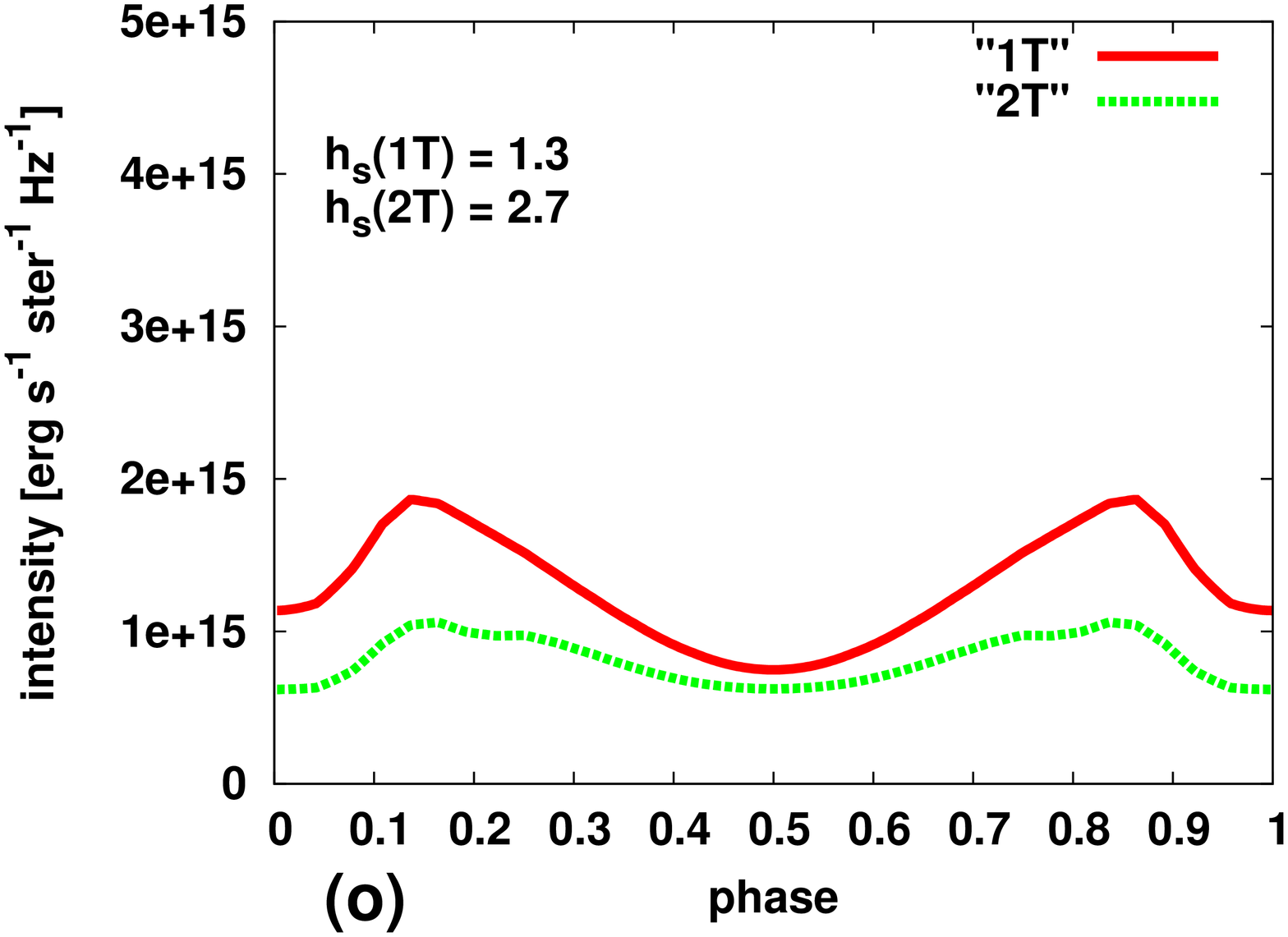}
\begin{picture}(25,100)(0,0)\put(0,-40){\scriptsize \bf \boldmath $\dot{M} = 10^{14}$ \unboldmath}
\end{picture}
\end{center}
\caption{Intensity orbital light curves comparing predictions for the
one-temperature model ({\tt "1T"}, red line) with the two-temperature ({\tt "2T"}, green line) model.
The figure is laid out in the same way as Fig.~\ref{linorb}. \label{intorb}}
\end{figure*}

\begin{figure*}
\begin{center}

\vspace*{-8em}
\begin{picture}(30,100)(0,0)\put(-5,-35){\scriptsize \bf \boldmath B$_{7} = 1$ \unboldmath}
\put(-5,-45){\scriptsize \bf J Filter}\end{picture}
\includegraphics[angle=-90,scale=0.16]{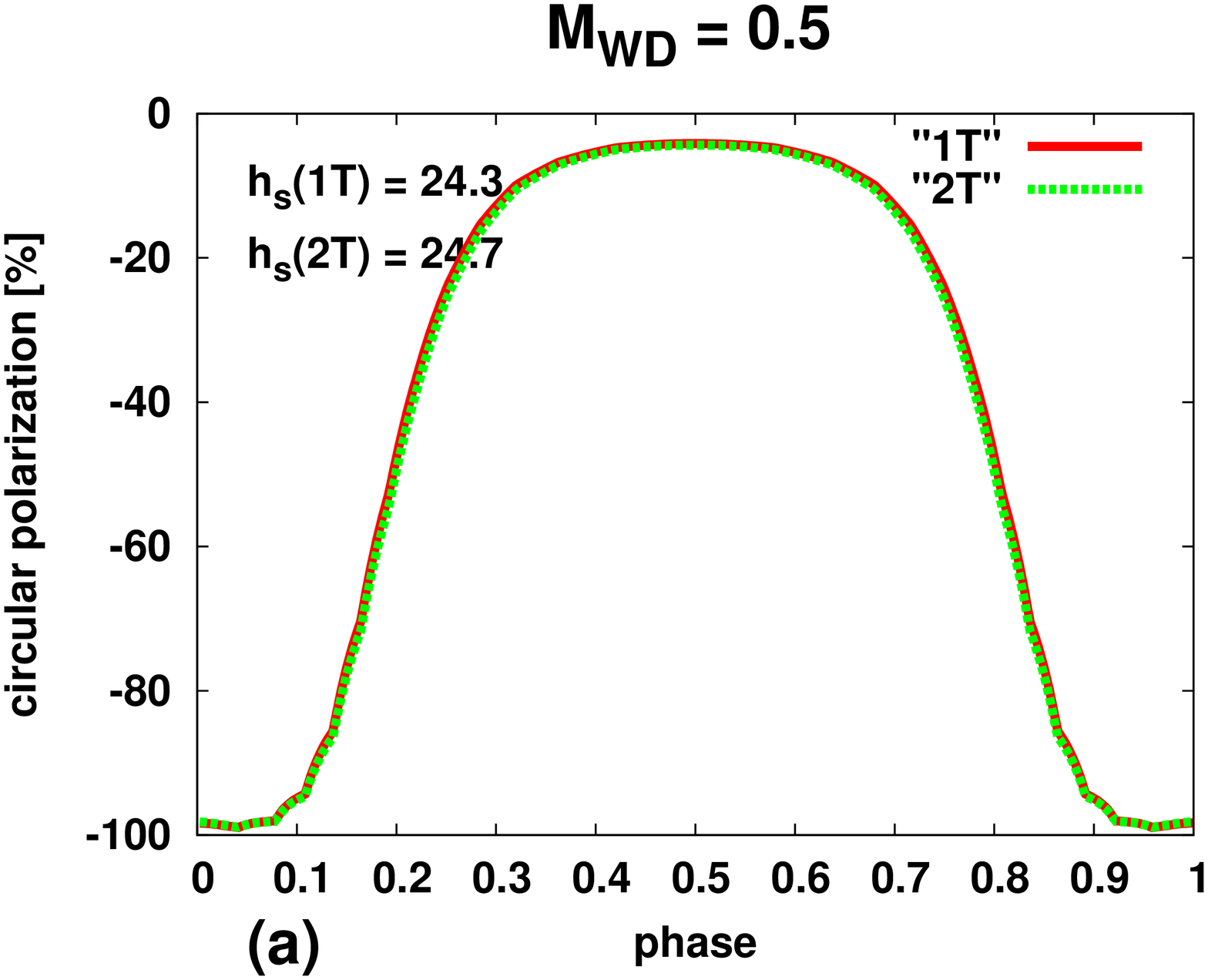}
\includegraphics[angle=-90,scale=0.16]{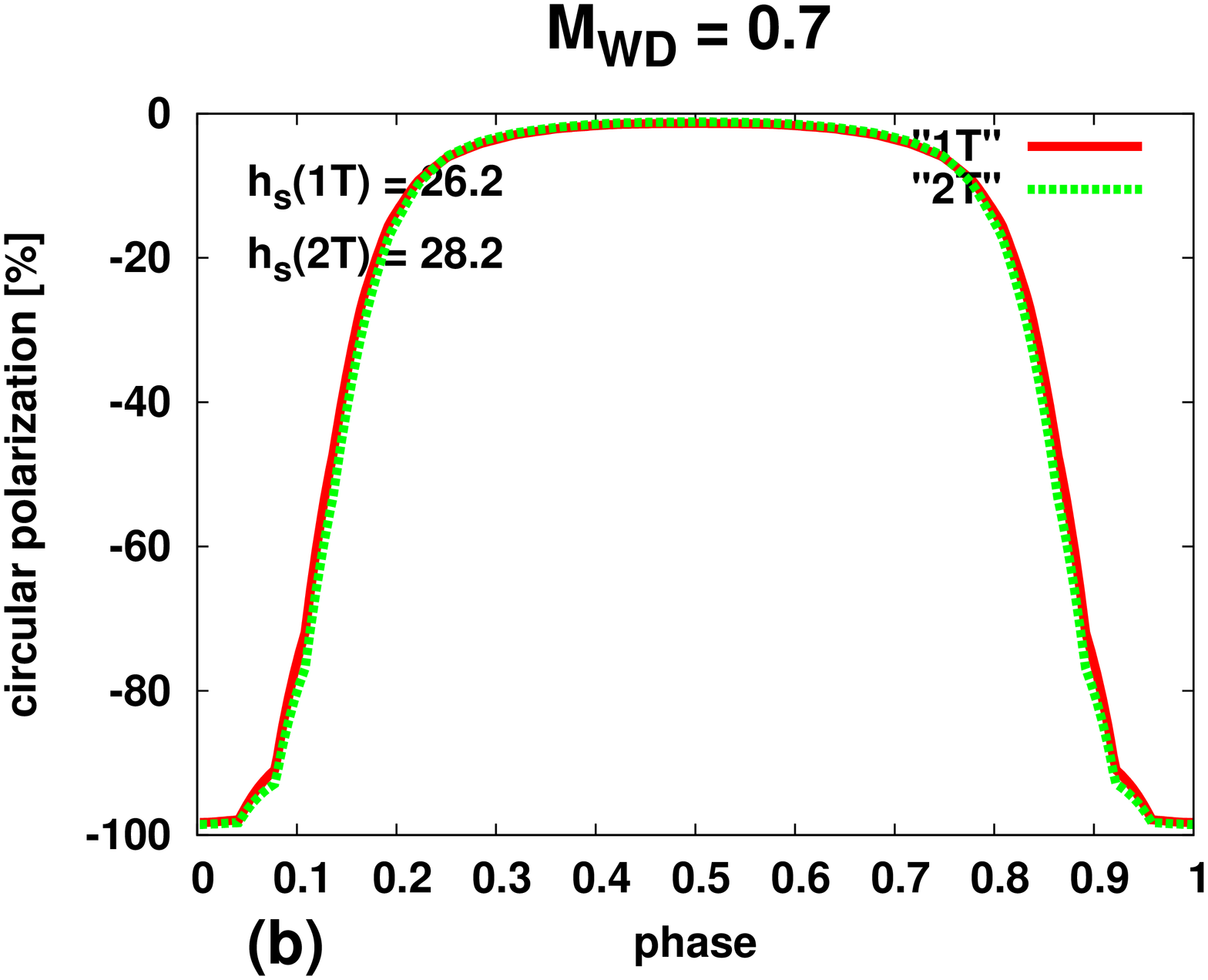}
\includegraphics[angle=-90,scale=0.16]{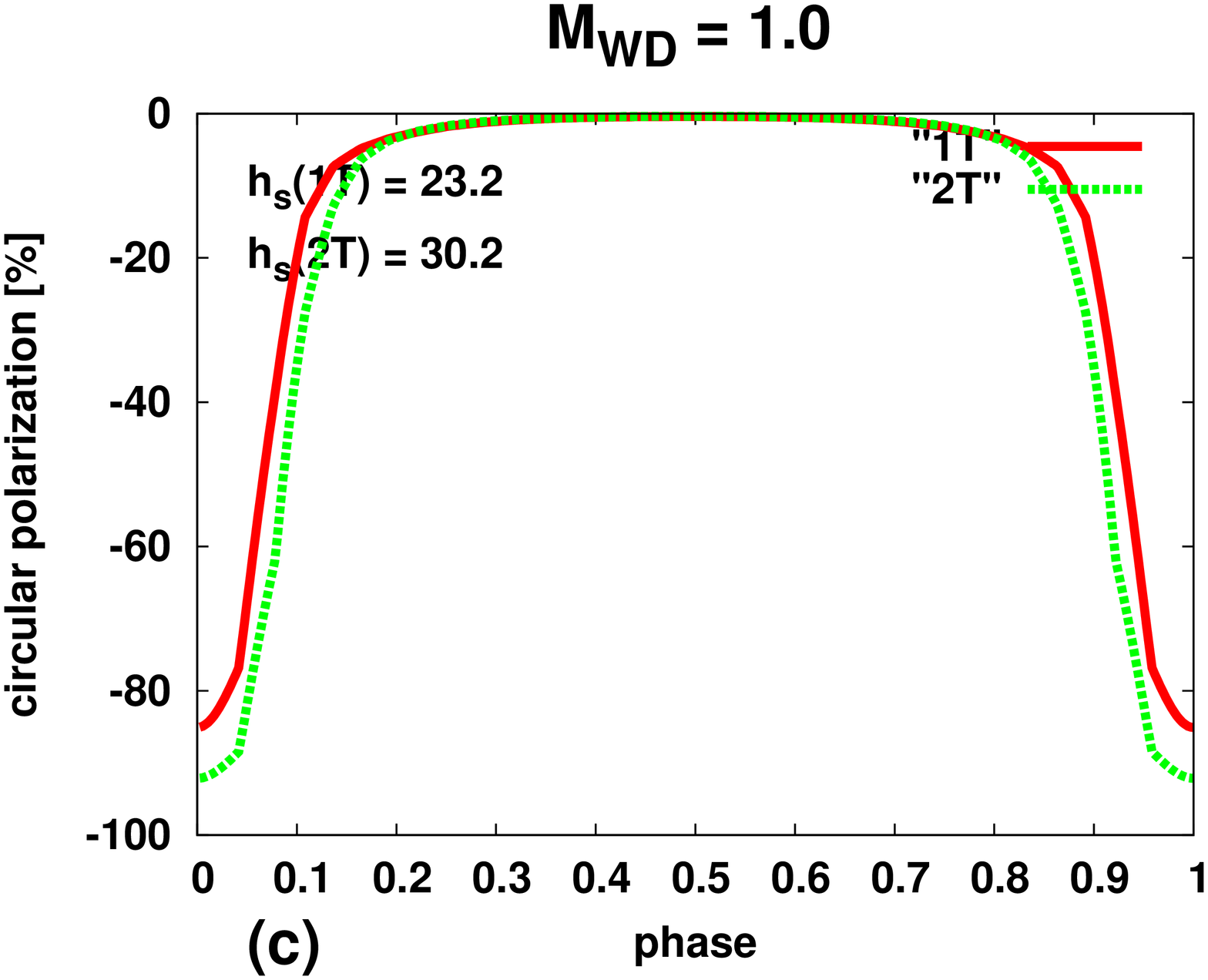}
\begin{picture}(25,100)(0,0)\put(0,-40){\scriptsize \bf \boldmath $\dot{M} = 10^{14}$ \unboldmath}
\end{picture}

\vspace*{-8em}
\begin{picture}(30,100)(0,0)\put(-5,-35){\scriptsize \bf \boldmath B$_{7} = 3$ \unboldmath}
\put(-5,-45){\scriptsize \bf U Filter}\end{picture}
\includegraphics[angle=-90,scale=0.16]{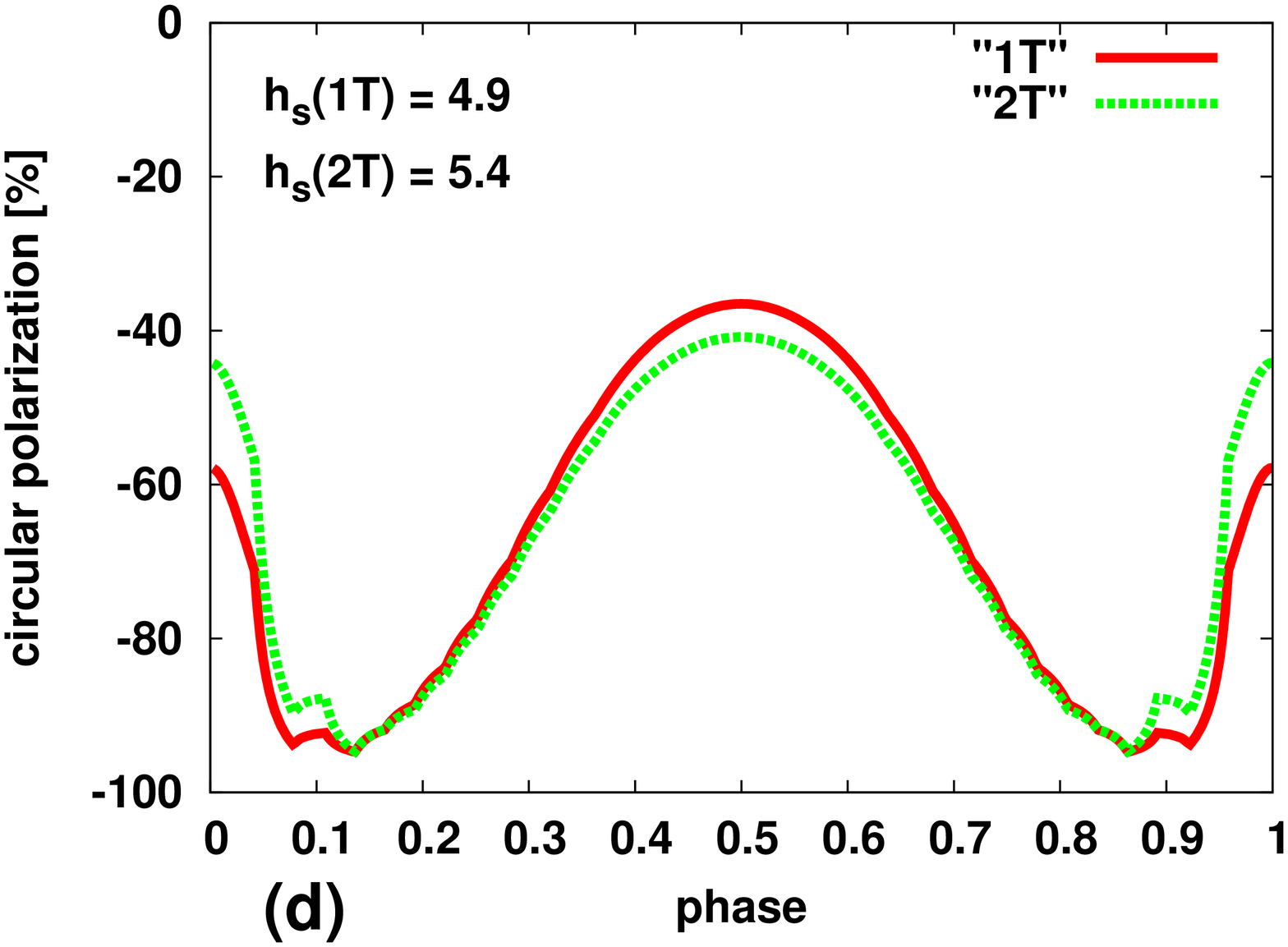}
\includegraphics[angle=-90,scale=0.16]{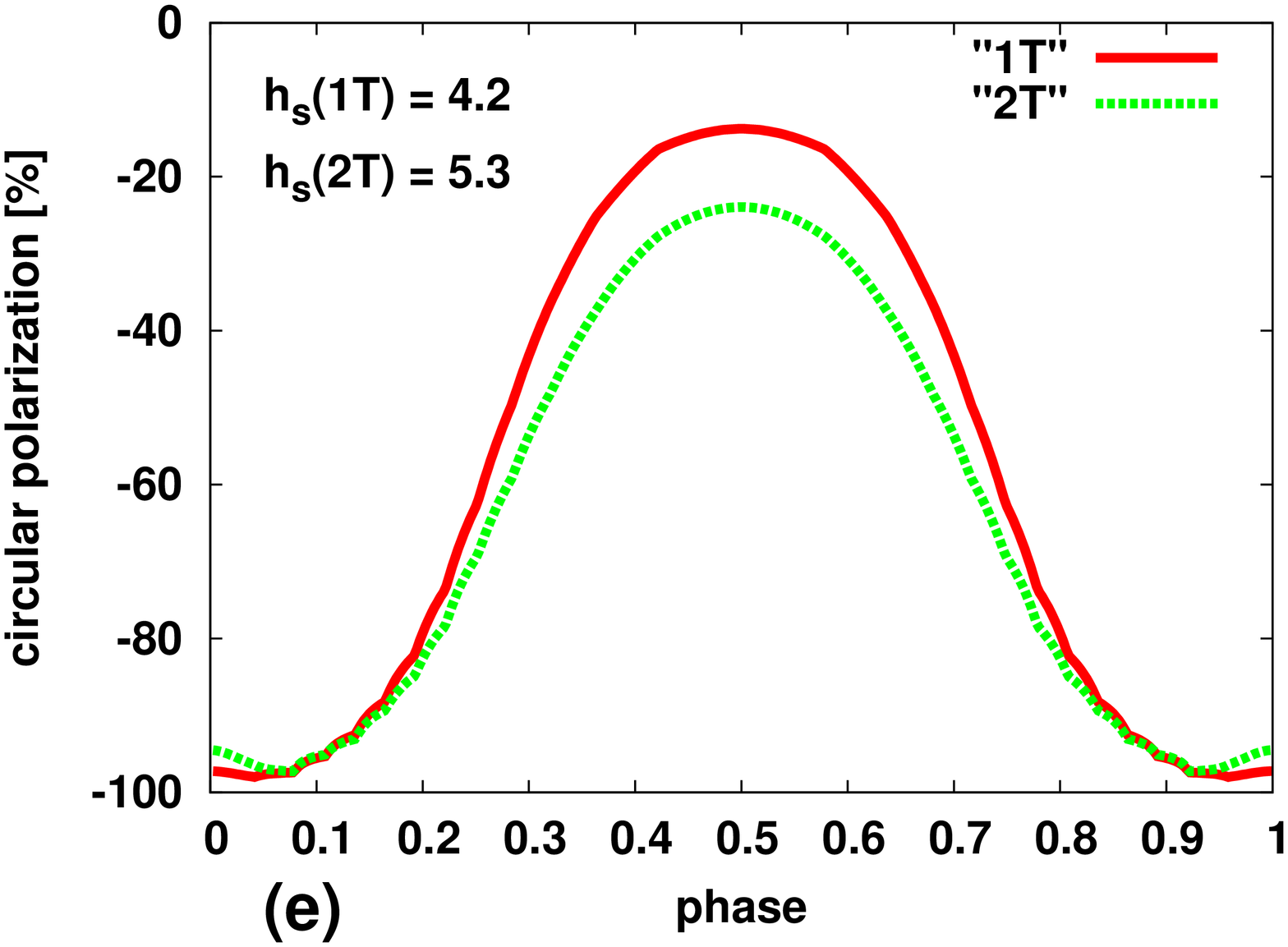}
\includegraphics[angle=-90,scale=0.16]{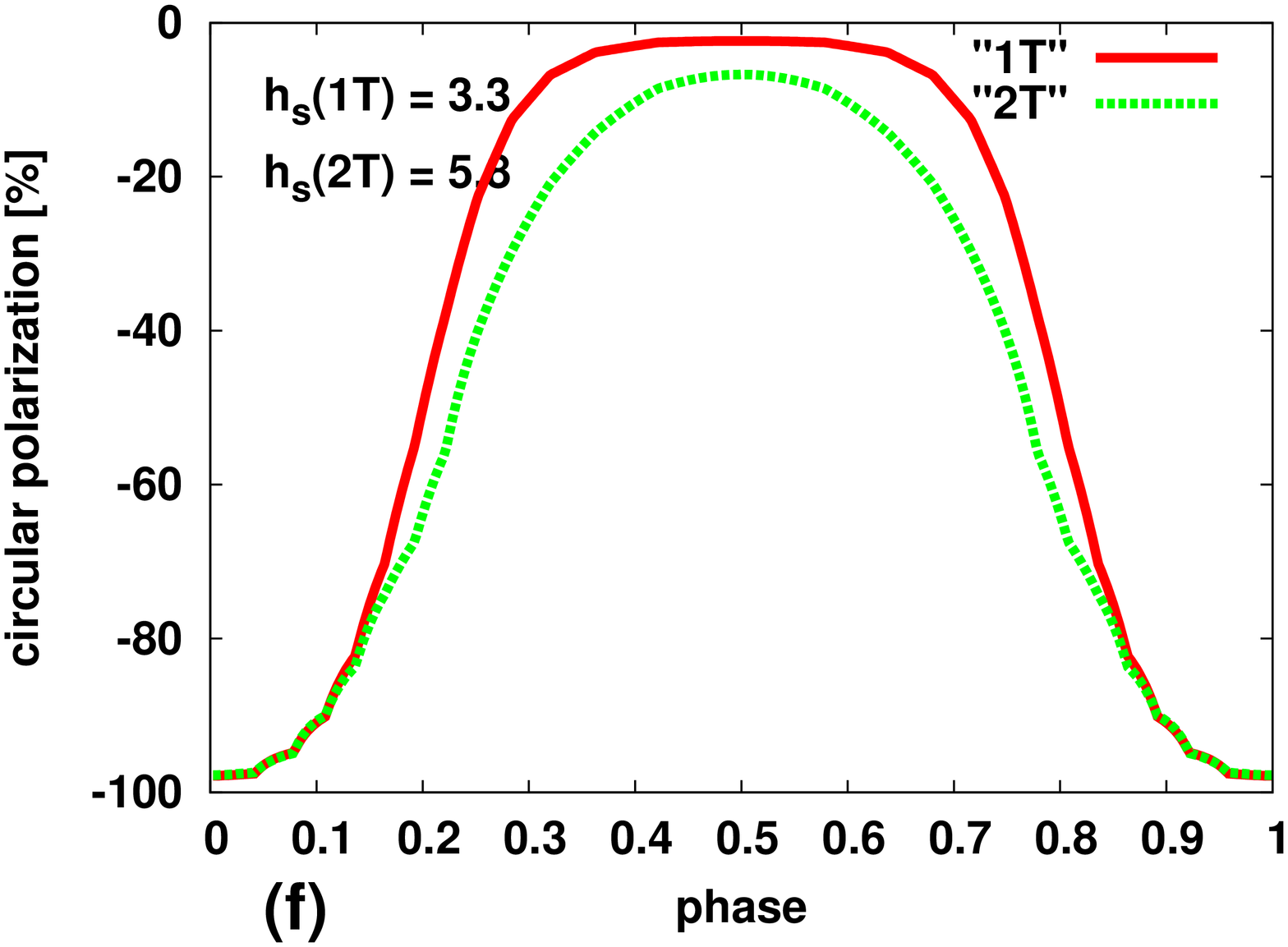}
\begin{picture}(25,100)(0,0)\put(0,-40){\scriptsize \bf \boldmath $\dot{M} = 10^{14}$ \unboldmath}
\end{picture}

\vspace*{-8em}
\begin{picture}(30,100)(0,0)\put(-5,-35){\scriptsize \bf \boldmath B$_{7} = 5$ \unboldmath}
\put(-5,-45){\scriptsize \bf H = 9.92}\end{picture}
\includegraphics[angle=-90,scale=0.16]{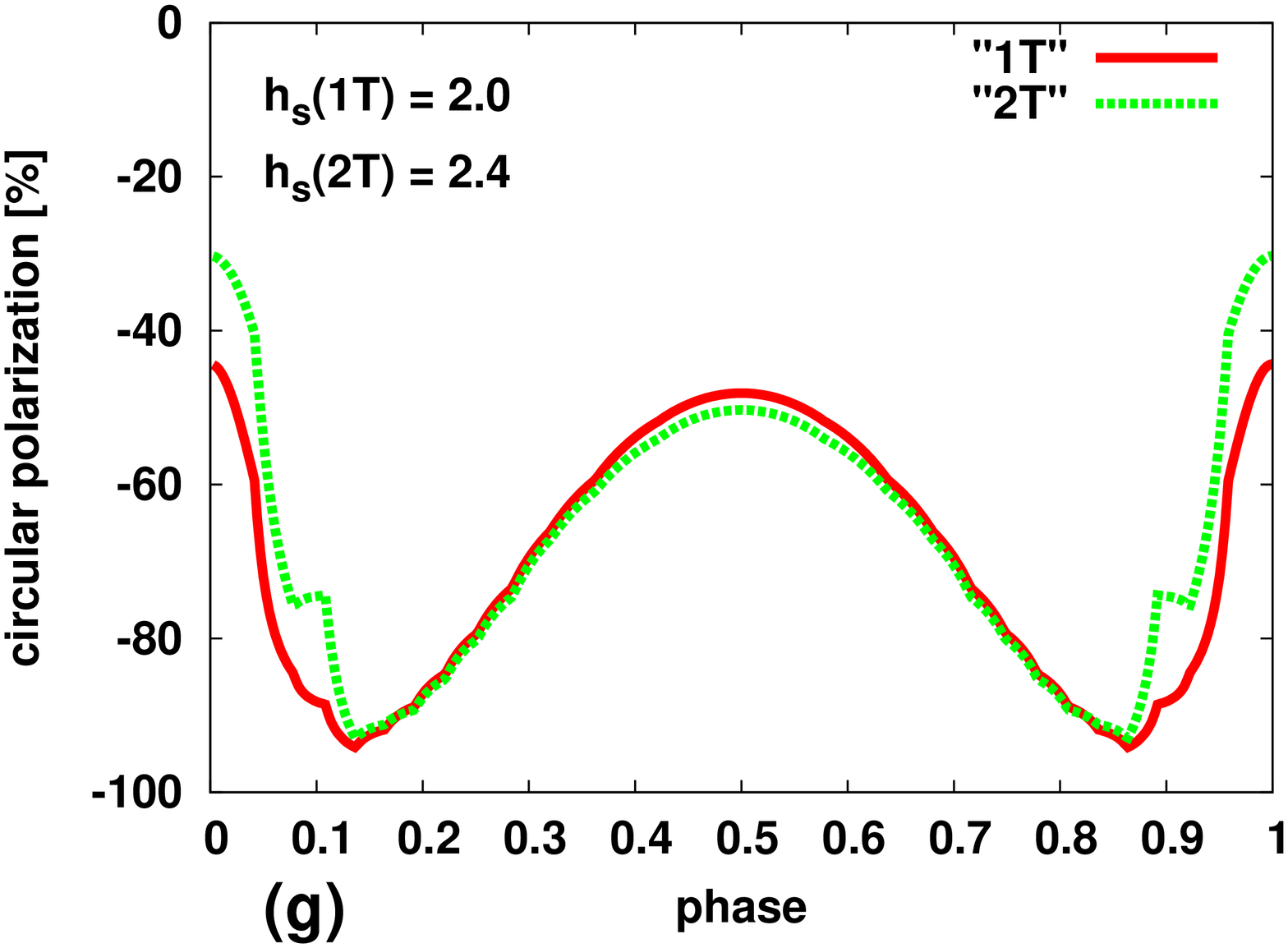}
\includegraphics[angle=-90,scale=0.16]{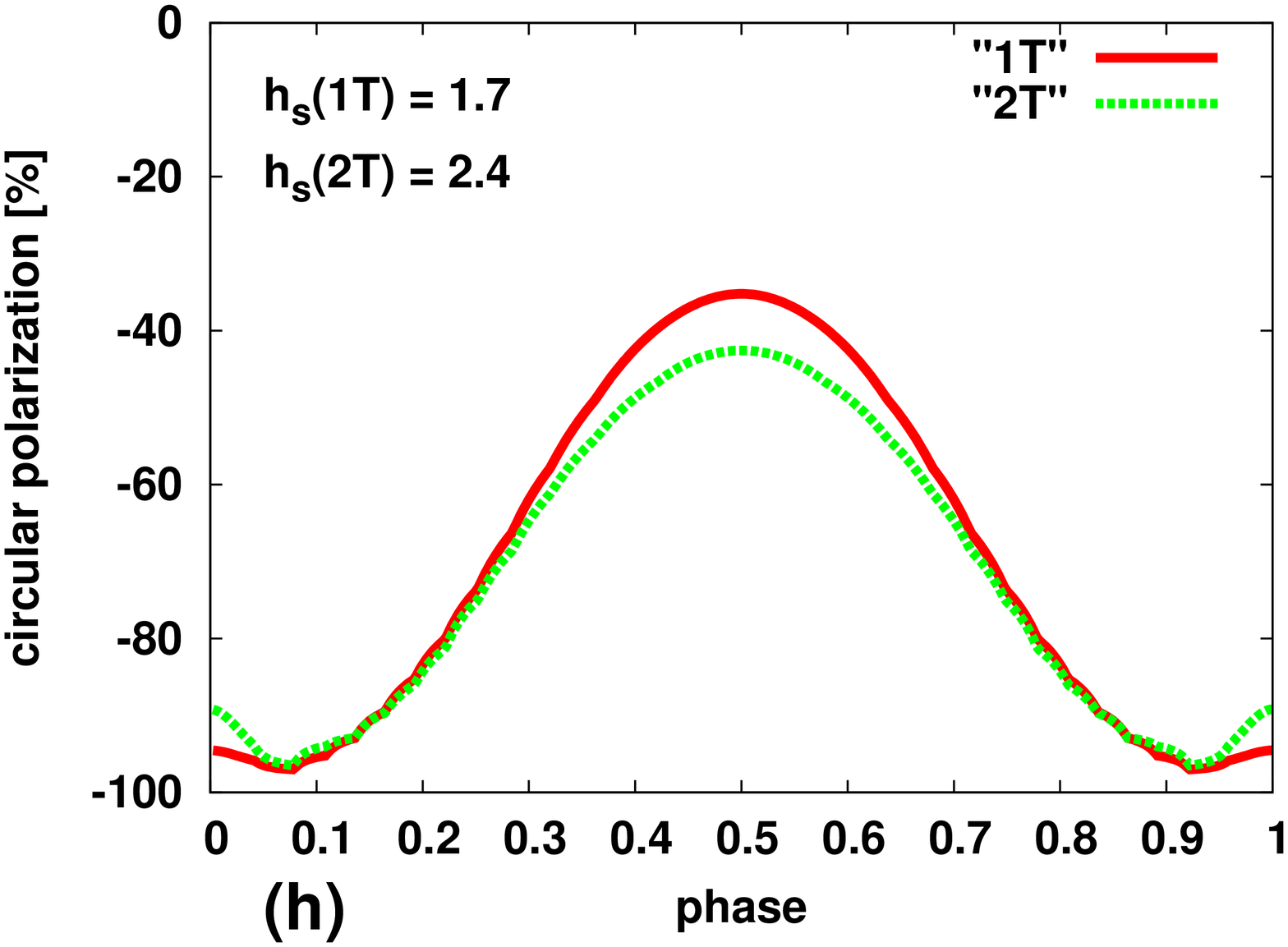}
\includegraphics[angle=-90,scale=0.16]{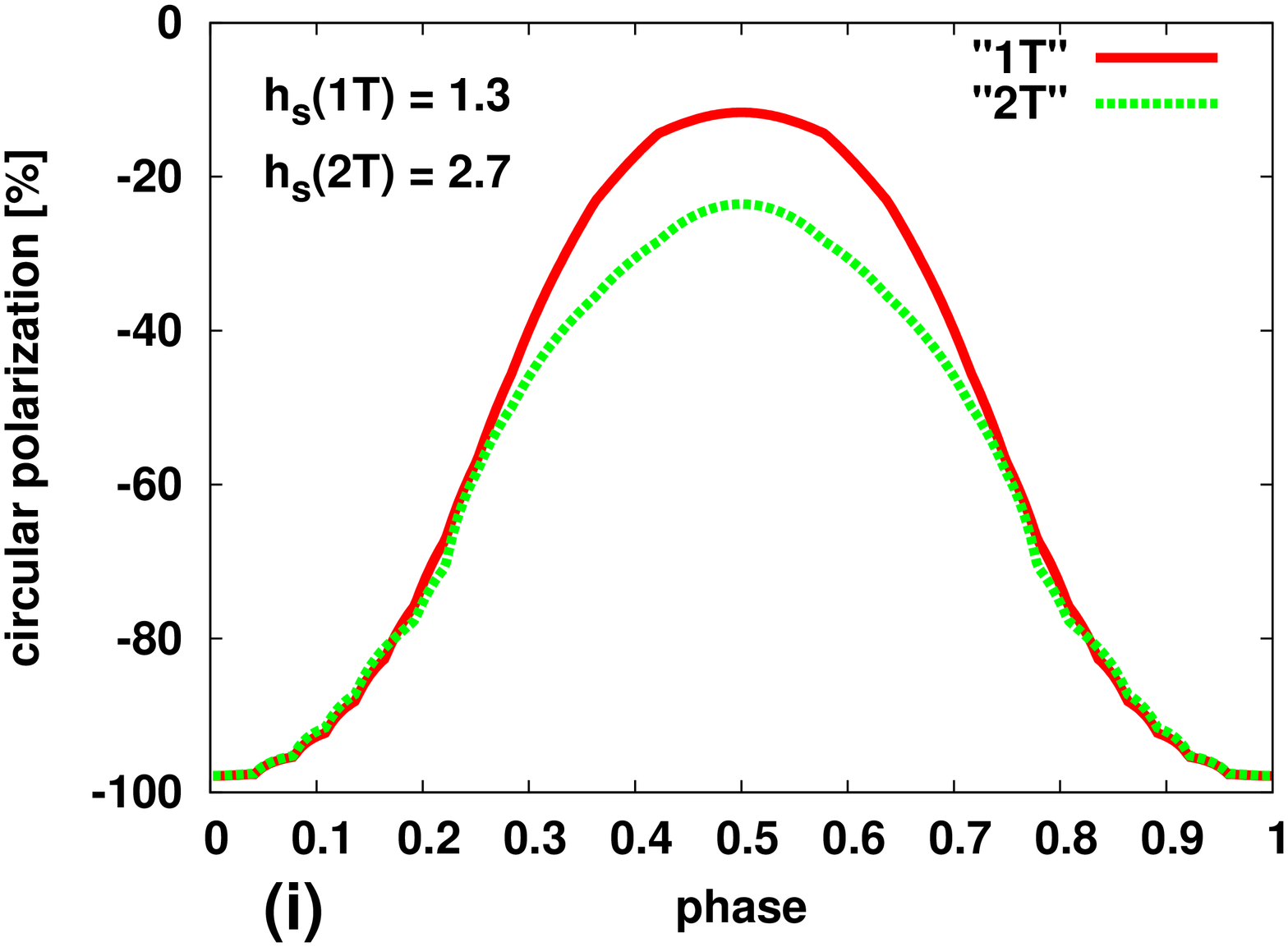}
\begin{picture}(25,100)(0,0)\put(0,-40){\scriptsize \bf \boldmath $\dot{M} = 10^{14}$ \unboldmath}
\end{picture}

\vspace*{-8em}
\begin{picture}(30,100)(0,0)\put(-5,-35){\scriptsize \bf \boldmath B$_{7} = 3$ \unboldmath}
\put(-5,-45){\scriptsize \bf V Filter}\end{picture}
\includegraphics[angle=-90,scale=0.16]{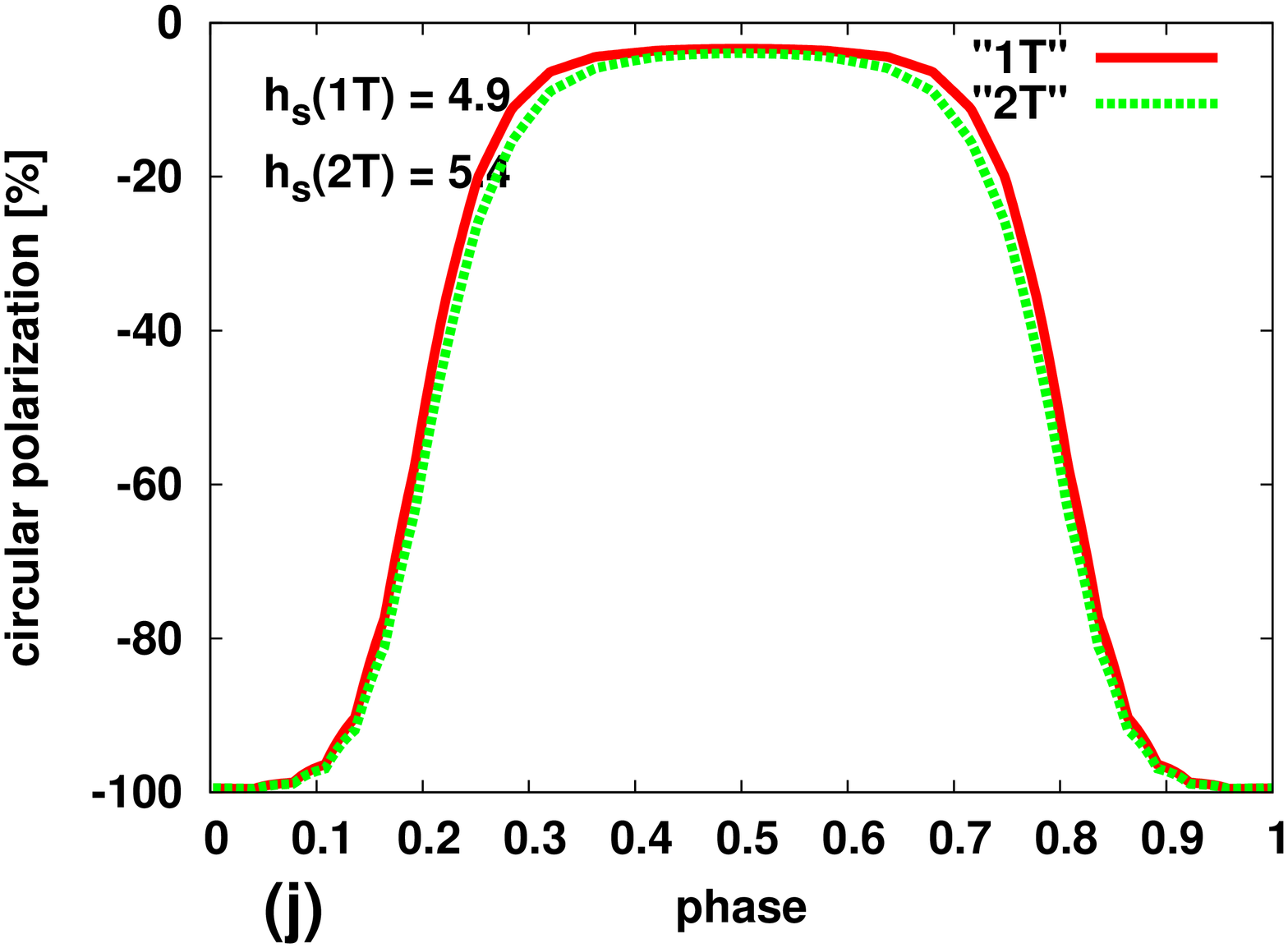}
\includegraphics[angle=-90,scale=0.16]{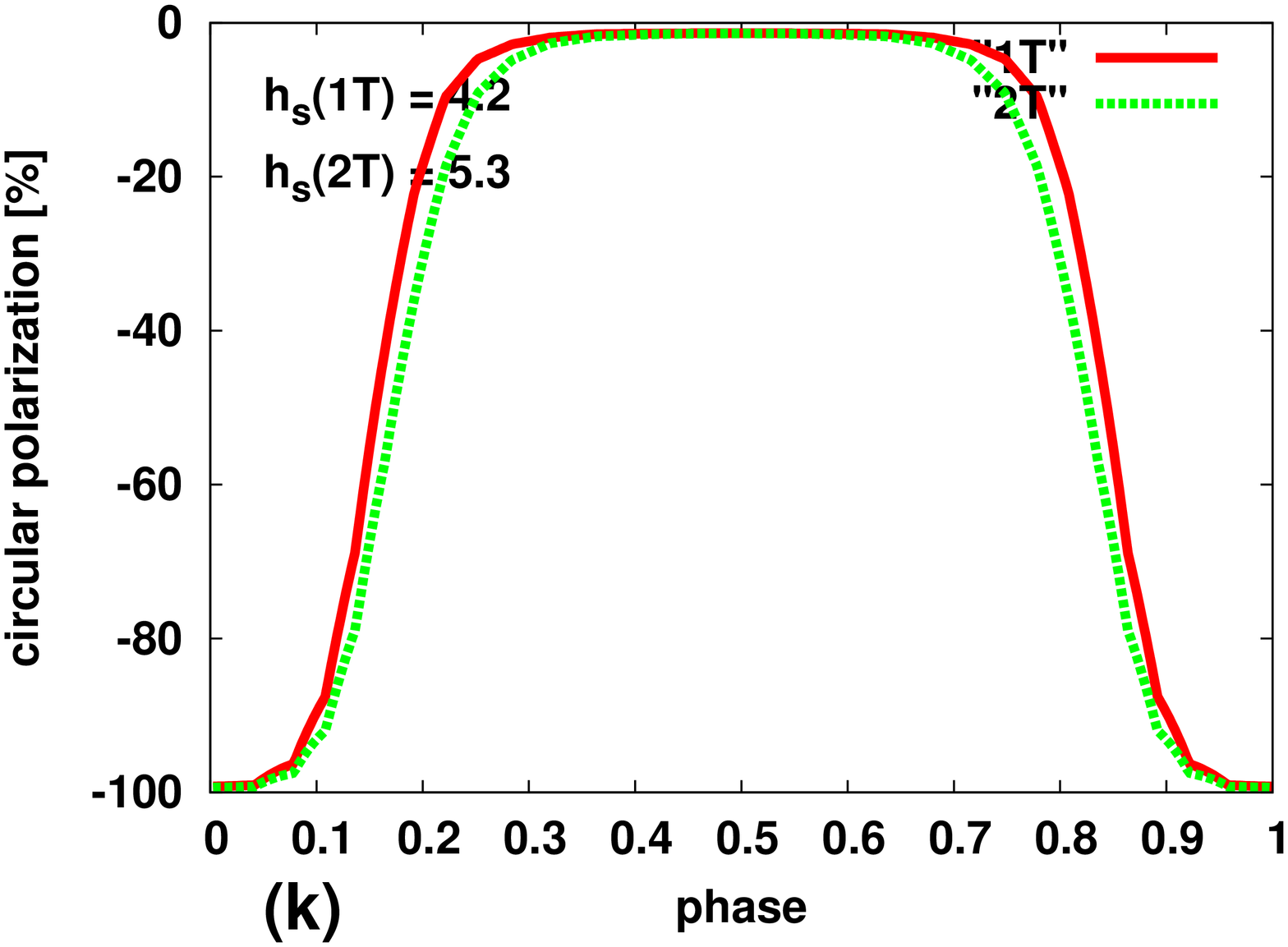}
\includegraphics[angle=-90,scale=0.16]{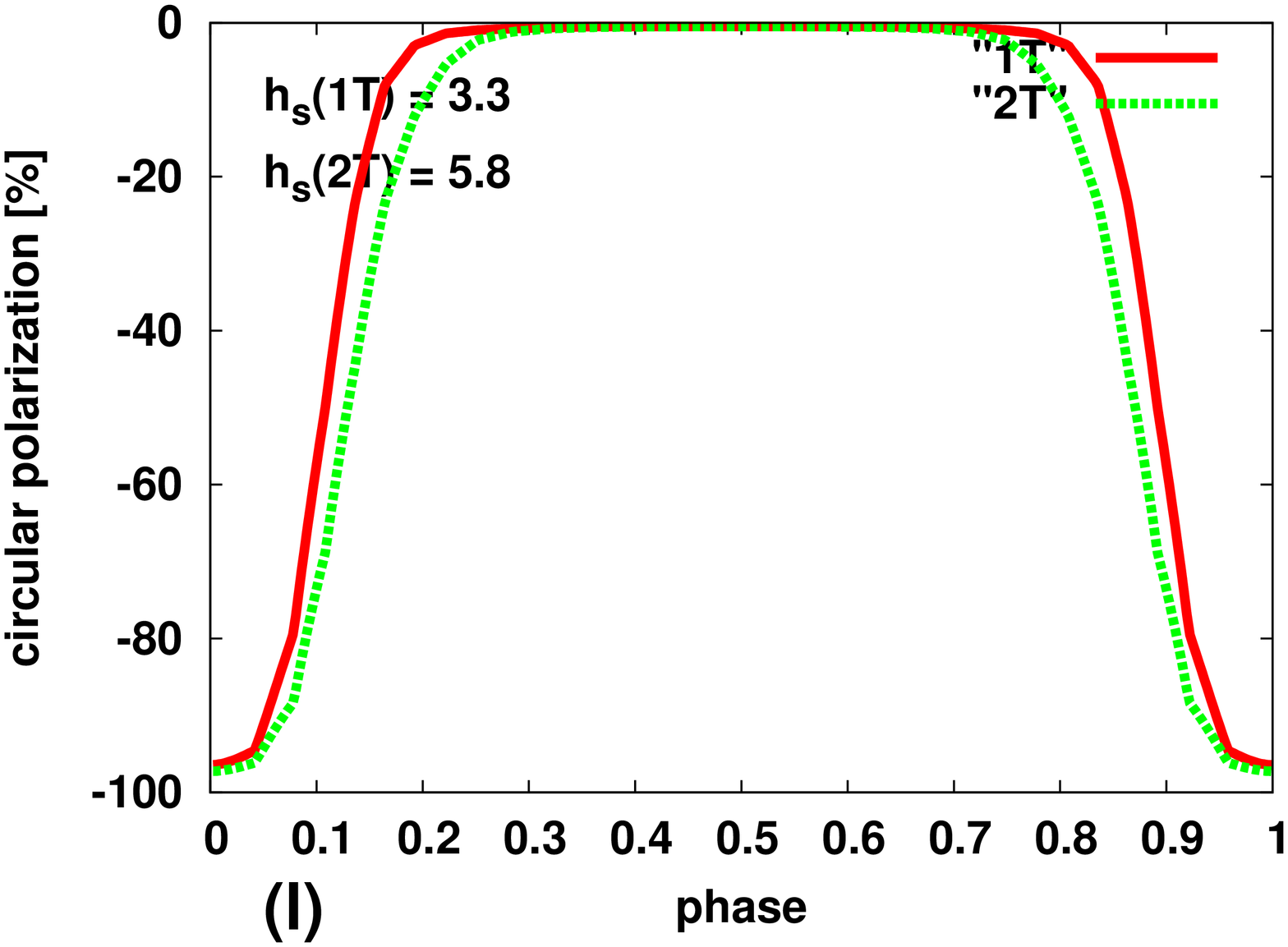}
\begin{picture}(25,100)(0,0)\put(0,-40){\scriptsize \bf \boldmath $\dot{M} = 10^{14}$ \unboldmath}
\end{picture}

\vspace*{-8em}
\begin{picture}(30,100)(0,0)\put(-5,-35){\scriptsize \bf \boldmath B$_{7} = 5$ \unboldmath}
\put(-5,-45){\scriptsize \bf V Filter}\end{picture}
\includegraphics[angle=-90,scale=0.16]{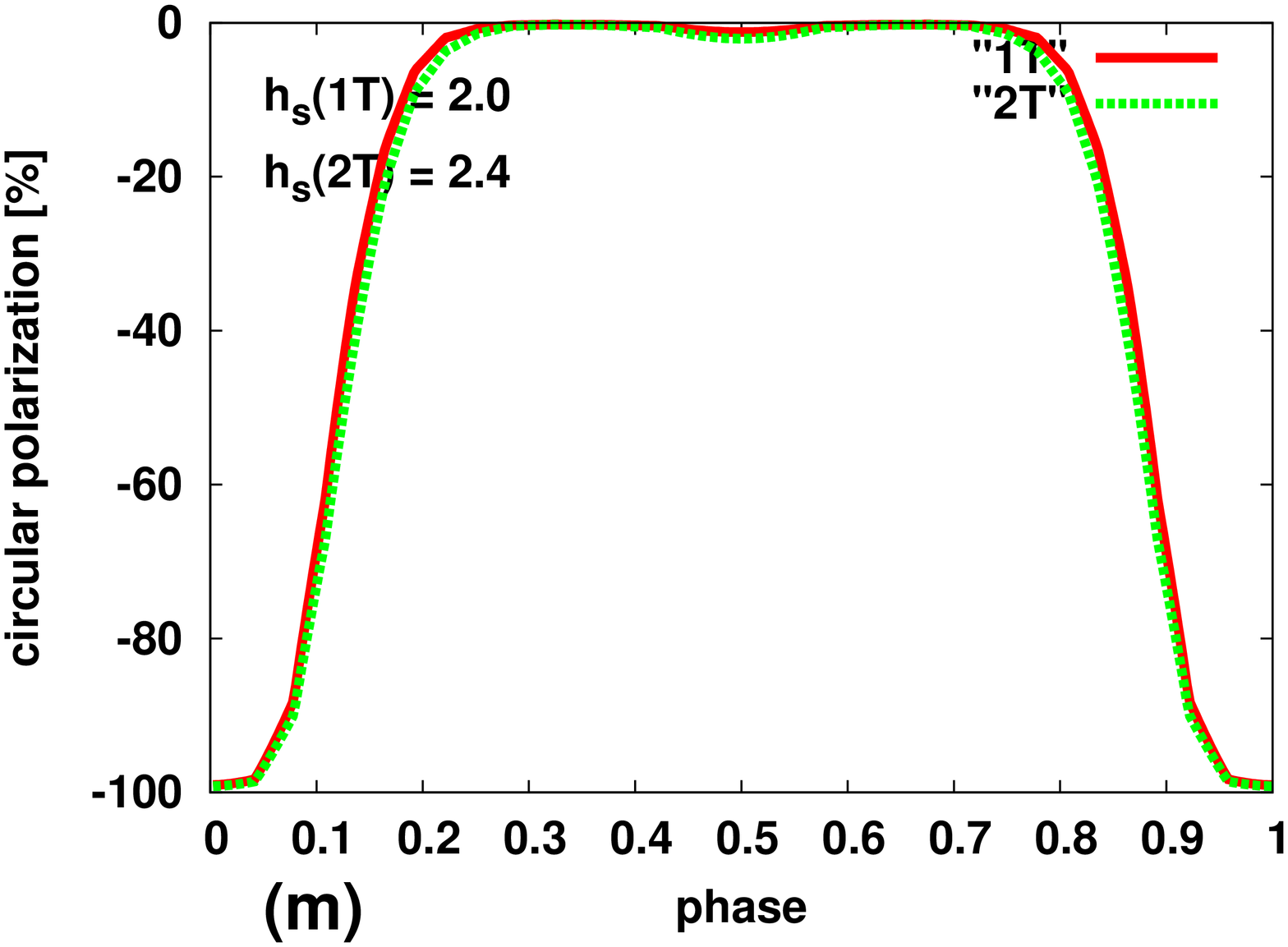}
\includegraphics[angle=-90,scale=0.16]{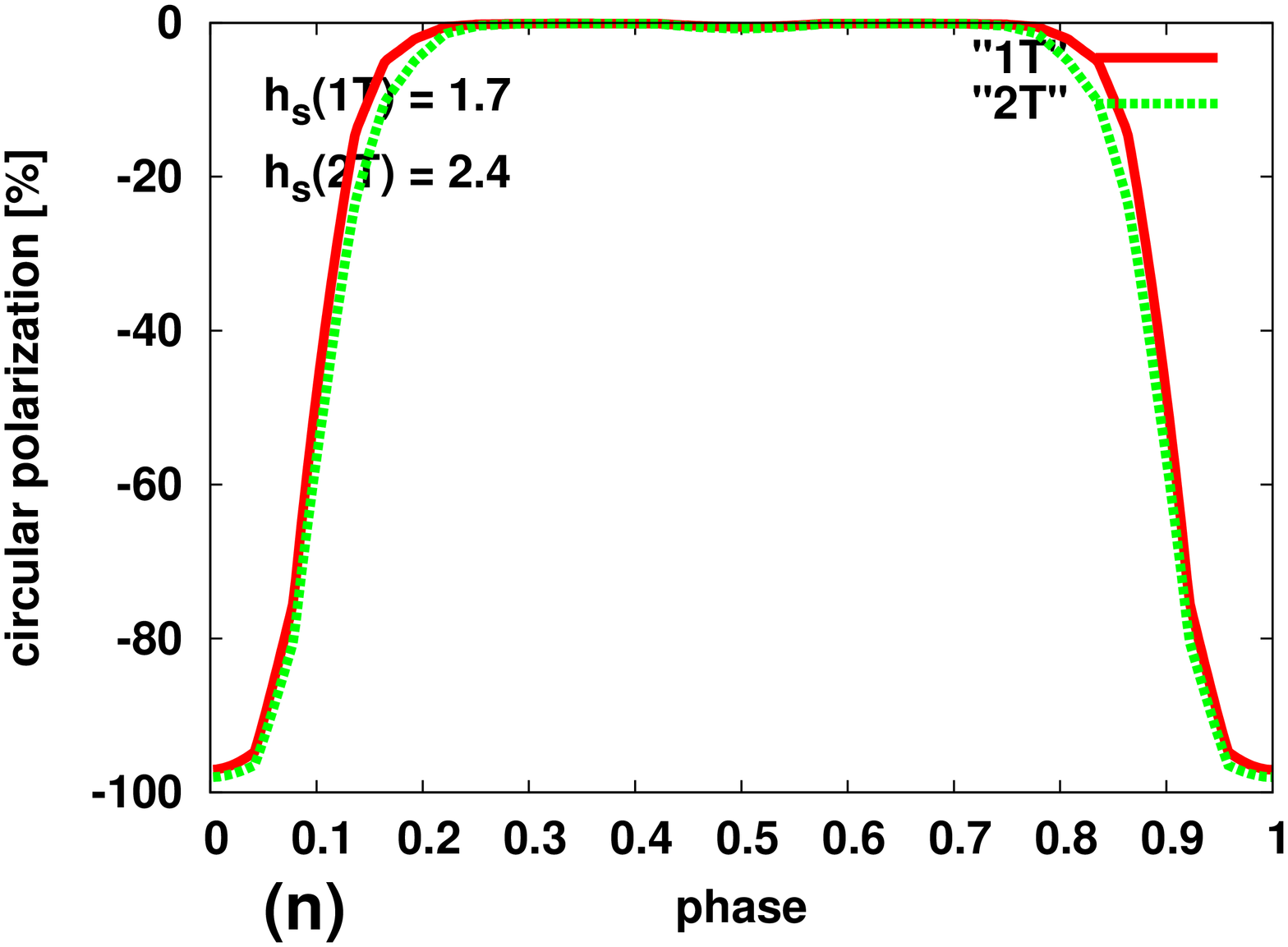}
\includegraphics[angle=-90,scale=0.16]{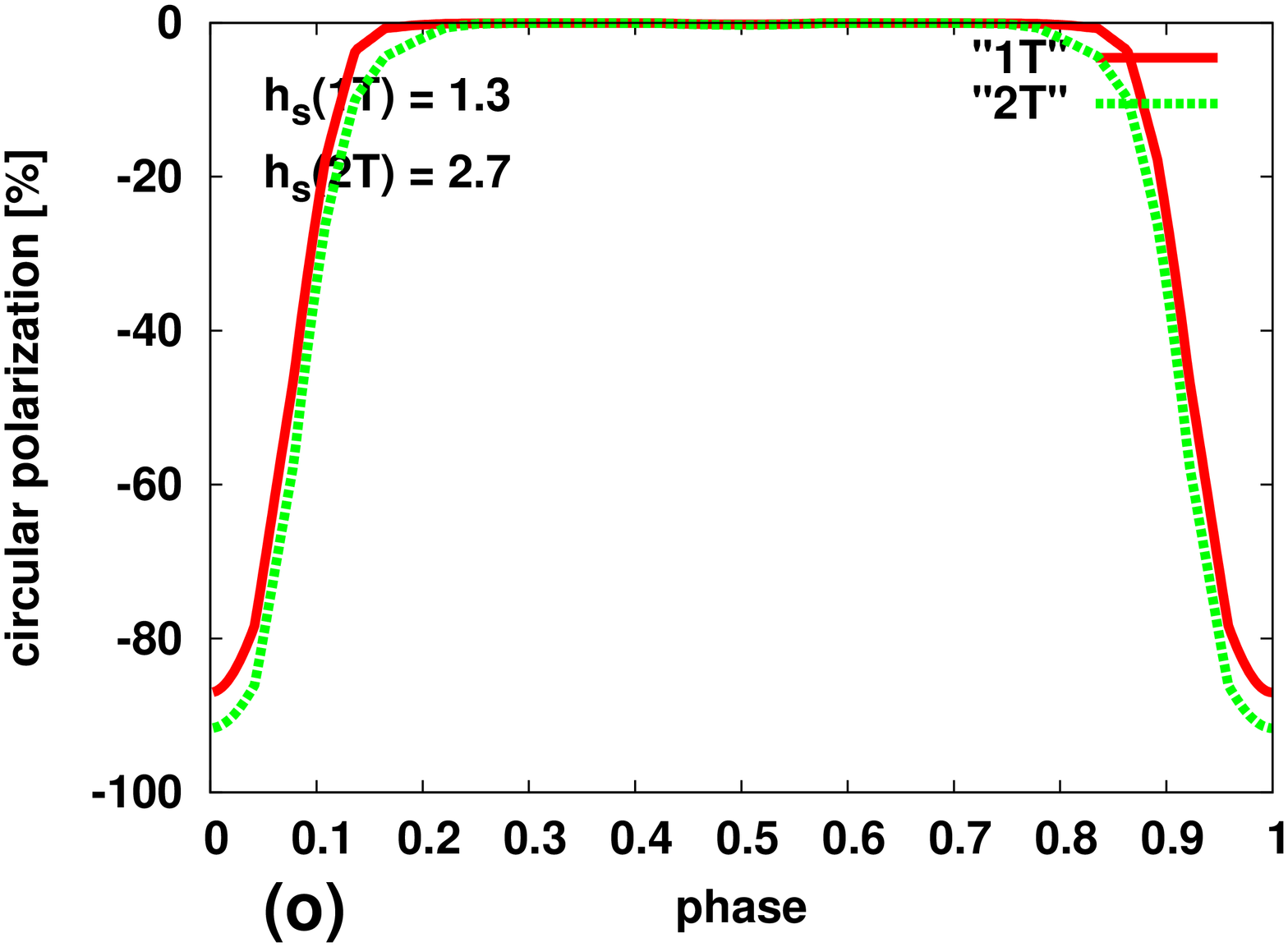}
\begin{picture}(25,100)(0,0)\put(0,-40){\scriptsize \bf \boldmath $\dot{M} = 10^{14}$ \unboldmath}
\end{picture}
\end{center}
\caption{Circular polarization orbital light curves comparing predictions for the
one-temperature model ({\tt "1T"}, red line) with the two-temperature ({\tt "2T"}, green line) model.
The figure is laid out in the same way as Fig.~\ref{linorb}. \label{cirorb}}
\end{figure*}

With the given choice of filters, the differences in linear polarization, linear polarization angle, intensity and circular polarization
    light curves are shown in Figs.~\ref{linorb} to \ref{cirorb}. 
The 1T and 2T predicted polarization angle and circular polarization light curves are generally the same, 
  but there are some differences for the high white dwarf mass, strong magnetic field cases in the ultraviolet. 
There are significant differences between the model predictions for the linearly polarized and intensity light curves in the ultraviolet. 
In two cases the 1T model predicts two pulses per orbit where the 2T model predicts one pulse, for both linear polarization and intensity, 
  with an increase in linear polarization and intensity at the time of superior conjunction 
   when we see the accretion column at the top of the white dwarf as seen from the observer. 
The two cases are: $B_{7} = 3, M_{\mbox{\tiny WD}} = 0.7 M_{\odot}$ and $B_{7} = 5, M_{\mbox{\tiny WD}} = 1.0 M_{\odot}$. 
However, in the visual V band there are no appreciable differences in the orbital light curves 
   predicted by the 1T and 2T models.

\section{Discussion}\label{discussion}

The emission of optical/infra-red polarized light from magnetic CVs is generally thought to arise 
  from the post-shock region of the magnetically funneled accretion flow onto the surface of the white dwarf. 
However the structure of that accretion column may be considerably more complex than the simple cylinder considered here. 
For example, it is probable that the threading of the gas stream from the $L_{1}$ point 
  onto the magnetic field lines at the magnetosphere boundary happens over a range of orbital longitudes. 
This spread of the threading region leads to a spread in the footprint of the accretion column into arcs  \citep{ferrario90}
    or possibly more complex patterns \citep{potter98a,potter98b,potter04} on the surface of the white dwarf. 
The complexity of the accretion column has undoubtedly contributed to the difficultly in modeling a good fit to observed polarimetric
   light curve data  (e.g. \citet{cropper89,piirola90,wickramasinghe91,potter97}\\ \citet{buckley00}), especially to linear polarization data. 
Here we have shown that the detail of modeling the post-shock flow as a 1T or 2T flow can have a substantial effect on the predicted polarization, particularly at lower mass flow rates and at magnetic fields strengths typical for polars \citep{warner95}. This
difference in predicted radiation would matter, in particular, for the lower specific mass flow rates required in models of extended accretion regions, especially at the edges of the accretion regions. 

For the lower mass flow rate of 0.5 g cm$^{-2}$s$^{-1}$, 
  we found that the flow was essentially 1T for $B_{7} = 1$ and 2T for the stronger fields, $B_{7} = 3$ and 5. 
This may be understood on the basis of the relative magnitudes of the cyclotron cooling times 
   and the electron-ion energy exchange times $t_{\mbox{\small cy}}$ and $t_{\mbox{\small ei}}$ \citep{lamb79}.
Two-temperature flows are expected to occur  
  for a larger magnetic field given a fixed mass flow rate or for smaller mass accretion rates given a fixed magnetic field 
  because the cyclotron cooling efficiency increases 
  relative to the electron-ion energy exchange efficiency in those circumstances. 
In line with those expectations, larger differences in the predicted orbitally-phased light curves, 
  between the 1T and 2T cases, are seen at stronger magnetic fields where the overall cyclotron cooling is more efficient.  

For differences in predicted linear polarization pulses, particularly note the cases of Fig.~\ref{linorb} (e) and (i), 
 which show one predicted linear polarization pulse per orbit for the ``correct'' 2T flow 
 versus two predicted linear polarization pulses per orbit when an inappropriate 1T flow is assumed. 
In that case one would incorrectly interpret an observation of two pulses per orbit as coming from one accretion column, 
  caused by viewing the accretion column sticking out from the white dwarf limb, if a 1T flow was used to model the data. 
In actuality, for a generic white dwarf of mass 0.7 $M_{\odot}$ and magnetic field of 30 MG, 
  two linear polarization pulses could be produced by two accretion columns, 
  one at each magnetic pole of the white dwarf. 
This consideration is important in indirect imaging cases where one is trying to model the white dwarf surface emission 
  and determine the magnetic field configuration  
  through a forward-modeling approach to the inversion of polarized light curve data 
  in a technique like `Stokes imaging' \citep{potter98a}. 
To date Stokes imaging has assumed a homogeneous temperature and density over the height of the column 
  but we have shown that the assumption of a 1T or 2T flow in the column 
  can significantly alter the amount linearly polarized light that a forward model would give, 
  especially at shorter wavelengths. 
More accurate Stokes imaging models may need to incorporate, somehow, 
 the fact that the flow will be 2T, especially at the rim of an accretion spot, 
 where the specific accretion rates may be low \citep{achilleos92}.  

On the positive side, for Stokes imaging, is that the differences in emission predicted from 1T and 2T columns is insignificant 
  in all cases for lower cyclotron harmonics in the infra-red and optical. 
So for infra-red and optical frequency light curves, as have been used for past applications of Stokes imaging, 
 it may not be necessary to model a two-temperature flow. 
However we should note that our grid of models does not cover field strengths between $B_{7}=1$ and $B_{7} = 3$ 
 where there may be small 2T effects at optical wavelengths. Also the variation of the accretion spot area, $A$, to larger values would increase 2T effects by reducing the specific accretion rate.
In any case, the addition of ultraviolet light curve data to optical light curve data for modeling the emission from polars 
  may give a truer image of the white dwarf surface. Such polarized UV data may become available from future space observatories like the proposed World Space Observatory for Ultraviolet (WSO-UV) \citep{pagano2007,uslenghi2008}.
We note, too, that there are essentially no differences in the prediction of circular polarization light curves between 1T and 2T models. 
Only linear polarization differences are apparent. 
Linear polarization light curves were more sensitive to assumptions made about the accretion flow than circular polarization light curves. 
This distinction between predicted linear and circularly polarized emission is consistent with past modeling of observational data,
   which either explicitly or implicitly assume one-temperature hydrodynamics,
   where the fits to linear polarization data are generally poorer than fits to the circular polarization data 
(e.g. \citet{barrett84,cropper89,potter97,buckley00}).

\section{Conclusion}\label{conclusion}

Assuming a one-temperature hydrodynamic post-shock accretion column as the source for polarized
radiation in models of magnetic cataclysmic variables can lead to erroneous predictions for the radiation
when the cyclotron cooling efficiency is greater than the electron-ion energy exchange efficiency. This effect shows up
at the lower mass flow rate modeled here ($\dot{m} = 0.5$ g cm$^{-2}$s$^{-1}$) at higher cyclotron harmonics for
the fairly generic white dwarf masses of 0.7 and 1.0 $M_{\odot}$ and magnetic fields of 30 MG or greater. So the interpretation
of light curve data obtained at higher frequencies, in the ultraviolet, needs to take into account the effect of two-temperature
flow on the production of polarized radiation. In particular, the number of linear polarization pulses observed in the
ultraviolet can be misinterpreted if a one-temperature accretion column flow is assumed.

\acknowledgments
GES is supported by a discovery grant  from the Natural Sciences and Engineering Research Council of Canada (NSERC) 
and he thanks the Mullard Space Science Laboratory for their hospitality during his visits in 2005 and 2006. 
The Beowulf cluster used for this study was constructed from recycled computers 
donated by the Arts and Science Computer Laboratory of the University of Saskatchewan.


\begin{thebibliography}{99} 

\bibitem[Achilleos, Wickramasinghe and Wu(1992)]{achilleos92}
Achilleos N., Wickramasinghe D.T., Wu K., MNRAS, {\bf 256}, 80 (1992) 

\bibitem[Barrett and Chanmugam(1984)]{barrett84}
Barrett P.E., Chanmugam G., ApJ,  {\bf 278}, 198 (1984) 

\bibitem[Bessell(1990)]{bessell90} 
Bessell M.S., PASP, {\bf 102}, 1181 (1990)

\bibitem[Buckley et al.(2000)]{buckley00}
Buckley D.A.H., Cropper M., van der Heyden K., Potter S.B., Wickramasinghe D.T., 
MNRAS, {\bf 318}, 187 (2000)


\bibitem[Chanmugam and Dulk(1981)]{chanmugam81}
Chanmugam G., Dulk G.A., ApJ, {\bf 244}, 569 (1981)


\bibitem[Cropper(1989)]{cropper89}
Cropper M., MNRAS, {\bf 236}, 935 (1989)

\bibitem[Fabian, Pringle and Rees(1976)]{fabian76}
Fabian A.C., Pringle J.E., Rees M.J., MNRAS, {\bf 173}, 43 (1976)

\bibitem[Ferrario and Wickramasinghe(1990)]{ferrario90}
Ferrario L., Wickramasinghe D.T., ApJ, {\bf 357}, 582 (1990) 

\bibitem[Imamua et al.(1996)]{imamura96} 
Imamura J., Abosaha A., Wolff M. T., Wood K. S., ApJ, {\bf 458}, 327 (1996)

\bibitem[Johnson(1965)]{johnson65}
Johnson H.L., ApJ, {\bf 141}, 923 (1965)

\bibitem[King and Lasota(1979)]{king79}
King A.R., Lasota J.P., MNRAS, {\bf 188}, 653 (1979)

\bibitem[Lamb and Masters(1979)]{lamb79} 
Lamb D.Q., Masters A.R., ApJ, {\bf 234}, L117 (1979)

\bibitem[Meggitt and Wickramasinghe(1982)] {meggitt82}
Meggitt S.M.A., Wickramasinghe D.T., MNRAS, {\bf 198}, 71 (1982)

\bibitem[Nauenberg(1972)]{nauenberg72} 
Nauenberg M., ApJ, {\bf 175}, 417 (1972) 

\bibitem[Pacholczyk and Swihart(1975)]{pacholczyk75} 
Pacholczyk A.G., Swihart T.L., ApJ, {\bf 196}, 125 (1975) 

\bibitem[Pagano et al.(2007)]{pagano2007}
Pagano I., Shustov B., Kappelmann, N., de Martino, D., Piotto, G., Scuderi, S., and Turatto M., ``WSO/UV: The World Space Observatory Project for the Ultraviolet'' in Proceedings Series of the Italian Physical Society,  F. Giovannelli and G. Mannocchi (eds.), {\bf 93}, 691 (2007) 

\bibitem[Piirola et al.(1990)]{piirola90} 
Piirola V., Coyne G.V., Reiz A., A\&A, {\bf 235}, 245 (1990) 

\bibitem[Potter et al.(1997)]{potter97}
Potter S.B., Cropper M., Mason K.O., Hough J.H., Bailey J.A., MNRAS, {\bf 285}, 82 (1997) 

\bibitem[Potter, Hakala and Cropper(1998)]{potter98a}
Potter S.B., Hakala P.J., Cropper M., MNRAS, {\bf 297}, 1261 (1998)

\bibitem[Potter, Hakala and Cropper(2000)]{potter98b}
Potter S.B., Hakala P.J., Cropper M., MNRAS, {\bf 315}, 423 (2000)

\bibitem[Potter et al.(2002)]{potter02}
Potter S., Ramsay G., Wu K., Cropper M., ASPC, {\bf 261}, 165 (2002)

\bibitem[Potter et al.(2004)]{potter04} 
Potter S.B., Romero-Colmenero E., Watson C.A., Buckley D.A.H., Phillips A., MNRAS, 
{\bf 348}, 316 (2004)

\bibitem[Rybicki and Lightman(1979)]{rybicki79} 
Rybicki G.B., Lightman A.P., Radiative Processes in Astrophysics, Wiley-Interscience, New York, 1979 

\bibitem[Saxton and Wu(2001)]{saxton01}
Saxton C.J., Wu K., MNRAS, {\bf 324}, 659 (2001) 

\bibitem[Saxton et al.(2005)]{saxton05} 
Saxton C.J., Wu K., Cropper M., Ramsay G., MNRAS, {\bf 360}, 1091 (2005)

\bibitem[Uslenghi et al.(2008)]{uslenghi2008}
Uslenghi M., Pagano I., Pontoni C., Scuderi S., Shustov B., ChJAA, in press (astro-ph/arXiv:0801.2080) (2008) 

\bibitem[Warner(1995)]{warner95}
Warner B., Cataclysmic Variable Stars, Cambridge University Press, Cambridge, 1985

\bibitem[Wickramasinghe et al.(1991)]{wickramasinghe91}
Wickramasinghe D.T., Bailey J., Meggitt S.M.A., Ferrario L., Hough J., Tuohy I.R., MNRAS, {\bf 251}, 
  28 (1991)

\bibitem[Wickramasinghe and Meggitt(1985)]{wickramasinghe85}
Wickramasinghe D.T., Meggitt S.M.A., MNRAS, {\bf 214}, 605 (1985) 

\bibitem[Woelk and Beuermann(1996)]{woelk96} 
Woelk U., Beuermann K., A\&A, {\bf 306}, 232 (1996) 


\bibitem[Wu(2000)]{wu00} 
Wu K., Space Sci. Rev., {\bf 93}, 611(2000) 

\bibitem[Wu and Chanmugam(1988)]{wu88} 
Wu K., Chanmugam G., ApJ, {\bf 331}, 861 (1988) 

\bibitem[Wu and Chanmugam(1989)]{wu89} 
Wu K., Chanmugam G., ApJ, {\bf 344}, 889 (1989) 

\bibitem[Wu et al.(2003)]{wu03}
Wu K., Cropper M., Ramsay G., Saxton C.J., Bridge C., ChJAA, {\bf 3} (Suppl.), 
   235 (2003) 

\bibitem[Wu and Wickramasinghe(1990)]{wu90} 
Wu K., Wickramasinghe D.T., MNRAS, {\bf 246}, 686 (1990)

\bibitem[Wu and Wickramasinghe(1992)]{wu92} 
Wu K., Wickramasinghe D.T., MNRAS, {\bf 256}, 329 (1992)

\end{thebibliography}
\end{document}